\documentclass[%
 reprint,
 amsmath,amssymb,
 aps,
]{revtex4-2}

\usepackage[mathlines]{lineno} % Enable numbering of text and display math
\usepackage{booktabs}
\usepackage{graphicx} % Include figure files
\usepackage{dcolumn} % Align table columns on decimal point
\usepackage{tikz}
\usepackage{amsmath} % for \boldsymbol

\usetikzlibrary{quantikz2}
\usepackage{enumitem}
\usepackage{multirow}
\usepackage{comment}
\usepackage{amsthm}

\usepackage{float}
\usepackage{mathrsfs}
\usepackage{dsfont}
\usepackage{tabularx}
\usepackage{adjustbox}
\usepackage{booktabs}

\newtheorem{prop}{Proposition}

\usepackage{mathtools}
\DeclarePairedDelimiter\bratwo{\langle}{\rvert}
\DeclarePairedDelimiter\kettwo{\lvert}{\rangle}
\DeclarePairedDelimiter\abstwo{\lvert}{\rvert}
\DeclarePairedDelimiter\normtwo{\lvert\lvert}{\rvert\rvert}
\DeclarePairedDelimiterX\ketbratwo[2]{\langle}{\rangle}{#1\,\delimsize\vert\,\mathopen{}#2}

\DeclareMathOperator*{\argmax}{argmax}
\DeclareMathOperator*{\argmin}{argmin}
\newcommand{\mathcb}[1]{{\boldsymbol{\mathcal{#1}}}}

\usepackage{bm} % bold math
\begin{document}
\preprint{APS/123-QED}
\title[Quantum generative classification with mixed states]{Quantum generative classification with mixed states}

\author{Diego H. Useche$^1$$^{,*}$ , Sergio Quiroga-Sandoval$^2$, Sebastian L. Molina$^1$, Vladimir Vargas-Calderón$^3$, Juan E. Ardila-García $^4$, Fabio A. González $^1$$^{,\dag}$}

\affiliation{$^1$ MindLab Research Group, Universidad Nacional de Colombia, 111321, Bogotá, Colombia}

\affiliation{$^2$ Departamento de Matemáticas, Universidad Nacional de Colombia, 111321, Bogotá, Colombia}

\affiliation{$^3$ D-Wave Systems, Burnaby, British Columbia, Canada}

\affiliation{$^4$ Grupo de Superconductividad y Nanotecnología, Departamento de Física, Universidad Nacional de Colombia, 111321, Bogotá, Colombia}

\email{$^{*}$diusecher@unal.edu.co, $^{\dag}$fagonzalezo@unal.edu.co}
\date{\today}

\begin{abstract}

Classification can be performed using either a discriminative or a generative learning approach. Discriminative learning consists of constructing the conditional probability of the outputs given the inputs, while generative learning consists of constructing the joint probability density of the inputs and outputs. Although most classical and quantum methods are discriminative, there are some advantages of the generative learning approach. For instance, it can be applied to unsupervised learning, statistical inference, uncertainty estimation, and synthetic data generation. In this article, we present a quantum generative multiclass classification strategy, called quantum generative classification (QGC). This model uses a variational quantum algorithm to estimate the joint probability density function of features and labels of a data set by means of a mixed quantum state. We also introduce a quantum map called quantum-enhanced Fourier features (QEFF), which leverages quantum superposition to prepare high-dimensional data samples in quantum hardware using a small number of qubits. We show that the quantum generative classification algorithm can be viewed as a Gaussian mixture that reproduces a kernel Hilbert space of the training data. In addition, we developed a hybrid quantum-classical neural network that shows that it is possible to perform generative classification on high-dimensional data sets. The method was tested on various low- and high-dimensional data sets including the 10-class MNIST and Fashion-MNIST data sets, illustrating that the generative classification strategy is competitive against other previous quantum models.

\end{abstract}

%\keywords{Suggested keywords}%Use showkeys class option if keyword
                              %display desired
\maketitle

%\tableofcontents

\section{Introduction\label{sec:introduction}}

Classification models can be trained using discriminative or generative learning. In the discriminative approach, we learn a function that approximates the conditional probability of the outputs given the inputs $p(\mathbf  y|\mathbf{x})$, while in the generative approach we approximate the joint probability of the inputs and the outputs $p(\mathbf x, \mathbf y)$, where $\mathbf x$ and $\mathbf y$ are random variables that correspond to characteristics and labels, respectively. Classification is made based on the class with the highest conditional probability for discriminative models and the highest joint probability for generative models. Most supervised learning algorithms are discriminative \cite{boser1992training, rumelhart1986learning, lecun1998gradient, Berkson1944Logistic, Breiman2001Forest}, generally reporting better predictive performance \cite{bernardo2007generative} and easier trainability compared to generative algorithms \cite{harshvardhan2020comprehensive}. Nonetheless, some advantages of generative models \cite{kingma2022autoencodingvariationalbayes, Ackley1985boltzmann, Papamakarios2021Normalizing, Rasmussen2004, 2014GoodfellowGAN, BaumMarkov1966} include the potential to generate new instances that resemble the original data \cite{harshvardhan2020comprehensive}, the ability to estimate the uncertainty of the predictions \cite{Jebara2004Generative}, and their feasibility to be trained with unlabeled data \cite{bernardo2007generative}.

Although most supervised learning algorithms have been developed for classical computers \cite{boser1992training, rumelhart1986learning, lecun1998gradient, Breiman2001Forest, Ackley1985boltzmann, Rasmussen2004}, there is growing interest in using quantum computers for classification. Most of these quantum models fall into the category of discriminative learning \cite{schuld2020circuit, farhi2018classification, watkins2023quantum, altares2021automatic, havlivcek2019supervised, park2023variational}, while there are only a few quantum algorithms for generative classification \cite{Bendetti2017quantumassisted, Zoufal2021, Chaudhary_2023, Useche2021QuantumQudits, vargas2022optimisation}. This recent enthusiasm for quantum machine learning is driven by the fact that it is possible to exploit the probabilistic nature of quantum mechanics for machine learning, and that some works \cite{harrow2009quantum, shor1994algorithms} have shown that quantum computers could potentially reduce the computational complexity of some classically hard problems.

In this article, we present a variational quantum algorithm for classification called quantum generative classification (QGC) that follows the generative learning approach. It uses a variational quantum circuit to represent the joint probability of inputs and outputs $p(\mathbf x, \mathbf y)$ through a mixed quantum state. The proposed algorithm maps the features and labels of a training data set $\mathscr{D} = \{(\boldsymbol{x}_j, \boldsymbol{y}_j)\}_{0 \cdots N-1}$  to quantum states $\{\boldsymbol{x}_j\} \mapsto \{\kettwo{\psi_{_\mathcb{X}}^{j}
}\}$,  $\{\boldsymbol y_j\} \mapsto \{\kettwo{\psi_{_\mathcb{Y}}^{j}}\}$ and uses these quantum features to train a variational quantum circuit that prepares a three component entangled pure state $\kettwo{q_{_{\mathcb{A}, \mathcb{X}, \mathcb{Y}}}({\boldsymbol \theta})}$, where the subscripts indicate the Hilbert spaces of outputs ($\mathcal{Y}$), inputs ($\mathcal{X}$) and auxiliary ($\mathcal{A}$) qubits, and $\boldsymbol \theta$ correspond to the parameters of the optimization, such that when tracing out the auxiliary qubits
\begin{align}
\rho_{{_\mathcb{X}}, {_\mathcb{Y}}}(\boldsymbol \theta) = \text{Tr}_{_\mathcb{A}}\big[\kettwo{q_{_{\mathcb{A}, \mathcb{X}, \mathcb{Y}}}({\boldsymbol \theta})}\bratwo{q_{_{\mathcb{A}, \mathcb{X}, \mathcb{Y}}}({\boldsymbol \theta})}\big],
\end{align}
we obtain a mixed state of the data set $\mathscr D$ that approximates the joint probability of features and labels $p(\mathbf x, \mathbf y)$.

The proposed quantum generative classification method can also be viewed as a model that represents the data in a reproducing a kernel Hilbert space (RKHS) \cite{gonzález2024kdm, useche2022quantum, Schuld2021quantumkernel}, whose basis function corresponds to the Gaussian kernel. Indeed, it prepares the data in a quantum computer using a one-hot basis encoding for the outputs and a proposed quantum embedding called quantum-enhanced Fourier features (QEFF) for the inputs. The QEFF mapping is based on random Fourier features (RFF) \cite{Rahimi2009RandomMachines}, which approximates a Gaussian kernel in the data space through an inner product in the quantum feature space.  Following Refs. \cite{Schuld2021quantumkernel, gonzález2024kdm}, we show that a variational quantum model trained with these quantum features results in a Gaussian mixture \cite{reynolds2009gaussian} which represents a function in a RKHS that builds the joint probability of inputs and outputs. This probabilistic feature enables the evaluation of the model's uncertainty, which has multiple applications, including the estimation of the confidence in the classification of medical images \cite{TOLEDOCORTES2022105472}.

Furthermore, we demonstrate the feasibility of integrating the variational quantum method with a classical deep neural network using a quantum simulator from the \textit{tensor circuit} library \cite{Zhang2023tensorcircuit}. This end-to-end quantum-classical strategy allows us to achieve competitive results on the classification of the 10-classes MNIST and 10-classes Fashion MNIST image data sets. In summary, the contributions of the article are as follows:

\begin{itemize}
	\item We present a variational quantum classification method that uses a generative learning approach.
	\item We present the quantum-enhanced Fourier features mapping, which can be viewed as a quantum analog of random Fourier features \cite{Rahimi2009RandomMachines}.
    \item We present an end-to-end quantum-classical strategy for classification that combines a quantum generative classifier with deep neural networks.
\end{itemize}

The structure of the article is as follows: in Sect. \ref{sec: Related works} we present the related works, in Sect. \ref{sec: Quantum generative classification}, we describe the proposed variational quantum generative classification model, in Sect. \ref{sec: D-QGC}, we describe the deep quantum classical classification model, in Sect. \ref{sec: Results}, we present the evaluation of the method and the comparison with other state-of-the-art methods, and in Sect. \ref{sec: Conclusions}, we discuss the conclusions and future work.

\section{\label{sec: Related works} Related work}

Generative machine learning has been widely studied in classical computers, of which some popular models include hidden Markov models \cite{BaumMarkov1966}, Gaussian mixture models \cite{Rasmussen2004}, variational autoencoders \cite{kingma2022autoencodingvariationalbayes}, Boltzmann machines \cite{Ackley1985boltzmann}, generative adversarial networks \cite{2014GoodfellowGAN}, and normalizing flows \cite{Papamakarios2021Normalizing}. Quantum generative learning is a more recent area that aims to leverage quantum computers to build interpretable and probabilistic generative machine learning models. Some of these quantum algorithms include quantum generative adversarial networks \cite{Chaudhary_2023, zoufal2019quantum, romero2021variational}, quantum circuit Born machines \cite{benedetti2019generative}, quantum Gaussian mixture models \cite{miyahara2023quantum, kerenidis2020quantum}, quantum Boltzmann machines \cite{amin2018quantum, Zoufal2021, Bendetti2017quantumassisted}, quantum measurement classification (QMC) \cite{Gonzalez2021Classification, gonzalez2022learning, Useche2021QuantumQudits, useche2022quantum, vargas2022optimisation}, and kernel density matrices (KDM) \cite{gonzález2024kdm}. 

Out of the previously presented methods, the proposed quantum generative classification algorithm shares some key similarities and differences with the models presented in Refs. \cite{amin2018quantum, Zoufal2021, Gonzalez2021Classification, gonzalez2022learning, Useche2021QuantumQudits, gonzález2024kdm, Schuld2021quantumkernel}. For instance, Refs. \cite{amin2018quantum, Zoufal2021} proposed a quantum Boltzmann machine that learns the joint probability $p(\mathbf x, \mathbf y)$ based on the preparation of a mixed Gibbs state in a quantum computer. One key feature of Boltzmann machines is that they were designed to learn discrete probability distributions, unlike our proposed QGC model, which can be applied to both continuous and discrete probability distributions. In addition, the presented work is closely related to the QMC and the QMC-SGD algorithms. Indeed, the QMC method \cite{Gonzalez2021Classification, Useche2021QuantumQudits, vargas2022optimisation} combines density matrices and RFF to learn a nonparametric probability density function of inputs and outputs in both classical and quantum computers. Although the QGC algorithm also combines density matrices and random features, it uses the variational quantum approach for classification, increasing its flexibility and allowing its integration with classical neural networks. In relation to the previous work, the classical QMC-SGD model by Gonzalez \textit{et al.} \cite{gonzalez2022learning} uses stochastic gradient descent to discriminately optimize the QMC algorithm. Instead of a classical discriminative model, our algorithm consists of a variational quantum circuit that can be used for both generative and discriminative learning. Additionally, the classical KDM method \cite{gonzález2024kdm} also combines kernels and density matrices for generative modeling. Besides being a classical algorithm, the KDM model uses an explicit representation of the kernel instead of an induced quantum feature space, which imposes a different type of constraints on the expressiveness of the model compared to the QGC strategy. Furthermore, we also present a connection between mixed states and kernels, and the representer theorem \cite{scholkopf2002learning} for inducing a function from the data in RKHS; this link was first made in Ref. \cite{Schuld2021quantumkernel} but not in the context of generative learning.

In this article, we also introduce the quantum-enhanced Fourier features strategy, which leverages quantum hardware to build a quantum embedding based on random Fourier features \cite{Rahimi2009RandomMachines}. RFF is a method that maps the original data into a feature space, such that the inner product in the feature space approximates a shift-invariant kernel in the original space; in particular, the QEFF approximates the Gaussian kernel. Random features have been applied to various areas of quantum machine learning, including quantum classification \cite{Gonzalez2021Classification, gonzalez2022learning, Useche2021QuantumQudits, vargas2022optimisation, sadowski2019machine}, quantum regression \cite{Peters2023generalization, landman2022classically, sweke2023potential}, and quantum density estimation \cite{gonzalez2022learning, Useche2021QuantumQudits, useche2022quantum, vargas2022optimisation, ardila2025memo}. Other studies have explored an opposite approach, e.g., Ref. \cite{landman2022classically} proposed the use of RFF to efficiently translate some QML models to classical computers. However, as noted in Ref. \cite{sweke2023potential}, they can not be used to dequantize some highly expressive quantum models. In addition, Refs. \cite{sadowski2019machine, useche2022quantum, Yamasaki2020, Gil-Fuster2024} have proposed some quantum strategies for improving the RFF's Monte Carlo sampling, of which Ref. \cite{Yamasaki2020} has presented a possible quantum advantage. Out of these methods, the QEFF mapping is closely related to the quantum random (QRFF) and quantum adaptive Fourier features (QAFF) introduced by Useche \textit{et al.} \cite{useche2022quantum}, see Sect. \ref{sec: QEFF}. In fact, the QRFF mapping is constructed using amplitude encoding \cite{mottonen2005transformation, shende2006synthesis}, while the QEFF uses the Pauli basis to enhance this quantum mapping. Accordingly, we use the QAFF quantum circuit to prepare the randomized QEFF embedding, in contrast to Ref. \cite{useche2022quantum}, where the circuit is used to construct the QAFF mapping as a parameterized quantum map.

\section{\label{sec: Quantum generative classification}Quantum generative classification}

\begin{figure*}
\includegraphics[scale=0.52]{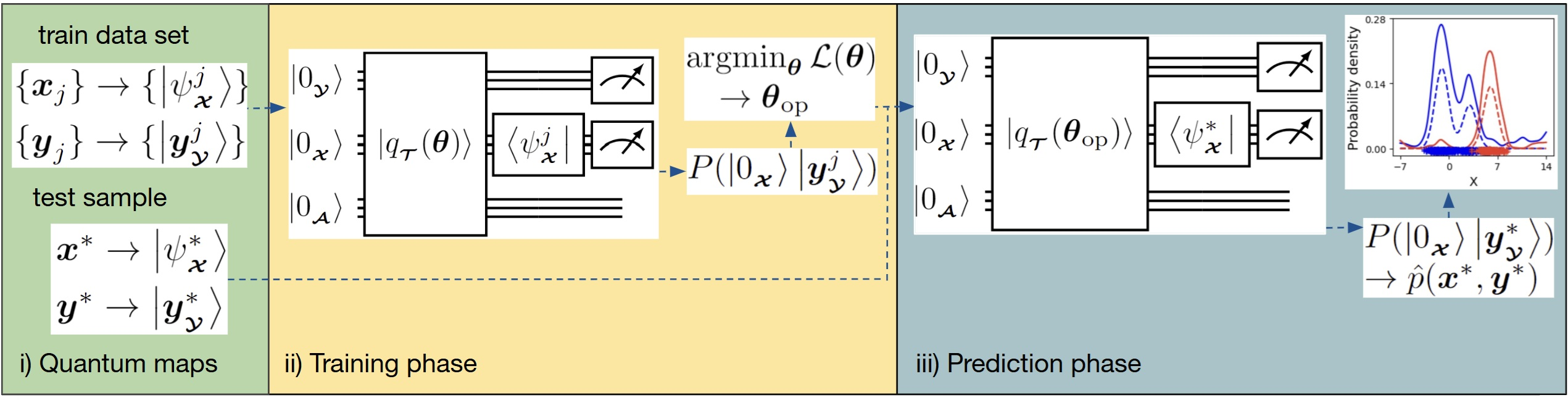}
\caption{QGC algorithm for quantum classification based on the joint probability density of features and labels. In step i), we perform a quantum map based on QEFF for the inputs and one-hot basis encoding for the outputs, in step ii), we use a variational quantum circuit to minimize the negative log-likelihood ${\mathcal{L}(\boldsymbol{\theta})}\sim -\sum_j\log{\bratwo{\psi_{_\mathcb{X}, _\mathcb{Y}}^j}\rho_{{_\mathcb{X}}, {_\mathcb{Y}}}(\boldsymbol\theta)\kettwo{\psi_{_\mathcb{X}, _\mathcb{Y}}^j}}$ of the projection between the training quantum states and a parametrized mixed state $\rho_{{_\mathcb{X}}, {_\mathcb{Y}}}(\boldsymbol \theta)= \text{Tr}_{_\mathcb{A}}\big[\kettwo{q_{_\mathcb{T}}(\boldsymbol \theta)}\bratwo{q_{_\mathcb{T}}(\boldsymbol \theta)}\big]$ (this state results from tracing out the ancilla qubits), in step iii), we estimate the joint probability density by projecting the optimized density matrix with the joint quantum state of the test input sample and a candidate test label $\hat p(\boldsymbol{x}^*, \boldsymbol y^*)\sim\bratwo{\psi_{_{\mathcb{X}}, _\mathcb{Y}}^*}\rho_{{_{\mathcb{X}}}, {_\mathcb{Y}}}(\boldsymbol\theta_{\text{op}})\kettwo{\psi_{_{\mathcb{X}}, _\mathcb{Y}}^*}$, finally, we make a prediction based on the class with largest joint probability density $\argmax_{{\boldsymbol{y}^*}}\hat p(\boldsymbol{x}^*,\boldsymbol y^*)$.}{\label{fig: QGC method}}
\end{figure*}

The proposed variational quantum algorithm for generative classification builds an estimator $\hat p(\mathbf x , \mathbf y) \approx p(\mathbf x, \mathbf y)$ of the joint probability density of inputs and outputs and makes a prediction based on the label with the highest joint probability density, that is, $\argmax_{\boldsymbol{y}^*}\hat p(\mathbf{x}, \mathbf{y} = \boldsymbol y^*)$. The method consists of the following three steps; see Fig. \ref{fig: QGC method}: 

\begin{enumerate}[label=(\roman*)]
    \item Quantum feature map: We map the training features and labels to quantum feature states, using the QEFF mapping for the inputs $\{\boldsymbol{x}_j\}_{0 \cdots N-1} \mapsto \{\kettwo{\psi_{_\mathcb{X}}^j}\}_{0 \cdots N-1}$ and the one-hot basis encoding for the outputs $\{\boldsymbol y_j\}_{0 \cdots N-1} \mapsto \{\kettwo{\psi_{_\mathcb{Y}}^j}= \kettwo{\boldsymbol{y}_{_\mathcb{Y}}^j}\}_{0 \cdots N-1}$. Consequently, we apply the same quantum mappings to the test sample $\boldsymbol{x}^*\mapsto \kettwo{\psi_{_\mathcb{X}}^*}$ and a possible test label $\boldsymbol{y}^*\mapsto \kettwo{\psi_{_\mathcb{Y}}^*} = \kettwo{\boldsymbol{y}_{_\mathcb{Y}}^*}$. We denote the joint quantum states of both the training and test samples by $\{\kettwo{\psi_{_\mathcb{X}, _\mathcb{Y}}^j} = \kettwo{\psi_{_\mathcb{X}}^j}\otimes\kettwo{\psi_{_\mathcb{Y}}^j}\}_{0 \cdots N-1}$ and $\kettwo{\psi_{_\mathcb{X}, _\mathcb{Y}}^*} = \kettwo{\psi_{_\mathcb{X}}^*}\otimes\kettwo{\psi_{_\mathcb{Y}}^*}$.
    \item Training phase: We build a variational quantum algorithm that minimizes the average negative log-likelihood ${\mathcal{L}(\boldsymbol{\theta})}\sim -\sum_j\log{\bratwo{\psi_{_\mathcb{X}, _\mathcb{Y}}^j}\rho_{{_\mathcb{X}}, {_\mathcb{Y}}}(\boldsymbol\theta)\kettwo{\psi_{_\mathcb{X}, _\mathcb{Y}}^j}}$, with parameters $\boldsymbol{\theta}$, of the expectation value between the variational mixed state $\rho_{{_\mathcb{X}}, {_\mathcb{Y}}}(\boldsymbol \theta) = \text{Tr}_{_\mathcb{A}}\big[\kettwo{q_{_\mathcb{T}}(\boldsymbol \theta)}\bratwo{q_{_\mathcb{T}}(\boldsymbol \theta)}\big]$ and the quantum features of the training data; this state results by tracing out the ancilla qubits ($\mathcal{A}$) of the pure state $\kettwo{q_{_\mathcb{T}}(\boldsymbol \theta)}$, where $(\mathcal{T})$ denotes the full Hilbert space of the ancilla ($\mathcal{A}$), inputs ($\mathcal{X}$), and outputs ($\mathcal{Y}$). To perform this operation, we build a quantum circuit that compiles the pure state $\kettwo{q_{_\mathcb{T}}({\boldsymbol \theta})}$ over all the qubits, see Fig. \ref{fig: QGC method}, followed by the preparation of the quantum features of the inputs $\{\kettwo{\psi_{_\mathcb{X}}^j}\}$ in the Hilbert space $\mathcal{X}$. By measuring the qubits of the inputs $\mathcal{X}$ and outputs $\mathcal{Y}$ in the computational basis, the probability of the state $P( \kettwo{0_{_\mathcb{X}}} \kettwo{\boldsymbol{y}^j_{_\mathcb{Y}}})$, where $\kettwo{0_{_\mathcb{X}}} = \kettwo{0}^{\otimes n_{_\mathcb{X}}}$ is the all-zeros state and $\kettwo{\boldsymbol{y}^j_{_\mathcb{Y}}}$ the quantum feature of the $j^{\text{}th}$  training label, corresponds to the expected value $\bratwo{\psi_{_\mathcb{X}, _\mathcb{Y}}^j}\rho_{{_\mathcb{X}}, {_\mathcb{Y}}}(\boldsymbol\theta)\kettwo{\psi_{_\mathcb{X}, _\mathcb{Y}}^j}$. For notational purposes, the rightmost qubit in the equations corresponds to the uppermost qubit in the circuit diagrams.
    \item Testing phase: Once the optimal values of the optimization are obtained $\boldsymbol{\theta}_{\text{op}} = \argmin_{\boldsymbol{\theta}}{\mathcal{L}(\boldsymbol{\theta})}$, we make use of the previous quantum circuit to prepare the state $\kettwo{q_{_\mathcb{T}}({\boldsymbol \theta_{\text{op}}})}$, followed by the test quantum state $\kettwo{\psi_{_\mathcb{X}}^*}$ over the Hilbert space of inputs. By performing a quantum measurement of the qubits of the inputs and outputs, we find that the probability $P(\kettwo{0_{_\mathcb{X}}}\kettwo{\boldsymbol{y}^*_{_\mathcb{Y}}})=\bratwo{\psi_{_\mathcb{X}, _\mathcb{Y}}^*}\rho_{{_\mathcb{X}}, {_\mathcb{Y}}}(\boldsymbol\theta)\kettwo{\psi_{_\mathcb{X}, _\mathcb{Y}}^*}$ is an estimator of the joint up to a normalization constant $\hat p(\boldsymbol{x}^*, \boldsymbol y^*) \approx p(\boldsymbol{x}^*, \boldsymbol y^*)$. The predicted class would then be obtained from $\argmax_{\boldsymbol{y}^*}\hat p(\mathbf x = \boldsymbol{x}^*, \mathbf y =  \boldsymbol y^*)$.
\end{enumerate}

\subsection{\label{sec: circuit details} Quantum circuit details}

We now describe in detail the construction of the variational quantum generative classification algorithm.

\textit{Quantum feature maps:} Given a data set of $N$ training data points and their corresponding categorical labels $\mathscr D = \{(\boldsymbol{x}_j, \boldsymbol y_j)\}_{0\cdots N-1}$ with $\boldsymbol{x}_j\in \mathbb X = \mathbb{R}^D$ and $\boldsymbol y_j \in \mathbb Y =  \{0, \cdots, L-1\}$, and a test sample $\boldsymbol{x^*}\in \mathbb X$ to be classified onto one of the $L$ classes, the algorithm starts by applying the proposed quantum-enhanced Fourier features mapping $\psi_{_\mathcb{X}}:\mathbb{X}\rightarrow\mathcal{X}$ to the input samples $\boldsymbol{x}_j \mapsto \kettwo{\psi_{_\mathcb{X}}^j}$, $\boldsymbol{x}^* \mapsto \kettwo{\psi_{_\mathcb{X}}^*}$ and the one-hot-basis encoding $\psi_{_\mathcb{Y}}:\mathbb{Y}\rightarrow\mathcal{Y}$  to the labels $\boldsymbol y_j \mapsto \kettwo{\psi_{_\mathcb{Y}}^j} = \delta_{{\boldsymbol y}_j,k}\kettwo{k_{_\mathcb{Y}}}$, where $\kettwo{\psi_{_\mathcb{X}}} \in {\mathcal{X}} = \mathbb{C}^d$, $\kettwo{\psi_{_\mathcb{Y}}}  \in {\mathcal{Y}} = \mathbb{R}^L$, $\delta_{{\boldsymbol y}_j,k}$ is a Kronecker's delta, and $\kettwo{k_{_\mathcb{Y}}}$ is the decimal representation of the one-hot vector. The QEFF and the one-hot-basis feature map, $\boldsymbol x' \mapsto \kettwo{\psi_{_{\mathcb{X}}}'}$ and $\boldsymbol y' \mapsto \kettwo{\psi_{_\mathcb{Y}}'}=\kettwo{\boldsymbol{y}_{_\mathcb{Y}}'}$ approximate and induce, respectively, the Gaussian kernel $\kappa_{_{\mathbb{X}}}(\boldsymbol{x}', \boldsymbol{x}'')=e^{-\frac{1}{2h^2} \normtwo{\boldsymbol{x}'-\boldsymbol{x}''}^2}$ with bandwidth parameter $h \in \mathbb{R}$ and the Kronecker's delta kernel $\kappa_{_{\mathbb{Y}}}(\boldsymbol{y}', \boldsymbol{y}')=\delta_{\boldsymbol{y}', \boldsymbol{y}''}$ through an inner product in the quantum features space $\abstwo{\ketbratwo{\psi_{_{\mathcb{X}}}'}{\psi_{_{\mathcb{X}}}''}}^2 \approx \kappa_{_{\mathbb X}}(\boldsymbol{x}', \boldsymbol{x}'')$, $\abstwo{\ketbratwo{\psi_{_\mathcb{Y}}'}{\psi_{_\mathcb{Y}}''}}^2 = \kappa_{_{\mathbb Y}}(\boldsymbol{y}', \boldsymbol{y}')$ for any $\boldsymbol{x}', \boldsymbol{x}''\in \mathbb X$ and any $\boldsymbol{y}', \boldsymbol{y}''\in \mathbb Y$.

%$\boldsymbol y^* \mapsto \kettwo{\psi_{_\mathcb{Y}}^*} = \delta_{{\boldsymbol y}^*,k}\kettwo{k_{_\mathcb{Y}}}$
% $\boldsymbol{y}^*\in \mathbb{Y}$

\begin{figure*}[!tbp]
  \centering
  \begin{minipage}{0.48\textwidth}
    \centering
    \includegraphics[width=0.7\linewidth]{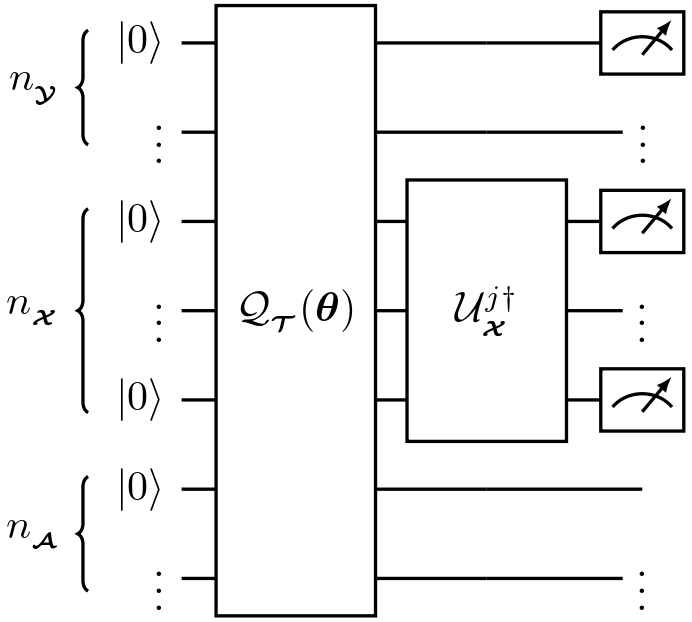}
    \caption{Training quantum circuit to estimate the expected value of the training density matrix and the $j^{\text{th}}$ training feature $\kettwo{\psi_{_\mathcb{X}, _\mathcb{Y}}^j} = \kettwo{\psi_{_\mathcb{X}}^j}\otimes\kettwo{\boldsymbol{y}^j_{_\mathcb{Y}}}$ by measuring the Hilbert spaces of inputs and outputs $P(\kettwo{0_{_\mathcb{X}}}\kettwo{\boldsymbol{y}^j_{_\mathcb{Y}}}) =\bratwo{\psi_{_\mathcb{X}, _\mathcb{Y}}^j}\rho_{{_\mathcb{X}}, {_\mathcb{Y}}}(\boldsymbol\theta)\kettwo{\psi_{_\mathcb{X}, _\mathcb{Y}}^j}$.}
    \label{fig: QMC training quantum circuit}
  \end{minipage}
  \hspace{0.02\textwidth} 
  \hfill
  \begin{minipage}{0.48\textwidth}
    \centering
    \includegraphics[width=0.7\linewidth]{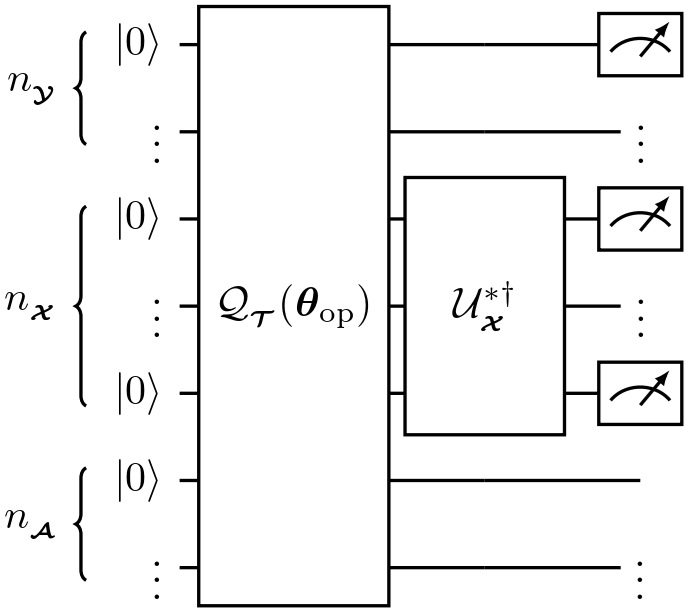}
    \caption{Test quantum circuit to estimate the joint probability of features and labels by computing the projection of the optimized density matrix and the test quantum sample $\kettwo{\psi_{_\mathcb{X}, _\mathcb{Y}}^*} = \kettwo{\psi_{_\mathcb{X}}^*}\otimes\kettwo{\boldsymbol{y}_{_{\mathcb Y}}^*}$ through a reading of the first $n_{_\mathcb{X}} + n_{_{\mathcb{Y}}}$ qubits $P(\kettwo{0_{_{\mathcb X}}}\kettwo{\boldsymbol{y}_{_{\mathcb Y}}^*})=\bratwo{\psi_{_{\mathcb{X}}, _\mathcb{Y}}^*}\rho_{{_{\mathcb{X}}}, {_\mathcb{Y}}}(\boldsymbol\theta_{\text{op}})\kettwo{\psi_{_{\mathcb{X}}, _\mathcb{Y}}^*}=M_h^{-1}\hat p(\boldsymbol{x}^*, \boldsymbol y^*)$.}
    \label{fig: QMC testing quantum circuit}
  \end{minipage}
\end{figure*}

\textit{Training quantum circuit:} The training step starts by preparing a variational pure quantum state $\kettwo{q_{_\mathcb{T}}(\boldsymbol{\theta})} =\mathcal{Q}_{_\mathcb{T}}(\boldsymbol{\theta})\kettwo{0_{_\mathcb{T}}}$ from the state of all zeros $\kettwo{0_{_\mathcb{T}}} = \kettwo{0}^{\otimes n_{_\mathcb{T}}}$, see Fig. \ref{fig: QMC training quantum circuit}. This state is constructed using a unitary matrix $\mathcal{Q}_{_\mathcb{T}}$ with parameters $\boldsymbol \theta$ and three entangled components: ancilla $(\mathcal{A}$), inputs $(\mathcal{X}$), and outputs ($\mathcal{Y}$), whose number of qubits $n_{_\mathcb{T}} = n_{_\mathcb{A}} + n_{_\mathcb{X}} + n_{_\mathcb{Y}}$ is given by $n_{_\mathcb{Y}} = \lceil \log{L} \rceil $, $n_{_\mathcb{X}} = \lceil \log{d} \rceil $, and $n_{_\mathcb{A}} \le n_{_\mathcb{Y}} + n_{_\mathcb{X}}$. It satisfies that its partial trace over the ancilla qubits results into mixed quantum state $\rho_{{_\mathcb{X}}, {_\mathcb{Y}}}(\boldsymbol \theta) = \text{Tr}_{_\mathcb{A}}\big[\kettwo{q_{_\mathcb{T}}(\boldsymbol \theta)}\bratwo{q_{_\mathcb{T}}(\boldsymbol \theta)}\big]$. Following the preparation of $\kettwo{q_{_\mathcb{T}}(\boldsymbol{\theta})}$, at each training iteration, we estimate the projection between the variational mixed state and the $j^{\text{th}}$ training feature $\bratwo{\psi_{_\mathcb{X}, _\mathcb{Y}}^j}\rho_{{_\mathcb{X}}, {_\mathcb{Y}}}(\boldsymbol\theta)\kettwo{\psi_{_\mathcb{X}, _\mathcb{Y}}^j}$, by applying the unitary matrix $\mathcal U_{_\mathcb{X}}^{j\dagger}$ on the $n_{_\mathcb{X}}$ qubits, such that $\mathcal U_{_\mathcb{X}}^j\kettwo{0_{_\mathcb{X}}} = \kettwo{\psi_{_\mathcb{X}}^j}$. By measuring the first $n_{_\mathcb{X}} + n_{_\mathcb{Y}}$ qubits the probability of the state $\kettwo{0_{_\mathcb{X}}}\kettwo{\boldsymbol{y}^j_{_\mathcb{Y}}}$ in the computational basis corresponds to tracing out the auxiliary qubits and estimating the expected value
\begin{align}
    & P(\ket{0_{_\mathcb{X}}}\ket{\boldsymbol{y}^j_{_\mathcb{Y}}}) = \notag \\
& \text{Tr}\Big[\text{Tr}_{_\mathcb{A}}\big(\ket{q_{_{\mathcb{A}, \mathcb{X}, \mathcb{Y}}}({\boldsymbol \theta})}\bra{q_{_{\mathcb{A}, \mathcb{X}, \mathcb{Y}}}({\boldsymbol \theta})}\big)\ket{\psi_{_\mathcb{X}, _\mathcb{Y}}^j}\bra{\psi_{_\mathcb{X}, _\mathcb{Y}}^j}\Big] = \notag \\
    & \bra{\psi_{_\mathcb{X}, _\mathcb{Y}}^j}\rho_{{_\mathcb{X}}, {_\mathcb{Y}}}(\boldsymbol\theta)\ket{\psi_{_\mathcb{X}, _\mathcb{Y}}^j}. \quad \label{eq: QGC training}
\end{align} 

We then use these estimations to minimize the average negative log-likelihood (ANL) loss function
\begin{equation}
   \mathcal L(\boldsymbol{\theta}) = -\frac{1}{N}\sum_{j=0}^{N-1}\log{\big(M_h\bratwo{\psi_{_{\mathcb{X}}, _{\mathcb {Y}}}^j}\rho_{_{\mathcb{X}}, {_{\mathcb{Y}}}}
(\boldsymbol\theta) \kettwo{\psi_{_{\mathcb{X}}, {_{\mathcb{Y}}}}^j}\big)},
\label{eq: negative log.likehoood}
\end{equation}
where $M_h = (2\pi h^2)^{-D/2}$ is a normalization constant related to the bandwidth of the Gaussian kernel $h$; for notational purposes, we alternatively use
\begin{equation}
    \hat{f}(\boldsymbol x', \boldsymbol{y}''|\boldsymbol \theta) = M_h\text{Tr}\big[\rho_{{_{\mathcb{X}}}, {_\mathcb{Y}}}(\boldsymbol{\theta})(\kettwo{\psi_{_{\mathcb{X}}}'}\bratwo{\psi_{_{\mathcb{X}}}'}\otimes\kettwo{\psi_{_{\mathcb Y}}''}\bratwo{\psi_{_{\mathcb Y}}''})\big],\label{eq: estimator of pdf}
\end{equation}
which normalizes and generalizes the parametrized estimator of Eq. \ref{eq: QGC training} for any $\boldsymbol x' \in \mathbb{X}$ and any $\boldsymbol y' \in \mathbb{Y}$.

In addition, we make use the parameter-shift rule \cite{mitarai2018quantum, schuld2019evaluating} to estimate the gradient of loss and obtain its optimal parameters $\boldsymbol{\theta}_{\text{op}}= \argmin_{\boldsymbol{\theta}} \mathcal L(\boldsymbol{\theta})$. These values then allows us to construct a mixed state  $\rho_{{_{\boldsymbol{\mathcal{X}}}}, {_\mathcb{Y}}}
(\boldsymbol \theta_{\text{op}})$ that summarizes the joint probability density of inputs and outputs of the training data set.

\textit{Testing quantum circuit:} To estimate the joint probability of the test sample $\boldsymbol{x}^* \mapsto \kettwo{\psi_{_{\mathcb X}}^*}$ and the class $ \boldsymbol{y}^* \mapsto \kettwo{\psi_{_{\mathcb Y}}^*} = \kettwo{\boldsymbol{y_{_{\mathcb Y}}}^*}$, we use the same structure of the previous quantum circuit, i.e., we initialize the state $\kettwo{q_{_\mathcb{T}}(\boldsymbol{\theta}_\text{op})} =\mathcal{Q}_{_\mathcb{T}}(\boldsymbol{\theta}_\text{op})\kettwo{0_{_\mathcb{T}}}$, see Fig. \ref{fig: QMC testing quantum circuit}, followed by the unitary matrix $\mathcal U_{_{\mathcb{X}}}^{*\dagger}$ that prepares the test state $\mathcal U_{_{\mathcb{X}}}^*\kettwo{0_{_\mathcb{X}}} = \kettwo{\psi_{_{\mathcb{X}}}^*}$ on the $n_{_{\mathcb{X}}}$ qubits and a quantum measurement over the $n_{_\mathcb{X}} + n_{_{\mathcb{Y}}}$ qubits. The probability $P(\kettwo{0_{_{\mathcb X}}}\kettwo{\boldsymbol{y}_{_{\mathcb Y}}^*}) = \text{Tr}\big[\rho_{{_{\mathcb{X}}}, {_\mathcb{Y}}}(\boldsymbol{\theta}_{\text{op}})(\kettwo{\psi_{_{\mathcb{X}}}^*}\bratwo{\psi_{_{\mathcb{X}}}^*}\otimes\kettwo{\boldsymbol{y}_{_{\mathcb Y}}^*}\bratwo{\boldsymbol{y}_{_{\mathcb Y}}^*})\big]=\bratwo{\psi_{_{\mathcb{X}}, _\mathcb{Y}}^*}\rho_{{_{\mathcb{X}}}, {_\mathcb{Y}}}(\boldsymbol\theta_{\text{op}})\kettwo{\psi_{_{\mathcb{X}}, _\mathcb{Y}}^*}$ builds an estimator of the joint probability density of inputs and outputs up to the normalization constant $ M_hP(\kettwo{0_{_{\mathcb X}}}\kettwo{\boldsymbol{y}_{_{\mathcb Y}}^*}) = \hat p(\boldsymbol{x}^*, \boldsymbol y^*)$. Hence, prediction is made based on the class with largest joint probability density
\begin{align}
         \argmax_{\kettwo{\boldsymbol{y}_{_{\mathcb Y}}^*}\in \mathcb{Y}}  P(\kettwo{0_{_{\mathcb X}}}\kettwo{\boldsymbol{y}_{_{\mathcb Y}}^*}) &= \argmax_{\kettwo{\boldsymbol{y}_{_{\mathcb Y}}^*}\in \mathcb{Y}}\bratwo{\psi_{_{\mathcb{X}}, _\mathcb{Y}}^*}\rho_{{_{\mathcb{X}}}, {_\mathcb{Y}}}(\boldsymbol\theta_{\text{op}})\kettwo{\psi_{_{\mathcb{X}}, _\mathcb{Y}}^*} \notag \\
       &= \argmax_{{\boldsymbol{y}^*}\in \mathbb{Y}}\hat p(\mathbf x = \boldsymbol{x}^*, \mathbf y = \boldsymbol y^*).
         \label{eq: VQGC prediction}
\end{align}

In the following sections, we further describe the theoretical details of the quantum generative classification algorithm.

\subsection{\label{sec: Classification with RKHS}Classification with reproducing kernel Hilbert spaces}

The presented QGC strategy belongs to the category of kernel methods \cite{boser1992training, vapnik1999support, rosenblatt1956remarks, parzen1962estimation}, whose relation to quantum algorithms has been explored in Refs. \cite{ Schuld2021quantumkernel, gonzález2024kdm}. In particular, our model approximates a function for classification in a reproducing kernel Hilbert space. Here, we describe a RKHS and some previous machine learning algorithms for classification and density estimation based on RKHS.

A RKHS is a Hilbert space consisting of functions $\textsl g(\boldsymbol x^*)$ whose domain is the data space $\textsl g : \mathbb X \rightarrow \mathbb R$ that result from combining linearly the functions of a kernel $\kappa_{_{\mathbb{X}}}: \mathbb X \times \mathbb X \rightarrow \mathbb R$ centered on the training data points $\{\boldsymbol{x}_j\}_{0\cdots N-1}$
\begin{equation}
	\textsl g(\boldsymbol x^*) = \sum_{j=0}^{N-1}\alpha_j\kappa_{_{\mathbb{X}}}(\boldsymbol x_j, \boldsymbol x^*), \label{eq: RKHS function}
\end{equation}
where $\alpha_j \in \mathbb R$ for all $j$. An example of a function in a RKHS is the kernel density estimation (KDE) method \cite{rosenblatt1956remarks, parzen1962estimation}.
\begin{equation}
	\textsl g_{\text{KDE}}(\boldsymbol{x}^*) = \frac{M_h}{N}\sum_{j=0}^{N-1}  e^{-\frac{1}{2h^2} \normtwo{\boldsymbol{x}_j-\boldsymbol{x^*}}^2}, \label{eq: kernel density estimation}
\end{equation}
where $\alpha_j = M_h/N$ for all $j$ and $\kappa_{_\mathbb{X}}(\boldsymbol{x}', \boldsymbol{x}'') = e^{-\frac{1}{2h^2} \normtwo{\boldsymbol{x}'-\boldsymbol{x}''}^2}$ is the Gaussian kernel; this algorithm corresponds to a non-parametric estimator of the probability density $p(\boldsymbol{x}^*)$ of the training data.

It is possible to extend these ideas to the problem of building a classification model from a training data set $\{(\boldsymbol{x}_j, \boldsymbol y_j)\}_{0 \cdots N-1}$ by defining a function $g : \mathbb{X}\times \mathbb{Y} \rightarrow \mathbb R$ with kernels $\kappa_{_{\mathbb{X}}}: \mathbb X \times \mathbb X \rightarrow \mathbb R$ and $\kappa_{_{\mathbb{Y}}}: \mathbb Y \times \mathbb Y \rightarrow \mathbb R$ of a test sample $\boldsymbol{x^*}$ and candidate label $\boldsymbol{y}^*$
\begin{equation}
g(\boldsymbol{x}^*, \boldsymbol{y}^*) = \sum_{j=0}^{N-1}\alpha_j\kappa_{_{\mathbb{Y}}}(\boldsymbol{y}_j, \boldsymbol{y}^*)\kappa_{_{\mathbb{X}}}(\boldsymbol{x}_j, \boldsymbol{x}^*).
\end{equation}
Setting the $\{\alpha_j\}$ as before we can build a classification model based on KDE, called kernel density classification (KDC) \cite{hastie2009elements}
\begin{equation}
g_{\text{KDC}}(\boldsymbol{x}^*, \boldsymbol{y}^*) = \frac{M_h}{N}\sum_{j=0}^{N-1}\delta_{\boldsymbol{y}_j, \boldsymbol{y}^*}e^{-\frac{1}{2h^2} \normtwo{\boldsymbol{x}_j-\boldsymbol{x^*}}^2}, \label{eq: kernel density classification}
\end{equation}
where $\kappa_{_{\mathbb{Y}}}(\boldsymbol{y}', \boldsymbol{y}'') = \delta_{\boldsymbol{y}', \boldsymbol{y}''} $ is the Kronecker's delta kernel. 

This algorithm corresponds to a nonparametric density estimator of the joint $g_{\text{KDC}}(\boldsymbol{x^*}, \boldsymbol{y}^*) \approx p(\boldsymbol x^*, \boldsymbol y^*)$, where the predicted class corresponds to the label with the largest joint probability density. It is worth mentioning that although the KDE and the KDC methods were originally developed for classical computers, they can be extended as quantum algorithms using the formalism of density matrices and kernels as described in Refs. \cite{Gonzalez2021Classification, gonzalez2022learning, gonzález2024kdm, Useche2021QuantumQudits, useche2022quantum, vargas2022optimisation, ardila2025memo}.

\subsection{Generative modeling with density matrices}

%$p(\boldsymbol x^*, \boldsymbol{y}^*) = \frac{1}{N}\sum_j\delta(\boldsymbol y^*- \boldsymbol{y}_j)\delta(\boldsymbol x^*-\boldsymbol{x}_j)$
Learning the probability distribution of a data set $\{(\boldsymbol{x}_j, \boldsymbol y_j)\}_{0 \cdots N-1}$ can be framed on the task of minimizing the KL-divergence between the true data distribution $p(\mathbf{x} = \boldsymbol x^*, \mathbf{y} = \boldsymbol{y}^*)$ and the model distribution $f(\mathbf{x} = \boldsymbol x^*, \mathbf{y} = \boldsymbol{y}^*|\boldsymbol \theta)$  with variational parameters $\boldsymbol \theta$

\begin{equation}
    \int_{\mathbb{X}, \mathbb{Y}} p(\boldsymbol x^*, \boldsymbol{y}^*)\log{\frac{p(\boldsymbol x^*, \boldsymbol{y}^*)}{f(\boldsymbol x^*, \boldsymbol{y}^*|\boldsymbol \theta)}} \,d \boldsymbol x^*d\boldsymbol y^*. 
\end{equation}

Since the term $\int_{\mathbb{X}, \mathbb{Y}}p(\boldsymbol x^*, \boldsymbol{y}^*)\log{p(\boldsymbol x^*, \boldsymbol{y}^*)}\,d \boldsymbol x^*d \boldsymbol y^*$ does not have any parameters this optimization is equivalent to minimizing 
\begin{equation}
    - \int_{\mathbb{X}, \mathbb{Y}} p(\boldsymbol x^*, \boldsymbol{y}^*)\log{f(\boldsymbol x^*, \boldsymbol{y}^*|\boldsymbol \theta)} \,d \boldsymbol x^*d \boldsymbol y^*,\label{eq: negative log likelihood 1.1}
\end{equation}
which in turn can be approximated using Monte Carlo integration by means of the average negative log-likelihood
\begin{align}
    \mathcal L(\boldsymbol \theta) = - \frac{1}{N}\sum_j \log{f(\boldsymbol x_j, \boldsymbol y_j| \boldsymbol \theta)},\label{eq: negative log likelihood 2}
\end{align}
owing to the fact that the training data points $\{(\boldsymbol{x}_j, \boldsymbol{y}_j)\}_{0 \cdots N-1}$ are samples drawn from $p(\mathbf{x}, \mathbf{y})$.

The model distribution $f(\boldsymbol x', \boldsymbol y'| \boldsymbol \theta)$ can then be constructed using a quantum feature map of the inputs and outputs $\boldsymbol x' \mapsto \kettwo{\bar\psi_{_{\mathcb{X}}}'}$, $\boldsymbol y' \mapsto \kettwo{\bar\psi_{_{\mathcb{Y}}}'}$, with $\kettwo{\bar\psi_{_\mathcb{X}, _\mathcb{Y}}'} = \kettwo{\bar\psi_{_\mathcb{X}}'}\otimes\kettwo{\bar\psi_{_\mathcb{Y}}'}$, and a variational density matrix $\bar\rho_{{_{\mathcb{X}}}, {_\mathcb{Y}}}(\boldsymbol \theta)$
%It is possible to build an estimator of $f(\boldsymbol x', \boldsymbol y'| \boldsymbol \theta) \approx \hat{f}(\boldsymbol x', \boldsymbol y'| \boldsymbol \theta)$ through a quantum feature map of inputs and outputs $\boldsymbol x' \mapsto \kettwo{\psi_{_{\mathcb{X}}}'}$, $\boldsymbol y' \mapsto \kettwo{\psi_{_{\mathcb{Y}}}'}$, and a variational density matrix $\rho_{{_{\mathcb{X}}}, {_\mathcb{Y}}}(\boldsymbol \theta)$
%\begin{equation}
%\hat{f}(\boldsymbol x', \boldsymbol y'| \boldsymbol \theta) = \mathcal{M}_{_{\mathbb{X}}}\mathcal{M}_{_{\mathbb{Y}}}\bratwo{\psi_{_\mathcb{X}, _\mathcb{Y}}'}\rho_{{_{\mathcb{X}}}, {_\mathcb{Y}}}
%(\boldsymbol\theta) \kettwo{\psi_{_\mathcb{X}, _\mathcb{Y}}'},
%\end{equation}
%$\kettwo{\psi_{_{\mathcb{X}}, _{\mathcb Y}}'} = \kettwo{\psi_{_{\mathcb{X}}}'} \otimes \kettwo{\psi_{_\mathcb{Y}}'}$, and
\begin{equation}
f(\boldsymbol x', \boldsymbol y'| \boldsymbol \theta) = \mathcal{M}_{_{\mathbb{X}}}\mathcal{M}_{_{\mathbb{Y}}}\bratwo{\bar\psi_{_\mathcb{X}, _\mathcb{Y}}'}\bar\rho_{{_{\mathcb{X}}}, {_\mathcb{Y}}}
(\boldsymbol\theta) \kettwo{\bar\psi_{_\mathcb{X}, _\mathcb{Y}}'},
\end{equation}
where $\mathcal{M}_{_{\mathbb{X}}}$ and $\mathcal{M}_{_{\mathbb{Y}}}$, correspond to the normalization constants of the shift-invariant kernels induced through an inner product in the quantum feature spaces $\abstwo{\ketbratwo{\bar\psi_{_{\mathcb{X}}}'}{\bar\psi_{_{\mathcb{X}}}''}}^2 = \kappa_{_{\mathbb X}}(\boldsymbol{x}', \boldsymbol{x}'')$, $\abstwo{\ketbratwo{\bar\psi_{_\mathcb{Y}}'}{\bar\psi_{_\mathcb{Y}}''}}^2 = \kappa_{_{\mathbb Y}}(\boldsymbol{y}', \boldsymbol{y}'')$, i.e.,
\begin{equation}
   \mathcal{M}_{_{\mathbb{X}}} = \frac{1}{\int_{\mathbb{X}}\kappa_{\mathbb{X}}(\boldsymbol x', \boldsymbol{x}^*)\,d \boldsymbol x^*}, \ \mathcal{M}_{_{\mathbb{Y}}} =\frac{1}{\int_{\mathbb{Y}}\kappa_{\mathbb{Y}}(\boldsymbol y', \boldsymbol{y}^*)\,d \boldsymbol y^*};
\end{equation}
the shift-invariance of the kernels guaranties that they are indeed constants. Note that we have used the bar notation $\kettwo{\bar\psi_{_{\mathcb{X}}}'}$ , $\kettwo{\bar\psi_{_{\mathcb{Y}}}'}$ to emphasize that these quantum maps explicitly induce the kernels $\kappa_{_{\mathbb X}}(\boldsymbol{x}', \boldsymbol{x}'')$ and $\kappa_{_{\mathbb Y}}(\boldsymbol{y}', \boldsymbol{y}'')$.

%By means of $\hat{f}(\boldsymbol x', \boldsymbol y'| \boldsymbol \theta)$ we can write the loss function as an estimator $\hat{\mathcal{L}}(\boldsymbol{\theta}) \approx \mathcal{L}(\boldsymbol{\theta})$. And Eq. \ref{eq: negative log likelihood 2} can then be approximated by means of
%\begin{equation}
  %    \hat{\mathcal L}(\boldsymbol \theta)  = - \frac{1}{N}\sum_j \log{\hat{f}(\boldsymbol x_j, \boldsymbol y_j| \boldsymbol \theta)},\label{eq: negative log likelihood 3}  
%\end{equation}
The optimal parameters of the minimization of Eq. \ref{eq: negative log likelihood 2} would then correspond to
\begin{equation}
    \boldsymbol \theta_{\text{op}} =  \argmin_{\boldsymbol{\theta}}\mathcal L(\boldsymbol \theta) .\label{eq: optimal theta}
\end{equation}

%Since we use the training quantum features $\{\kettwo{\psi_{_{\mathcb{X}}, _{\mathcb Y}}^j}\}_{0\cdots N-1}$ to learn the estimator $\hat{f}(\boldsymbol x', \boldsymbol y'| \boldsymbol \theta)$, representer theorem \cite{scholkopf2002learning, Schuld2021quantumkernel} guarantees that its evaluation over some test sample $(\boldsymbol{x}^* , \boldsymbol{y}^* )\mapsto \kettwo{\psi_{_{\mathcb X}}^*}\otimes\kettwo{\psi_{_{\mathcb Y}}^*}=\kettwo{\psi_{_{\mathcb X, \mathcb Y}}^*}$ at its optimal values $\boldsymbol \theta_{\text{op}}$ approximates a function in a RKHS 
Moreover, following the representer theorem \cite{scholkopf2002learning, Schuld2021quantumkernel} we have that the solution of the previous minimization problem can be written as a linear combination of the kernel functions centered at the training data points
\begin{align}
f(\boldsymbol x^*, \boldsymbol y^*| \boldsymbol{\theta}_{\text{op}}) &=  \mathcal{M}_{_{\mathbb{X}}}\mathcal{M}_{_{\mathbb{Y}}}\bratwo{\bar\psi_{_{\mathcb{X}}, _{\mathcb{Y}}}^*}\bar\rho_{{_{\mathcb{X}}}, {_\mathcb{Y}}}(\boldsymbol{\theta}_{\text{op}})\kettwo{\bar\psi_{_{\mathcb{X}}, _\mathcb{Y}}^*}  \notag \\ 
&= \sum_{j=0}^{N-1}\alpha_j\kappa_{_{\mathbb{Y}}}(\boldsymbol{y}_j, \boldsymbol{y}^*)\kappa_{_{\mathbb{X}}}(\boldsymbol{x}_j, \boldsymbol{x}^*),\label{eq: variational RKHS}
\end{align}
with each $\alpha_j \in \mathbb{R}$. This expression corresponds to a function in a RKHS, which can also be written using the formalism of the density matrices as follows 
\begin{equation}
f(\boldsymbol x^*, \boldsymbol y^*| \boldsymbol{\theta}_{\text{op}}) 
= \mathcal{M}_{_{\mathbb{X}}}\mathcal{M}_{_{\mathbb{Y}}}\sum_{j=0}^{N-1}q_j(\boldsymbol{\theta}_{\text{op}})\kappa_{_{\mathbb{Y}}}(\boldsymbol{y}_j, \boldsymbol{y}^*)\kappa_{_{\mathbb{X}}}(\boldsymbol{x}_j, \boldsymbol{x}^*),
\end{equation}
with $q_j(\boldsymbol{\theta}_{\text{op}}) \in \mathbb{R}$ and $q_j(\boldsymbol{\theta}_{\text{op}})\ge 0$ for all $j$, and $\sum_jq_j(\boldsymbol{\theta}_{\text{op}})=1$; we formalize and prove this result in Proposition \ref{prop: proposition 1}, illustrating the representer theorem in the context of density matrices, see \ref{sec: representer theorem for ms}. 

For the present QGC algorithm, we have that the QEFF mapping $\kettwo{\psi_{_{\mathcb{X}}}'}$ approximates the Gaussian kernel $\abstwo{\ketbratwo{\psi_{_{\mathcb{X}}}'}{\psi_{_{\mathcb{X}}}''}}^2 \approx e^{-\frac{1}{2h^2}\normtwo{\boldsymbol{x}'- \boldsymbol{x}''}^2}$, while the one-hot basis encoding induce explicitly the Kronecker's delta kernel  $\abstwo{\ketbratwo{\psi_{_\mathcb{Y}}'}{\psi_{_\mathcb{Y}}''}}^2 = \delta_{\boldsymbol{y}', \boldsymbol{y}''}$, and  $\mathcal{M}_{_{\mathbb{X}}}=M_h=(2\pi h^2)^{-D/2}$ and $\mathcal{M}_{_{\mathbb{Y}}}=1$. Hence, the estimator of the probability density of a test sample $(\boldsymbol x^*, \boldsymbol y^*)$, see Eq. \ref{eq: estimator of pdf},  approximates a function in a RKHS of the form 

\begin{align}
&\hat{f}(\boldsymbol x^*, \boldsymbol y^*| \boldsymbol \theta_{\text{op}})  =  \mathcal{M}_{_{\mathbb{X}}}\mathcal{M}_{_{\mathbb{Y}}}\bra{\psi_{_{\mathcb{X}}, _{\mathcb{Y}}}^*}\rho_{{_{\mathcb{X}}}, {_\mathcb{Y}}}(\boldsymbol{\theta}_{\text{op}})\ket{\psi_{_{\mathcb{X}}, _\mathcb{Y}}^*} \approx \notag \\ 
&\sum_{j=0}^{N-1}q_j(\boldsymbol \theta_{\text{op}})\delta_{\boldsymbol{y}_j, \boldsymbol{y}^*}\frac{1}{(2\pi h^2)^{D/2}}e^{-\frac{1}{2h^2} \normtwo{\boldsymbol{x}_j-\boldsymbol{x^*}}^2} = \notag \\ & \sum_{j=0}^{N-1}q_j(\boldsymbol \theta_{\text{op}})\delta_{\boldsymbol{y}_j, \boldsymbol y^*}\mathcal{N}(\boldsymbol{x}^*|\boldsymbol{x}_j, h^2\mathbf{I}_D). \label{eq: variational Gaussian}
\end{align}
This estimator is an approximation of a Gaussian mixture classifier, where $\mathcal{N}:\mathbb{X}\times\mathbb{X}\rightarrow\mathbb{R}$ denotes the Gaussian probability density function. In particular, the mixture components are centered on the training data points and they all share the same covariance matrix $h^2\mathbf{I}_D$, where $\mathbf{I}_D$ is the $D$-dimensional identity matrix. In summary,  the function $\hat f(\boldsymbol x^*, \boldsymbol y^*| \boldsymbol \theta_{\text{op}}) \approx f(\boldsymbol x^*, \boldsymbol y^*| \boldsymbol{\theta}_{\text{op}})$ would be an approximation of the joint probability density of inputs and outputs $p(\boldsymbol{x}^*, \boldsymbol{y}^*)$.

\subsection{\label{sec: QEFF}Quantum-enhanced Fourier features}

To build the quantum features in the input space, we propose the quantum enhanced Fourier features mapping, which can be viewed as a quantum-enhanced implementation of the quantum random Fourier features \cite{useche2022quantum} and a quantum analog of random Fourier features \cite{Rahimi2009RandomMachines}. The QEFF embedding maps a sample $\boldsymbol{x}' \in \mathbb X = \mathbb{R}^D$ to a quantum state $\kettwo{\psi_{_{\mathcb{X}}}'} \in {\mathcal{X}} = \mathbb{C}^d$, such that for any $\boldsymbol{x}', \boldsymbol{x}'' \in \mathbb X$, the inner product in the Hilbert space approximates a Gaussian kernel in the input space $\abstwo{\ketbratwo{\psi_{_{\mathcb{X}}}'}{\psi_{_{\mathcb{X}}}''}}^2 \approx \kappa_{_{\mathbb X}}(\boldsymbol{x}', \boldsymbol{x}'') = e^{-\frac{1}{2h^2}\normtwo{\boldsymbol{x}'-\boldsymbol x''}^2}$, where $h$ is termed the kernel bandwidth.

The QEFF encoding leverages quantum superposition for implementing the QRFF mapping \cite{useche2022quantum} in quantum hardware
\begin{equation}
          	\kettwo{\psi_{_{\mathcb{X}}}'} = \sqrt{\frac{1}{2^{n_{_{\mathcb{X}}}}}}\sum_{k=0}^{2^{n_{_{\mathcb{X}}}} - 1}e^{i\sqrt{\frac{1}{2h^2}}\boldsymbol{w}_k \cdot\boldsymbol{x}'}\kettwo{k_{_{\mathcb{X}}}}.\label{eq: QRFF map}
\end{equation}
In this expression the data sample $\boldsymbol{x}'$ is encoded on the phases of a quantum state, denoting $(\cdot)$ the standard dot product and $n_{_{\mathcb{X}}} = \log{d}$ the number of qubits. In addition, the weights $\{\boldsymbol{w}_k\}_{0 \cdots 2^{n_{_{\mathcb{X}}}} - 1}\in \mathbb R^D$ result from sampling a $D\text{-dimensional} $ standard normal distribution $\boldsymbol{w}_k \sim \mathcal{N}(\mathbf{0}, \mathbf I_D)$.

\begin{figure*}
\centering
\includegraphics[scale=0.77]{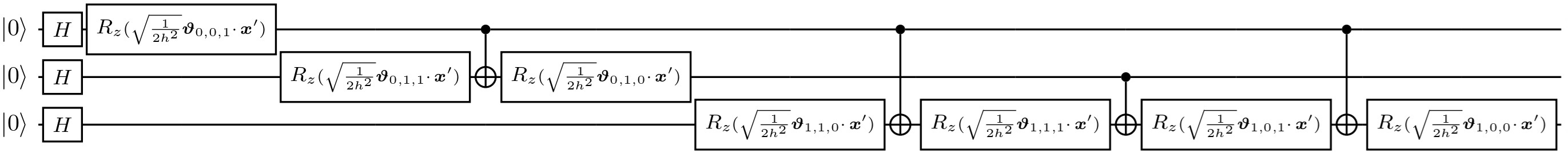}
\caption{QAFF quantum circuit strategy \cite{useche2022quantum} for preparing the three-qubit QEFF mapping. In this example, we map a $D$-dimensional input sample $\boldsymbol{x}'$ to an eight-dimensional pure quantum state $\kettwo{\psi_{_{\mathcb{X}}}'}$ whose weights $\{\boldsymbol\vartheta_{\alpha_{3}, \alpha_2, \alpha_1}\}_{\alpha_{3}, \alpha_2, \alpha_1 \in \{0, 1\}}$, excluding the all zeros index, are sampled from the same arbitrary probability distribution $\Phi$ with mean $\mathbf{0}\in \mathbb{R}^D$ and covariance matrix $4/(2^3-1)\mathbf{I}_D$. See Ref. \cite{useche2022quantum} for the extension of the QAFF quantum circuit to a higher number of qubits.}
\label{fig: QAFF quantum circuit}
\end{figure*}

Ref. \cite{useche2022quantum} proposed the use of amplitude encoding protocols \cite{shende2006synthesis, mottonen2005transformation} to prepare the QRFF mapping on a quantum computer. Specifically, the dot products $\{\boldsymbol{w}_k\cdot \boldsymbol{x}\}$ are encoded in the phases $\{\phi_k\}$ of the canonical basis $\{\kettwo{k_{_{\mathcb{X}}}}\}_{0 \cdots 2^{n_{_{\mathcb{X}}}} - 1}
$, i.e., $\kettwo{\psi_{_{\mathcb{X}}}'} = (1/\sqrt{2^{n_{_{\mathcb{X}}}}})\sum_{k}e^{i\phi_k}\kettwo{k_{_{\mathcb{X}}}}$. In contrast, in this article we present the QEFF embedding which compiles the QRFF mapping using a basis based on the Pauli-z $\sigma^z = \kettwo{0}\bratwo{0}-\kettwo{1}\bratwo{1}$ and the identity $\mathbf I_2 = \kettwo{0}\bratwo{0}+\kettwo{1}\bratwo{1}$ matrix. 

The QEFF embedding starts from the $n_{_{\mathcb{X}}}$ zeros state $\kettwo{0_{_{\mathcb{X}}}}$ and then applies a Hadamard gate $H$ to each qubit resulting in the superposition state $H^{\otimes n_{_{\mathcb{X}}}}\kettwo{0_{_{\mathcb{X}}}} = \kettwo{+}^{\otimes n_{_{\mathcb{X}}}}$ with $\kettwo{+} = \frac{1}{\sqrt{2}}(\kettwo{0}+\kettwo{1})$, followed by a unitary matrix $U_{_\mathcb X}(\boldsymbol{x}'|{\boldsymbol \Theta})$ on $n_{_{\mathcb{X}}}$ qubits given by
\begin{align}
    &U_{_\mathcb X}(\boldsymbol{x}'|{\boldsymbol \Theta}) = \notag \\
    &\exp \bigg\{-\frac{i}{2}\sum_{\alpha_1, \alpha_2, \cdots, \alpha_{n_{_{\mathcb{X}}}} \in \{0, 1\}}\bigg(\sqrt{\frac{1}{2h^2}}\boldsymbol{\vartheta}_{ \alpha_{n_{_{\mathcb{X}}}}, \cdots, \alpha_2, \alpha_1} \cdot\boldsymbol{x}'\bigg) \notag \\
    &\times \bigg(\sigma^{\alpha_{n_{_{\mathcb{X}}}}} \otimes \cdots \otimes \sigma^{\alpha_2}\otimes \sigma^{\alpha_1}\bigg)\bigg\}, \quad \quad
    \label{eq: unitary QEFF}
\end{align}
where $\sigma^{0} = \mathbf{I}_2$, $\sigma^{1} = \sigma^{z}$, and $\boldsymbol\Theta = \{\boldsymbol\vartheta_{\alpha_{n_{_{\mathcb{X}}}}, \cdots, \alpha_1}\}_{\alpha_{n_{_{\mathcb{X}}}}, \cdots, \alpha_1 \in \{0, 1\}}$ is a set $2^{n_{_{\mathcb{X}}}}$ vectors in $\mathbb{R}^D$, with $\boldsymbol\vartheta_{0, 0, \cdots, 0} = \boldsymbol 0$ to account for the global phase and the remaining $2^{n_{_{\mathcb{X}}}}-1$ weights sampled i.i.d. from an arbitrary symmetric probability distribution $\Phi(\mathbf{0}, \frac{4}{(2^{n_{_{\mathcb{X}}}})-1}\mathbf{I}_D)$ with zero mean $\boldsymbol{\mu}=\mathbf{0}\in\mathbb R^D$ and covariance matrix $\Sigma = \frac{4}{(2^{n_{_{\mathcb{X}}}})-1}\mathbf{I}_D$.

The proposed method can then be summarized by a unitary matrix $\mathcal{U}'_{_{\mathcb{X}}} = U_{_\mathcb X}(\boldsymbol{x}'|{\boldsymbol \Theta})H^{\otimes n_{_{\mathcb{X}}}}$, which build from the zeros state the quantum feature map 
\begin{align}
    \kettwo{\psi_{_{\mathcb{X}}}'} & = \mathcal{U}'_{_{\mathcb{X}}}\kettwo{0_{_{\mathcb{X}}}} \notag \\
    & = U_{_\mathcb X}^{\boldsymbol \Theta}(\boldsymbol{x}')H^{\otimes n_{_{\mathcb{X}}}}\kettwo{0_{_{\mathcb{X}}}}.
\end{align}

In \ref{sec: QRFF-QEFF}, we use the multivariate central limit theorem to show that the QEFF sampling strategy in the Pauli basis $\{\boldsymbol\vartheta_{\alpha_{n_{_{\mathcb{X}}}}, \cdots, \alpha_1}\}\sim\Phi(\mathbf{0}, \frac{4}{(2^{n_{_{\mathcb{X}}}})-1}\mathbf{I}_D)$, effectively prepares the QRFF mapping, whose weights are identically distributed from a normal distribution $\{\boldsymbol{w}_i\}_{0, \cdots, (2^{n_{_{\mathcb{X}}}}) - 1} \sim \mathcal{N}(\mathbf{0}, \mathbf I_D)$. 

Furthermore, to implement the QEFF mapping as a quantum circuit, we use the QAFF ansatz proposed in Ref. \cite{useche2022quantum}. In that article, in addition of presenting the QRFF method, the authors propose a variational feature embedding called QAFF where the Fourier weights are optimized using a variational quantum circuit. In our work, we also utilize the QAFF quantum circuit, but the weights are fixed and are obtained from the QEFF sampling procedure. Fig \ref{fig: QAFF quantum circuit} shows an example of the QAFF quantum circuit for three qubits that prepares the eight-dimensional QEFF feature mapping; this circuit uses Hadamard, CNOT, and $R_z(\beta) = e^{-i\beta/2}\kettwo{0}\bratwo{0}+e^{i\beta/2}\kettwo{1}\bratwo{1}$ quantum gates. It is worth noticing that the weight vector $\boldsymbol\vartheta_{0, \cdots, 0} = \boldsymbol 0$ associated with the global phase does not need to be compiled in the quantum circuit.

\subsection{Purification ansatz}

As previously discussed, the QGC algorithm builds the variational density matrix by tracing out the auxiliary qubits of a three-component pure state with joint Hilbert space of inputs, outputs, and ancilla $\rho_{{_\mathcb{X}}, {_\mathcb{Y}}}(\boldsymbol \theta) = \text{Tr}_{_\mathcb{A}}\big[\kettwo{q_{_{\mathcb{A}, \mathcb{X}, \mathcb{Y}}}({\boldsymbol \theta})}\bratwo{q_{_{\mathcb{A}, \mathcb{X}, \mathcb{Y}}}({\boldsymbol \theta})}\big]$. Indeed, any mixed state can be prepared from a partial measurement over a pure quantum state, also called a purification of a mixed quantum state. Although a purification of a mixed state is not unique, it has been shown that it is optimal when $n_{\mathcb{A}} = \lceil \log{r} \rceil$, where $r$ is the rank of the density matrix \cite{useche2022quantum, Ping2013OptimalPO}. Henceforth, in the proposed QGC method, the size of the Hilbert space of the pure state $\kettwo{q_{_{\mathcb{A}, \mathcb{X}, \mathcb{Y}}}}$ could be at least equal to the size of the training density matrix and at most twice its size, i.e., $n_{\mathcb{X}}+n_{\mathcb{Y}}
\ge n_{\mathcb{A}} \ge 0$.

To construct a purification of a quantum machine learning algorithm  one must choose  an appropriate quantum circuit ansatz. This selection usually involves a trade-off between expressibility and trainability \cite{HolmesExpressibility2022}. Namely, the use of a more expressive ansatz that explores a larger region of the Hilbert space can lead to trainability issues due to the barren plateau problem \cite{mcclean2018barren}. Motivated from these challenges, we propose the use of the hardware efficient ansatz (HEA) \cite{Kandala2017HEA} for compiling the unitary matrix $\mathcal Q_{_\mathcb{T}}(\boldsymbol \theta)$ of the purification $\kettwo{q_{_\mathcb{T}}(\boldsymbol \theta)} = \mathcal Q_{_\mathcb{T}}(\boldsymbol \theta) \kettwo{0_{_\mathcb{T}}}$ (recall that $\mathcal{T}$ is the Hilbert space of all the qubits and $n_{_\mathcb{T}} ={n_{_\mathcb{Y}} + n_{_{\mathcb{X}}} + n_{_\mathcb{A}}}$). This ansatz has been shown to be highly expressive, to have a manageable circuit depth, and to avoid the barren plateau problem when applied to certain quantum machine learning tasks \cite{Leone2024practicalusefulness}. 

\begin{figure}
\centering
\includegraphics[scale=1.0]{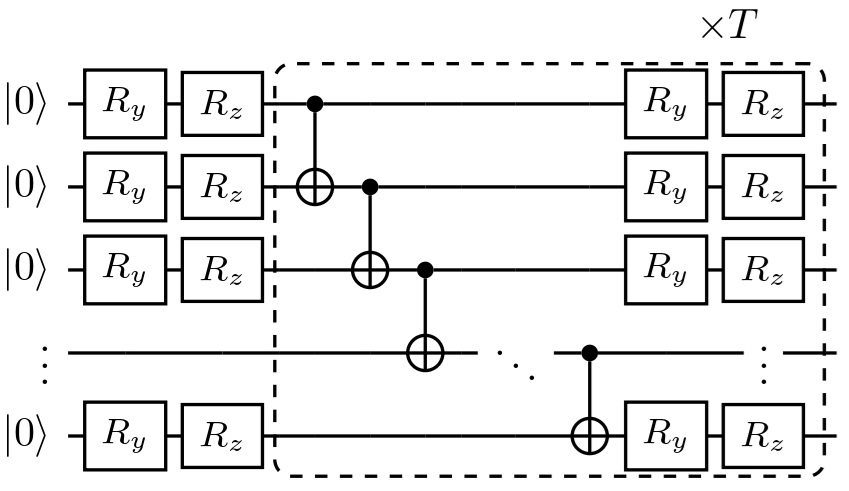}
\caption{Hardware-efficient ansatz for preparing the purification   $\kettwo{q_{_\mathcb{T}}(\boldsymbol \theta)}$ of the training density matrix $\rho_{_{\mathcb{X}}, _\mathcb{Y}}(\boldsymbol \theta) = \text{Tr}_{_\mathcb{A}}(\kettwo{q_{_\mathcb{T}}(\boldsymbol \theta})\bratwo{q_{_\mathcb{T}}(\boldsymbol \theta))})$. The $R^y$ and $R^z$ are parametrized by $\boldsymbol{\theta}$ which corresponds to rotation angles in $[0, 2\pi)$.}{\label{fig: HEA circuit}}
\end{figure}

The circuit of HEA ansatz \cite{Kandala2017HEA} starts from the all-zeros state $\kettwo{0_{_\mathcb{T}}}$ and then it sequentially applies to each qubit the variational $R^y$ and $R^z$ quantum gates. Next, $T$ quantum layers are constructed, each consisting of a cascade of $n_{_\mathcb{T}} - 1$ CNOT gates that entangle adjacent qubits, followed by the qubit-wise $R^y$ and $R^z$ unitaries, see Fig. \ref{fig: HEA circuit}. The single $R^y$ and $R^z$ quantum rotations are parameterized by $\boldsymbol{\theta} \in [0, 2\pi)^{2n_{_\mathcb{T}}(T+1)}$, corresponding to $2n_{_\mathcb{T}}(T+1)$ trainable parameters. Furthermore, the depth of the ansatz compromises $T(n_{_\mathcb{T}}-1)$ CNOT gates.

\subsection{\label{sec: Conditional pred and error estimation} Posterior probability density and variance estimation} 

Following the quantum generative classification strategy, we can apply Bayes's rule to estimate the conditional probability of the outputs given an input sample from the joint probability density,
\begin{equation}
    \hat p(\mathbf y = \boldsymbol y^*|\mathbf x = \boldsymbol{x}^*) = \frac{\hat p(\mathbf x = \boldsymbol{x}^*, \mathbf y = \boldsymbol y^*)}{\sum_{{\boldsymbol{y}'}\in \mathbb{Y}}\hat p(\mathbf x = \boldsymbol{x}^*, \mathbf y = \boldsymbol y')}.
\end{equation}
This can be calculated using a classical computer that normalizes the output of the QGC testing quantum circuit with the probabilities of all possible labels $\{P(\kettwo{0_{_\mathcb{X}}}\kettwo{\boldsymbol{y}'_{_\mathcb{Y}}})\}_{\kettwo{\boldsymbol{y}'_{_\mathcb{Y}}} \in \mathcal{Y}}$, see Eq. \ref{eq: VQGC prediction},

\begin{align}
&\hat p(\mathbf y = \boldsymbol y^*|\mathbf x = \boldsymbol{x}^*) = \frac{P(\ket{0_{_\mathcb{X}}}\ket{\boldsymbol{y}^*_{_\mathcb{Y}}})}{\sum_{\ket{\boldsymbol{y}_{_{\mathcb Y}}'}\in \mathcal{Y}}P(\ket{0_{_\mathcb{X}}}\ket{\boldsymbol{y}'_{_\mathcb{Y}}})} \notag \\
 &= \frac{\bra{\psi_{_\mathcb{X}, _\mathcb{Y}}^*}\rho_{{_\mathcb{X}}, {_\mathcb{Y}}}(\boldsymbol\theta_{\text{op}})\ket{\psi_{_\mathcb{X}, _\mathcb{Y}}^*}}{\sum_{\ket{\boldsymbol{y}_{_{\mathcb Y}}'}\in \mathcal{Y}}\text{Tr}\big[\rho_{{_\mathcb{X}}, {_\mathcb{Y}}}(\boldsymbol\theta_{\text{op}})|\psi_{_\mathcb{X}}^*\rangle\langle\psi_{_\mathcb{X}}^*|\otimes\ket{\boldsymbol y'_{_\mathcb{Y}}}\bra{\boldsymbol y'_{_\mathcb{Y}}}\big]}.
 \label{eq: dsicriminative prediction}
\end{align}
What's more, we can leverage the probabilistic features of quantum computers to approximate the variance of the predictions of both the joint and the posterior probabilities. As a matter of fact, a single prediction of the joint estimator $\hat p(\mathbf x = \boldsymbol{x}^*, \mathbf y = \boldsymbol y^*)$ can be obtained by performing $S$ measurements of the test quantum circuit and counting the number of activations $S^*$ of the state $\kettwo{0_{_\mathcb{X}}}\kettwo{\boldsymbol{y}^*_{_\mathcb{Y}}}$, 
\begin{equation}
 \hat p (\boldsymbol{x}^*, \boldsymbol y^*) \approx \frac{S^*}{S}.
\end{equation}
We can then perform this computation $R$ times, leading a series of predictions $\{\hat p_r(\boldsymbol{x}^*, \boldsymbol y^*)\}_{r=0\cdots R-1}$, from which we can calculate the expected value $\mathbb{E}[\hat p(\boldsymbol{x}^*, \boldsymbol y^*)] =  (1/R)\sum_{r=0}^{R-1}\hat p_r(\boldsymbol{x}^*, \boldsymbol y^*)$ and the variance of the joint probability density,
\begin{equation}
 \text{Var}(\hat p(\boldsymbol{x}^*, \boldsymbol y^*)) =  \frac{1}{R}\sum_{r=0}^{R-1}\bigg(\hat p_r(\boldsymbol{x}^*, \boldsymbol y^*)-\frac{1}{R}\sum_{r'=0}^{R-1}\hat p_{r'}(\boldsymbol{x}^*, \boldsymbol y^*)\bigg)^2.\label{eq: QGC variance estimation joint}
\end{equation}

Following the same line of though, we may also estimate a single prediction of the conditional probability density
\begin{equation}
 \hat p(\boldsymbol y^*|\boldsymbol{x}^*) \approx \frac{S^*}{\sum_{\kettwo{\boldsymbol{y}_{_{\mathcb Y}}'}\in \mathcal{Y}} S'},
\end{equation}
where $S'$ equates the number of counts of $\kettwo{0_{_\mathcb{X}}}\kettwo{\boldsymbol{y}'_{_\mathcb{Y}}}$ for all $\kettwo{\boldsymbol{y}_{_{\mathcb Y}}'}\in \mathcal{Y}$; noticing that $\sum_{\kettwo{\boldsymbol{y}_{_{\mathcb Y}}'}\in \mathcal{Y}} S' \leq S$. And similarly, we can find the expected value $\mathbb{E}[\hat p(\boldsymbol y^*|\boldsymbol{x}^*)] =  (1/R)\sum_{r=0}^{R-1}\hat p_r(\boldsymbol y^*|\boldsymbol{x}^*)$ and the variance of the conditional by performing $R$ quantum demonstrations,
\begin{equation} \text{Var}(\hat p(\boldsymbol y^*|\boldsymbol{x}^*)) =  \frac{1}{R}\sum_{r=0}^{R-1}\bigg(\hat p_r(\boldsymbol y^*|\boldsymbol{x}^*)-\frac{1}{R}\sum_{r'=0}^{R-1}\hat p_{r'}(\boldsymbol y^*|\boldsymbol{x}^*)\bigg)^2.\label{eq: QGC variance estimation conditional}
\end{equation}

\subsection{Quantum discriminative learning \label{sec: quantum discriminative learning}}

The presented quantum generative algorithm can also be adapted for discriminative learning, whose goal is to construct a function that maximizes the conditional probability of the outputs given the inputs $p(\mathbf{y}|\mathbf{x})$. Unlike the previous sections, this section describes how the algorithm is trained in a discriminative manner rather than a generative manner.

The proposed quantum discriminative classification approach shares some similarities with the quantum generative strategy, that is, we use the training quantum features $\{\kettwo{\psi_{_\mathcb{X}, _\mathcb{Y}}^j}= \kettwo{\psi_{_\mathcb{X}}^j}\otimes\kettwo{\boldsymbol y_{_\mathcb{Y}}^j}\}$ to learn a variational density matrix $\rho_{{_\mathcb{X}}, {_\mathcb{Y}}}(\boldsymbol{\theta})$. Nevertheless, in this new approach we estimate the conditional probability by performing a normalization of the circuit outputs. To achieve this, we extract from the training quantum circuit, see Fig. \ref{fig: QMC training quantum circuit}, the probabilities of each input state $\kettwo{\psi_{_\mathcb{X}}^j}$ with all classes by measuring the probabilities $\{P(\kettwo{0_{_\mathcb{X}}}\kettwo{\boldsymbol{y}'_{_\mathcb{Y}}})\}$ for all $\kettwo{\boldsymbol{y}'_{_\mathcb{Y}}} \in \mathcal{Y}$; notice that one of these estimations corresponds to the probability associated with the true label $P(\kettwo{0_{_\mathcb{X}}}\kettwo{\boldsymbol{y}^j_{_\mathcb{Y}}})$. We then use a classical computer to normalize the circuit outputs
\begin{align}
\hat h(\boldsymbol{y}_j|\boldsymbol{x}_j;\boldsymbol{\theta})& = \frac{\hat{f}(\boldsymbol x_j, \boldsymbol y_j| \boldsymbol \theta)}{\sum_{\boldsymbol{y}'\in \mathbb{Y}}\hat{f}(\boldsymbol x_j, \boldsymbol y'| \boldsymbol \theta)}\notag \\
 &= \frac{P(\kettwo{0_{_\mathcb{X}}}\kettwo{\boldsymbol{y}^j_{_\mathcb{Y}}})}{\sum_{\kettwo{\boldsymbol{y}_{_{\mathcb Y}}'}\in \mathcal{Y}}P(\kettwo{0_{_\mathcb{X}}}\kettwo{\boldsymbol{y}'_{_\mathcb{Y}}})} \notag \\
 &= \frac{\bratwo{\psi_{_\mathcb{X}, _\mathcb{Y}}^j}\rho_{{_\mathcb{X}}, {_\mathcb{Y}}}(\boldsymbol\theta)\kettwo{\psi_{_\mathcb{X}, _\mathcb{Y}}^j}}{\sum_{\kettwo{\boldsymbol{y}_{_{\mathcb Y}}'}\in \mathcal{Y}}\text{Tr}\big[\rho_{{_\mathcb{X}}, {_\mathcb{Y}}}(\boldsymbol\theta)|\psi_{_\mathcb{X}}^j\rangle\langle\psi_{_\mathcb{X}}^j|\otimes\kettwo{\boldsymbol y'_{_\mathcb{Y}}}\bratwo{\boldsymbol y'_{_\mathcb{Y}}}\big]}.
 \label{eq: dsicriminative training}
\end{align}

Just as in the generative case, we estimate the optimal parameters $\boldsymbol{\theta}_{\text{op}} = \argmin_{\boldsymbol \theta}L(\boldsymbol{\theta})$ by performing a minimization of the average negative log-likelihood
\begin{equation}
	\hat{L}(\boldsymbol{\theta}) = -\frac{1}{N}\sum_{j=0}^{N-1}\log{\hat h(\boldsymbol{y}_j|\boldsymbol{x}_j;\boldsymbol{\theta})};
\end{equation}
notice that in this new scenario we are maximizing the conditional probability of the true label $\boldsymbol y_j$ given the input sample $\boldsymbol x_j$. 

In a similar fashion, for the prediction stage, we use the optimized density matrix $\rho_{{_\mathcb{X}}, {_\mathcb{Y}}}(\boldsymbol{\theta}_{\text{op}})$ to build an estimator of the conditional probability $p(\boldsymbol{y}^*|\boldsymbol{x}^*)$ of the candidate test label $\boldsymbol y^*\mapsto\kettwo{\boldsymbol y_{_\mathcb{Y}}^*}$ given the test input $\boldsymbol x^*\mapsto\kettwo{\psi_{_\mathcb{X}}^*}$ by a normalization of the outputs of the test quantum circuit, see Fig \ref{fig: QMC testing quantum circuit},

\begin{align} 
&\hat h(\boldsymbol{y}^*|\boldsymbol{x}^*;\boldsymbol\theta_{\text{op}})= \frac{P(\ket{0_{_\mathcb{X}}}\ket{\boldsymbol{y}^*_{_\mathcb{Y}}})}{\sum_{\ket{\boldsymbol{y}_{_{\mathcb Y}}'}\in \mathcal{Y}}P(\ket{0_{_\mathcb{X}}}\ket{\boldsymbol{y}'_{_\mathcb{Y}}})} \notag \\
& = \frac{\bra{\psi_{_\mathcb{X}, _\mathcb{Y}}^*}\rho_{{_\mathcb{X}}, {_\mathcb{Y}}}(\boldsymbol\theta_{\text{op}})\ket{\psi_{_\mathcb{X}, _\mathcb{Y}}^*}}{\sum_{\ket{\boldsymbol{y}_{_{\mathcb Y}}'}\in \mathcal{Y}}\text{Tr}\big[\rho_{{_\mathcb{X}}, {_\mathcb{Y}}}(\boldsymbol\theta_{\text{op}})|\psi_{_\mathcb{X}}^*\rangle\langle\psi_{_\mathcb{X}}^*|\otimes\ket{\boldsymbol y'_{_\mathcb{Y}}}\bra{\boldsymbol y'_{_\mathcb{Y}}}\big]}.\notag\\
\end{align}
Prediction is then made from $\argmax_{{\boldsymbol{y}^*}\in \mathbb{Y}}\hat h(\boldsymbol y^*|\boldsymbol{x}^*;\boldsymbol \theta_{\text{op}})$.

This discriminative learning strategy was first proposed as a classical algorithm called QMC-SGD by Gonzalez \textit{et al.} \cite{gonzalez2022learning, Gonzalez2021Classification}. In this paper, we extend this model for generative learning and present both generative and discriminative approaches as variational quantum algorithms.

\subsection{Computational complexity analysis \label{sec: complexity analysis}}

We present the computational complexity of the proposed quantum generative classification strategy on both classical and quantum computers.

\subsubsection{Complexity of the quantum-enhanced Fourier features \label{sec: QEFF complexity analysis}}

To review the quantum complexity of the QEFF, recall that in this method we map a data sample $\boldsymbol x'\in \mathbb R^D$ to a quantum state $\kettwo{\psi_{_\mathcb{X}}'}\in \mathbb{C}^d$, where $d = 2^{n_{_\mathcb{X}}}$, and sample ${d-1}$ weights $\{\boldsymbol{\vartheta}_j\}\in\mathbb R^D$ from $\Phi(\mathbf{0}, \frac{4}{(2^{n_{_{\mathcb{X}}}})-1}\mathbf{I}_D)$. First, assuming that it takes $O(1)$ to sample one component of a single Fourier weight, the complexity of sampling all QEFF weights would be $O(Dd)$. Furthermore, to prepare the quantum circuit of the QEFF, it is also required $O(Dd)$ to estimate the angles $\{\boldsymbol \vartheta_j\cdot \boldsymbol{x}'\}_{1\cdots d-1}$ of the gates $R_z$ in the Pauli basis and $O(d)$ to compile the QAFF circuit; in fact, the QAFF circuit \cite{useche2022quantum} is analogous to a circuit for the preparation of an arbitrary quantum state \cite{shende2006synthesis, mottonen2005transformation}, see Fig. \ref{fig: QAFF quantum circuit}. In short, the total quantum complexity to prepare the Fourier state $\kettwo{\psi_{_\mathcb{X}}'}$ would be $O(Dd)$.

On the contrary, to prepare the QEFF mapping on a classical computer one must use the canonical basis $\{\kettwo{k_{_\mathcb X}}\} $ instead of the Pauli basis. However, in the classical case, we do not have access to the Fourier weights on the canonical basis since these weights depend on the weights on the basis of Pauli matrices; see Eq. \ref{eq: Pauli-to-canonical QEFF weights} and \ref{sec: QRFF-QEFF}.  Alternatively, to classically implement the QEFF, one could sample for each Fourier vector $\boldsymbol w_j$ on the canonical basis, $d$ i.i.d. weights $\{\boldsymbol{\upsilon}_{j k}\}_{k=0\cdots d-1}$ from $\Phi(\mathbf{0}, \frac{4}{2^{n_{_{\mathcb{X}}}}}\mathbf{I}_D)$ and perform the summation $\boldsymbol w_j = \sum_{k=0}^{d-1} \boldsymbol \upsilon_{j k}$, resulting in $\boldsymbol{w}_j \sim \mathcal{N}(\mathbf{0}, \mathbf I_D)$. Henceforth, following the arguments of the quantum case, it is required  $O(Dd^2)$ to sample the weights $\{\boldsymbol{w}_j\}_{0\dots d-1}$, $O(Dd)$ to estimate the dot products with the data sample $\boldsymbol{x}'$, and $O(d)$ to construct the vector $\kettwo{\psi_{_\mathcb X}'}$. Therefore, the total complexity for preparing the QEFF mapping in a classical computer would be $O(Dd^2)$.
\begin{table*}
  \centering
  \caption{Computational complexity on a classical and a quantum computer for both the training and test phases of the quantum generative classification algorithm. We denote \(N\) the number of training samples, \(K\) the number of test samples, \(D\) the dimension of the data, \(\epsilon\) and \(\epsilon^*\) the target estimation errors, \(\wp_j\) and \(\wp_k^*\) the minimum estimated joint probability density for training sample \(j\) and test sample \(k\), and \(H\), \(d\), and \(L\) the sizes of the Hilbert spaces of the ancilla, input, and output qubits respectively.}
  \label{tab: complexity QGC algorithm}
 \begin{tabular}{@{}lll}
\midrule
Phase           & Classical complexity                                   & Quantum complexity                                                      \\
\midrule
Training phase  & \(O\big(N(HdL + Dd) + Dd^2\big)\)                        & \(O\!\Big(\sum_{j=0}^{N-1}\tfrac{\log(HdL)\big(\log(HdL) + Dd\big)}{\epsilon^2 \wp_j (1-\wp_j)}\Big)\) \\
Test phase      & \(O\big(K(HdL + Dd) + Dd^2\big)\)                        & \(O\!\Big(\sum_{k=0}^{K-1}\tfrac{\log(HdL) + Dd}{(\epsilon^*)^2 \wp_k^* (1-\wp_k^*)}\Big)\)         \\
\midrule
\end{tabular}

\end{table*}

\subsubsection{Complexity of the training and test quantum circuits\label{sec: QGC quantum circuit complexity analysis}}

We now study the complexity of the training and test quantum circuits of the QGC algorithm. These circuits can be divided into three distinct parts: the preparation of the purification ansatz $\kettwo{q_{_{\mathcb A,\mathcb X,\mathcb Y}}(\boldsymbol{\theta})}$, the compilation of the prediction sample $\kettwo{\psi_{_{\mathcb X}}'}$, and a quantum measurement of the probabilities $\{P(\kettwo{\mathbf 0_{_{\mathcb X}}}\kettwo{\boldsymbol y'_{_{\mathcb Y}}})\}$ for all $\kettwo{\boldsymbol y'_{_{\mathcb Y}}} \in \mathcal Y$. 

To begin with, the complexity of compiling the purification using the HEA ansatz is linear in the number of qubits $O(n_{_{\mathcb A}} +n_{_{\mathcb X}}+n_{_{\mathcb Y}}) = O(\log{(HLd))}$, defining $H=2^{n_{_{\mathcb A}}}$ as the size of the Hilbert space of the ancilla, along with $2^{n_{_{\mathcb Y}}-1}<L\le2^{n_{_{\mathcb Y}}}$ and $d =2^{n_{_\mathcb{X}}}$. Moreover, to prepare the QEFF prediction state $\kettwo{\psi_{_\mathcb X}'}$ takes time $O(Dd)$, see previous subsection. And the estimation the probabilities of the $L$ classes $\{P(\kettwo{0_{_{\mathcb{X}}}}\kettwo{y_{_{\mathcb{Y}}}}) \} $ requires to perform $S$ measurements, which from probability theory scales as
\begin{equation}
	S \sim \frac{1}{\epsilon^2\wp(1-\wp)},
\end{equation}
where $\epsilon$ is the error of the approximation and $\wp = \min_{\kettwo{\boldsymbol{y}_{_{\mathcb Y}}'}} P(\kettwo{\mathbf 0_{_{\mathcb X}}}\kettwo{\boldsymbol y'_{_{\mathcb Y}}})=\min_{\kettwo{\boldsymbol{y}_{_{\mathcb Y}}'}}\bratwo{\psi_{_{\mathcb{X}}, _{\mathcb{Y}}} '}\rho_{{_{\mathcb{X}}}, {_\mathcb{Y}}}(\boldsymbol{\theta})\kettwo{\psi_{_{\mathcb{X}}, _\mathcb{Y}}' }$ is the minimum value over all classes of the unnormalized joint probability density. Therefore, the total quantum complexity of both the training and test quantum circuits is
\begin{equation}
	O\bigg( \frac{\log{(HdL)} + Dd}{\epsilon^2\wp(1-\wp)}\bigg).
    \label{eq: complexity QGC training circuit}
\end{equation}

In contrast, the complexity of calculating the projection $\bratwo{\psi_{_{\mathcb{X}}, _{\mathcb{Y}}} '}\rho_{{_{\mathcb{X}}}, {_\mathcb{Y}}}(\boldsymbol{\theta})\kettwo{\psi_{_{\mathcb{X}}, _\mathcb{Y}}' }$ in a classical computer is $O(HdL+Dd^2)$. In particular, it takes $O(HdL)$ to build the training density matrix, $O(Dd^2)$ to sample the classical QEFF weights, $O(Dd)$ to prepare the quantum state of the QEFF (assuming that the weights were already sampled), and $O(HdL)$ to compute the expected value between the training density matrix and the prediction state.

\subsubsection{\label{sec: QGC complexity gradient estimation}Complexity of estimating the gradients}

The parameters of the loss  (Eq. \ref{eq: negative log.likehoood}) of the QGC algorithm are updated based on the gradient descent rule
\begin{equation}
\theta_k^{(t+1)} = \theta_k^{(t)} - \eta \frac{\partial \mathcal{L}(\boldsymbol \theta^{(t)})}{\partial \theta_k^{(t)}}, 
\end{equation}
being $\eta \in \mathbb{R}$ the learning rate, $t$ the time of the iteration, and $\theta_k$ the $k^{\text{th}}$ component of $\boldsymbol{\theta}$. These gradients can be estimated in a quantum computer using the parameter-shift rule \cite{mitarai2018quantum, schuld2019evaluating} 
\begin{align}
&\frac{\partial \mathcal{L}(\boldsymbol \theta^{(t)})}{\partial \theta_k^{(t)}} = \notag \\
&-\frac{1}{N}\sum_{j=0}^{N-1}\frac{1}{2\hat{f}(\boldsymbol x_j, \boldsymbol y_j| \boldsymbol \theta^{(t)})}\Big\{\hat{f}(\boldsymbol x_j, \boldsymbol y_j| \boldsymbol \theta^{(t)})_{\theta_{k}\rightarrow\theta_{k}+\pi/2}\notag \\
&\times -\hat{f}(\boldsymbol x_j, \boldsymbol y_j| \boldsymbol \theta^{(t)})_{\theta_{k}\rightarrow\theta_{k}-\pi/2}\Big\},
\end{align}
where the subscript indicates that the parameter $\theta_k$ in $\hat f$ is shifted to  $\theta_{k} \pm \pi/2$. Therefore, for each $k$ and each training sample, one must estimate the projections $\hat{f}$, whose complexity is given by Eq. \ref{eq: complexity QGC training circuit}. Assuming also that the number of parameters of the HEA ansatz scales linearly with the number of qubits $|\boldsymbol \theta| \sim \log{(HdL)}$, the total complexity for estimating the gradients would then be
\begin{equation}
O\Bigg(\sum_{j=0}^{N-1}\frac{\log{(HdL)}\big(\log{(HdL)} + Dd\big)}{\epsilon^2\wp_j(1-\wp_j)}\Bigg),
\end{equation}
with
\begin{equation}
   \wp_j =\min_{\kettwo{\boldsymbol{y}_{_{\mathcb Y}}'}\in \mathcal{Y}}\text{Tr}\big[\rho_{{_\mathcb{X}}, {_\mathcb{Y}}}(\boldsymbol\theta)|\psi_{_\mathcb{X}}^j\rangle\langle\psi_{_\mathcb{X}}^j|\otimes\kettwo{\boldsymbol y'_{_\mathcb{Y}}}\bratwo{\boldsymbol y'_{_\mathcb{Y}}}\big], 
\end{equation} 
is the minimum value of the joint probability density of the training sample $j$ over all possible labels.

In contrast, it has been shown that in a classical neural network, a training step and making a prediction have similar computational complexity \cite{livni2014computational}. Henceforth, the complexity of estimating the gradients of a data set of $N$ samples on a classical computer would be $O\big(N(HdL+Dd)+Dd^2\big)$. Note that the complexity of the classical sampling of the QEFF does not depend on the number of training data, since all data samples have the same QEFF weights, and therefore it is sufficient to perform this operation only once.

\subsubsection{\label{sec: Complexity QGC algorithm}Complexity of the quantum generative classification algorithm}

In Table \ref{tab: complexity QGC algorithm}, we summarize the classical and quantum complexity of the training and test phases of the proposed quantum generative classification algorithm. We assume $N$ training and $K$ test samples. And, unlike $\epsilon$ and $\wp_j$, we denote $\epsilon^*$ the error in the estimation of the test data set and $\wp_k^*$ the minimum value of the joint of the test sample $k$ over all possible labels. It is worth highlighting that for both classical and quantum computers, the training complexity is dominated by the complexity of estimating the gradients, see previous subsection, and the test complexity corresponds to the complexity of estimating the projection $\bratwo{\psi_{_{\mathcb{X}}, _{\mathcb{Y}}} '}\rho_{{_{\mathcb{X}}}, {_\mathcb{Y}}}(\boldsymbol{\theta_{\text{op} } })\kettwo{\psi_{_{\mathcb{X}}, _\mathcb{Y}}' }$  over all the test samples.

\subsubsection{\label{sec: Complexity discussion}Complexity discussion}

The classical and quantum complexity of the QEFF mapping are $O(Dd^2)$ and $O(Dd)$, respectively. Thus, there is a theoretical advantage of constructing the QEFF mapping in quantum hardware. However, this possible quadratic advantage may not be an actual advantage if one considers the additional load of the error-correction code \cite{babbush2021errorcorrection}; we leave a comprehensive analysis of this complexity issue for future work.

In addition, comparing the classical and quantum complexities for the training and test phases of the QGC algorithm, see Table \ref{tab: complexity QGC algorithm},  we observe that to achieve a quantum advantage the error of the approximation of the training  $\epsilon$  and test $\epsilon^*$ steps must satisfy
\begin{equation}
\epsilon^2 > \frac{\log{(HdL)}\big(\log{(HdL)} + Dd\big)}{\big(N(HdL+Dd)+Dd^2\big)\sum_{j}\wp_j(1-\wp_j)}
\end{equation}
and
\begin{equation}
(\epsilon^*)^2 > \frac{\log{(HdL)} + Dd}{\big(K(HdL+Dd)+Dd^2\big)\sum_{k}\wp^*_k(1-\wp^*_k)}.
\end{equation}

Henceforth, the advantage is highly dependent on the unnormalized  joint probability densities (without the $M_h$ scaling value) of the training and test samples $\{\wp_j\}_{0\cdots N-1}$ and $\{\wp_k^*\}_{0\cdots K-1}$.  However, we have no prior knowledge of these values because the objective of the QGC algorithm is to estimate them, and therefore it is not possible to conclude a possible quantum advantage. Still, the previous inequalities indicate that to attain a lower error in the approximation the data must not be very sparse (the samples should not have low unnormalized joint density values).

\begin{figure*}
\centering
\includegraphics[scale=0.53]{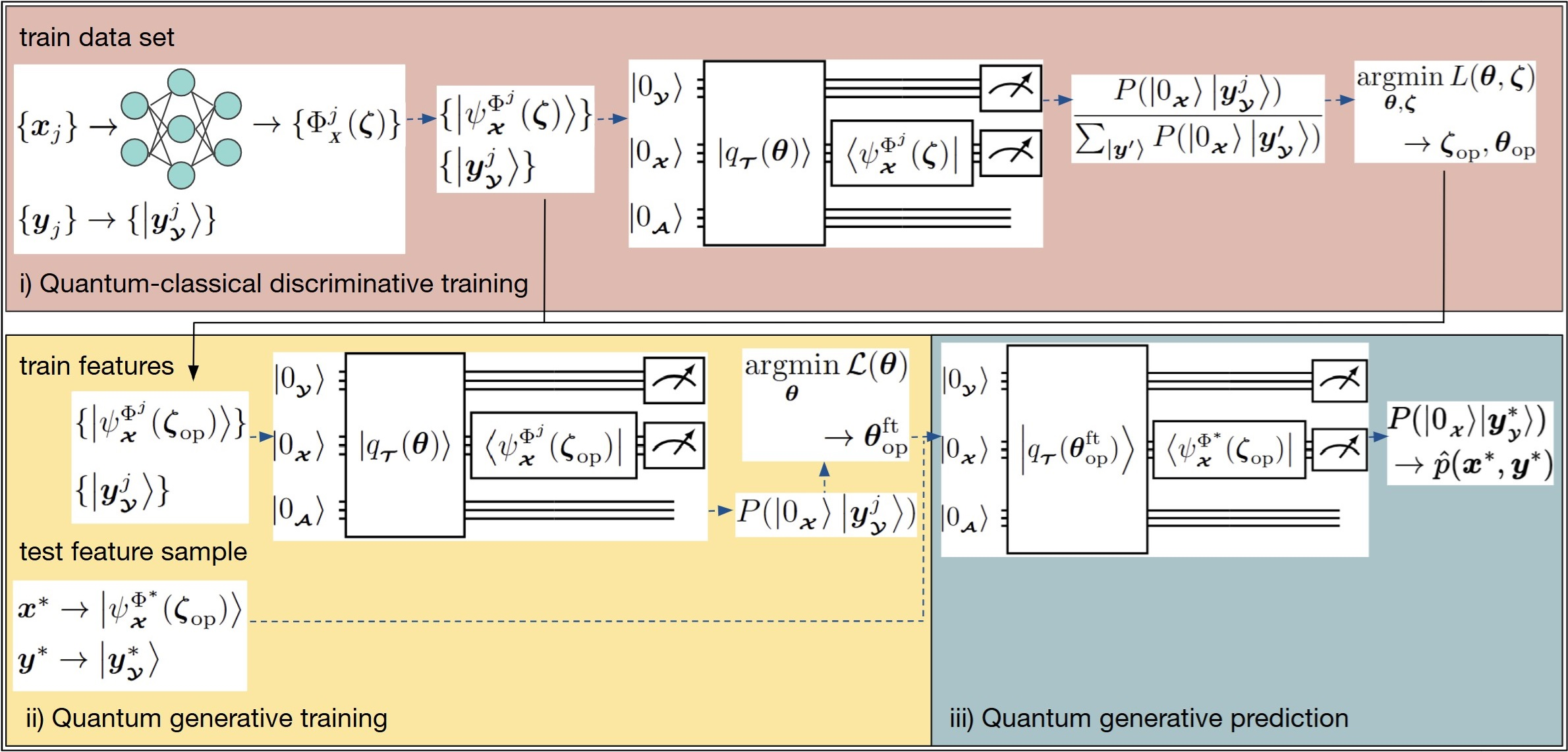}
\caption{D-QGC algorithm for hybrid classification of high-dimensional data based on the joint probability density of features and labels. In step (i), we use a deep neural network with parameters $\boldsymbol{\zeta}$ to map the training features to a latent space followed by the QEFF quantum map $\boldsymbol x_j \mapsto  \Phi_{_{\boldsymbol{\mathit{X}}}}^j(\boldsymbol\zeta)\mapsto \big|\psi_{_\mathcb{X}}^{\Phi^j}(\boldsymbol\zeta)\big\rangle$, the labels are mapped to a one-hot basis encoding $\boldsymbol y_j \mapsto \big|\boldsymbol y_{_\mathcb{Y}}^{j}(\boldsymbol\zeta)\big\rangle$; we also perform discriminative learning by maximizing the likelihood of the normalized training probabilities over all possible labels $\mathbb Y$ using the loss $L(\boldsymbol{\theta}, \boldsymbol{\zeta}) \sim -\sum_j\log{\{\hat f(\boldsymbol{x}_j, \boldsymbol{y}_j|\boldsymbol\zeta;\boldsymbol{\theta})/\sum_{\boldsymbol y'\in \mathbb Y} \hat f(\boldsymbol{x}_j, \boldsymbol{y}'|\boldsymbol\zeta;\boldsymbol{\theta})\}}$, where $\hat f(\boldsymbol{x}', \boldsymbol{y}''|\boldsymbol\zeta;\boldsymbol{\theta}) = \text{Tr}\big[\rho_{{_\mathcb{X}}, {_\mathcb{Y}}}(\boldsymbol\theta)\big|\psi_{_\mathcb{X}}^{\Phi'}(\boldsymbol\zeta)\big\rangle\big\langle\psi_{_\mathcb{X}}^{\Phi'}(\boldsymbol\zeta)\big|\otimes\kettwo{\boldsymbol y''_{_\mathcb{Y}}}\bratwo{\boldsymbol y''_{_\mathcb{Y}}}\big]$ and $\rho_{{_\mathcb{X}}, {_\mathcb{Y}}}(\boldsymbol \theta) = \text{Tr}_{_\mathcb{A}}\big[\kettwo{q_{_\mathcb{T}}(\boldsymbol \theta)}\bratwo{q_{_\mathcb{T}}(\boldsymbol \theta)}\big]$, in step (ii), we perform generative learning by extracting the optimal training states $\boldsymbol x^j \mapsto \big|\psi_{_\mathcb{X}}^{\Phi^j}(\boldsymbol\zeta_{\text{op}})\big\rangle$ and using them to maximize likelihood of the unnormalized probabilities $\mathcal {L}(\boldsymbol{\theta}) \sim -\sum_j\log{\hat f(\boldsymbol{x}_j, \boldsymbol{y}_j|\boldsymbol\zeta_{\text{op}};\boldsymbol{\theta})}$, which results in the finetuning of the training circuit parameters $\boldsymbol{\theta}_{\text{op}}\rightarrow\boldsymbol{\theta}_{\text{op}}^{\text{ft}}$, in step iii), we obtain an estimator of the joint probability density by projecting the finetuned density matrix with the joint test state $\hat{p}(\boldsymbol x^*,\boldsymbol y^*) \sim \hat f(\boldsymbol{x}^*, \boldsymbol{y}^*|\boldsymbol\zeta_{\text{op}};\boldsymbol{\theta}_{\text{op}}^{\text{ft}}) = \big\langle\psi_{_\mathcb{X}, _\mathcb{Y}}^{\Phi^*}(\boldsymbol\zeta_{\text{op}})\big|\rho_{{_\mathcb{X}}, {_\mathcb{Y}}}(\boldsymbol{\theta}_{\text{op}}^{\text{ft}})\big|\psi_{_\mathcb{X}, _\mathcb{Y}}^{\Phi^*}(\boldsymbol\zeta_{\text{op}})\big\rangle$, lastly, prediction is made from $\argmax_{{\boldsymbol{y}^*}\in \mathbb Y}\hat p(\boldsymbol{x}^*,\boldsymbol y^*)$.} {\label{fig: Deep QGC method}}
\end{figure*}

\section{\label{sec: D-QGC}Deep quantum-classical generative classification}

A challenging task when building a machine learning model based on kernel density estimation \cite{rosenblatt1956remarks, parzen1962estimation, gonzalez2022learning} consists of its application to high-dimensional data samples. In particular, the sparsity of the data increases in proportion to its dimensionality, resulting in poor pointwise estimates of the probability density. Previous works \cite{gonzalez2022learning, gonzález2024kdm, gallegomejia2024latentanomalydetectiondensity} have proposed the use of deep neural networks to reduce the dimensionality of the samples, and then apply the corresponding kernel-based method in this lower-dimensional space, also called latent space.

To overcome the curse of dimensionality, we propose the deep quantum-classical generative classification (D-QGC) algorithm. The method performs classification of high-dimensional data samples by combining classical and quantum neural networks to learn the joint probability density of inputs and outputs $p(\mathbf x, \mathbf y)$. The D-QGC model uses a deep neural network to map the data to a latent space and the QGC algorithm to learn the joint probability density in this lower-dimensional space. This strategy is also based on the combination of discriminative and generative learning for generative classification. Briefly, the variational quantum-classical algorithm, see Fig. \ref{fig: Deep QGC method}, can be divided into two training phases and one testing phase, as follows:

(i) Quantum-classical discriminative training: Given a high-dimensional data set of features and labels  $\{(\boldsymbol x_j, \boldsymbol y_j)\}_{0\cdots N-1}$ with $\boldsymbol x_j \in \mathbb{X} = \mathbb{R}^C$ and $\boldsymbol y_j \in \mathbb{Y} = \{0, \cdots, L-1\}$ and a test input sample $\boldsymbol x^* \in \mathbb{X}$ and candidate test label  $\boldsymbol y^* \in \mathbb{Y}$ whose density we aim to estimate. We start by preparing an end-to-end quantum-classical differentiable model designed to maximize the conditional probability $p(\mathbf y|\mathbf x)$ of the training data. The algorithm combines a deep neural network with the quantum discriminative training strategy presented in Sect. \ref{sec: quantum discriminative learning}. It consists of building for the input space a two-stage variational embedding of the form $\mathbb{X} \rightarrow \boldsymbol{\mathit{X}} \rightarrow \mathcal{X}$, with $\boldsymbol{\mathit{X}} = \mathbb{R}^D$ and $\mathcal{X} = \mathbb{C}^d$, where the function $\Phi: \mathbb{X} \rightarrow \boldsymbol{\mathit{X}}$ corresponds to the deep neural network that maps the data to the latent space and the function $\psi_{_\mathcb{X}}: \boldsymbol{\mathit{X}} \rightarrow \mathcal{X}$ to the QEFF mapping. In addition, we apply the one-hot-basis quantum encoding $\psi_{_\mathcb{Y}}: \mathbb{Y} \rightarrow \mathcal{Y} = \mathbb{R}^L$ to the label space. Specifically, an input sample is mapped to $\boldsymbol x_j \mapsto \Phi(\boldsymbol x_j|\boldsymbol\zeta) = \Phi_{_{\boldsymbol{\mathit{X}}}}^j(\boldsymbol\zeta)$, where $\boldsymbol\zeta$ are the trainable parameters of the classical network, followed by the QEFF encoding $\Phi_{_{\boldsymbol{\mathit{X}}}}^j(\boldsymbol\zeta) \mapsto \big|\psi_{_\mathcb{X}}^{\Phi^j}(\boldsymbol\zeta)\big\rangle$, while the outputs are mapped to $\boldsymbol y_j \mapsto \kettwo{\psi_{_\mathcb{Y}}^j}= \kettwo{\boldsymbol{y}_{_\mathcb{Y}}^j}$. We then use the training quantum circuit, see Fig. \ref{fig: QMC training quantum circuit}, to project the quantum features with the trainable density matrix $\rho_{{_\mathcb{X}}, {_\mathcb{Y}}}(\boldsymbol{\theta})$ with parameters $\boldsymbol{\theta}$. And finally we estimate the conditional probability by normalizing the outputs of the circuit, as in Eq. \ref{eq: dsicriminative training},  
\begin{align}
&\hat h(\boldsymbol{y}_j|\boldsymbol{x}_j;\boldsymbol{\theta}, \boldsymbol{\zeta}) = \notag \\
&\frac{\big\langle\psi_{_\mathcb{X}, _\mathcb{Y}}^{\Phi^j}(\boldsymbol\zeta)\big|\rho_{{_\mathcb{X}}, {_\mathcb{Y}}}(\boldsymbol\theta)\big|\psi_{_\mathcb{X}, _\mathcb{Y}}^{\Phi^j}(\boldsymbol\zeta)\big\rangle}{\sum_{\ket{\boldsymbol y'_{_\mathcb{Y}}}\in \mathcal{Y}}\text{Tr}\big[\rho_{{_\mathcb{X}}, {_\mathcb{Y}}}(\boldsymbol\theta)\big|\psi_{_\mathcb{X}}^{\Phi^j}(\boldsymbol\zeta)\big\rangle\big\langle\psi_{_\mathcb{X}}^{\Phi^j}(\boldsymbol\zeta)\big|\otimes\ket{\boldsymbol y'_{_\mathcb{Y}}}\bra{\boldsymbol y'_{_\mathcb{Y}}}\big]}
\end{align}
where $\big|\psi_{_\mathcb{X}, _\mathcb{Y}}^{\Phi^j}(\boldsymbol\zeta)\big\rangle = \big|\psi_{_\mathcb{X}}^{\Phi^j}(\boldsymbol\zeta)\big\rangle\otimes\big|\psi_{_\mathcb{Y}}^{j} \big\rangle$. 

As previously, we solve the minimization problem 
\begin{align}
    \boldsymbol{\theta}_{\text{op}}, \boldsymbol{\zeta}_{\text{op}} &= \argmin_{\boldsymbol \theta, \boldsymbol \zeta}L(\boldsymbol{\theta}, \boldsymbol{\zeta}) \notag \\
    &= \argmin_{\boldsymbol \theta, \boldsymbol \zeta}\bigg\{-\frac{1}{N}\sum_{j=0}^{N-1}\log{\hat h(\boldsymbol{y}_j|\boldsymbol{x}_j;\boldsymbol{\theta}, \boldsymbol{\zeta})}\bigg\}.
\end{align}

(ii) Quantum generative training: The previous discriminative training strategy allows the extraction of the training features in the latent space $\boldsymbol x_j \mapsto \Phi_{_{\boldsymbol{\mathit{X}}}}^j(\boldsymbol\zeta_{\text{op}})$ that maximize the classification accuracy. To build an estimator of the joint probability $\hat{p}(\mathbf x = \boldsymbol x^*, \mathbf y = \boldsymbol y^*)$, we fix these lower-dimensional features and fine-tune the quantum model using the generative learning approach. Indeed, we now neglect the normalization of the output of the training quantum circuit, see Eq. \ref{eq: QGC training}. And therefore, at this stage, we apply the proposed quantum generative algorithm to the quantum features $\{\Phi_{_{\boldsymbol{\mathit{X}}}}^j(\boldsymbol\zeta_{\text{op}}) \mapsto \big|\psi_{_\mathcb{X}}^{\Phi^j}(\boldsymbol\zeta_{\text{op}})\big\rangle\}_{0\cdots N-1}$, setting the parameters $\boldsymbol\zeta_{\text{op}}$ fixed, and fine-tuning the starting variational parameters $\boldsymbol\theta_{\text{op}}$. The estimator of the joint probability density then corresponds to 
\begin{align}
\hat f(\boldsymbol{x}_j, \boldsymbol{y}_j|\boldsymbol\zeta_{\text{op}};\boldsymbol{\theta}) =
M_h\big\langle\psi_{_\mathcb{X}, _\mathcb{Y}}^{\Phi^j}(\boldsymbol\zeta_{\text{op}})\big|\rho_{{_\mathcb{X}}, {_\mathcb{Y}}}(\boldsymbol\theta)\big|\psi_{_\mathcb{X}, _\mathcb{Y}}^{\Phi^j}(\boldsymbol\zeta_{\text{op}})\big\rangle,
\end{align}
where $\big|\psi_{_\mathcb{X}, _\mathcb{Y}}^{\Phi^j}(\boldsymbol\zeta_{\text{op}})\big\rangle =  \big|\psi_{_\mathcb{X}}^{\Phi^j}(\boldsymbol\zeta_{\text{op}})\big\rangle\otimes\big|\psi_{_\mathcb{Y}}^{j} \big\rangle$.
Finally, the fine-tuning of $\boldsymbol{\theta}_{\text{op}}\rightarrow\boldsymbol{\theta}_{\text{op}}^{\text{ft}}$ follows the minimization procedure 
\begin{align}
    \boldsymbol{\theta}_{\text{op}}^{\text{ft}} &= \argmin_{\boldsymbol \theta}\mathcal{L}(\boldsymbol{\theta})\notag \\
    &= \argmin_{\boldsymbol \theta}\bigg\{-\frac{1}{N}\sum_{j=0}^{N-1}\log{\hat f(\boldsymbol{x}_j, \boldsymbol{y}_j|\boldsymbol\zeta_{\text{op}};\boldsymbol{\theta})}\bigg\}.
\end{align}

%In Appendix [Ref?], we present the motivation of using the intermediate quantum-classical discriminative strategy for building the quantum-classical generative model.

(iii) Quantum generative testing: To estimate the joint probability of the high-dimensional test sample $\boldsymbol x^*\in \mathbb{X}$ and the class $\boldsymbol y^*\in \mathbb{Y}$, we estimate the expected value of the quantum test state $\boldsymbol x^* \mapsto \Phi_{_{\boldsymbol{\mathit{X}}}}^*(\boldsymbol\zeta_{\text{op}}) \mapsto \big|\psi_{_\mathcb{X}}^{\Phi^*}(\boldsymbol\zeta_{\text{op}})\big\rangle$ with the optimized training density matrix $\rho_{_{\mathcb{X},\mathcb{Y}}}(\boldsymbol{\theta}_{\text{op}}^{\text{ft}})$ using the test quantum circuit; see Fig. \ref{fig: QMC testing quantum circuit}. The estimator of the joint probability density $p(\boldsymbol{x}^*, \boldsymbol{y}^*)$ would then be given by 
\begin{align} 
\hat p(\boldsymbol{x}^*, \boldsymbol{y}^*) &= \hat f(\boldsymbol{x}^*, \boldsymbol{y}^*|\boldsymbol\zeta_{\text{op}};\boldsymbol{\theta}_{\text{op}}^{\text{ft}})  \notag \\ &= M_h\big\langle\psi_{_\mathcb{X}, _\mathcb{Y}}^{\Phi^*}(\boldsymbol\zeta_{\text{op}})\big|\rho_{{_\mathcb{X}}, {_\mathcb{Y}}}(\boldsymbol{\theta}_{\text{op}}^{\text{ft}})\big|\psi_{_\mathcb{X}, _\mathcb{Y}}^{\Phi^*}(\boldsymbol\zeta_{\text{op}})\big\rangle,
\end{align}
with $\big|\psi_{_\mathcb{X}, _\mathcb{Y}}^{\Phi^*}(\boldsymbol\zeta_{\text{op}})\big\rangle =  \big|\psi_{_\mathcb{X}}^{\Phi^*}(\boldsymbol\zeta_{\text{op}})\big\rangle\otimes\big|\psi_{_\mathcb{Y}}^{*} \big\rangle$ and $M_h = (2\pi h^2)^{-D/2}$. Finally, the predicted label would result from $\argmax_{{\boldsymbol{y}^*}\in \mathbb{Y}}\hat{p}(\mathbf x = \boldsymbol x^*, \mathbf y = \boldsymbol y^*)$. 

\section{\label{sec: Results}Method evaluation}

\subsection{One- and two-dimensional binary classification\label{sec: Results two-dimensional data sets}}

We demonstrate the proposed quantum generative method for the binary classification of one- and two-dimensional data sets. The algorithm obtains the joint probability density of the inputs and outputs and predicts the class with the largest joint probability density. To evaluate the performance of the proposed QGC model with the QEFF mapping, we also compare it with two other state-of-the-art quantum feature maps, the QRFF embedding \cite{useche2022quantum}, and the ZZ feature map (ZZFM) \cite{havlivcek2019supervised}; a description of the latter is presented in the \ref{app: baselines ZZFM}.

\subsubsection{Data sets\label{sec: Two-dimensional data sets}}

We worked with a one-dimensional data set and three two-dimensional data sets. Each consisted of a training, test, and out-of-distribution (OOD) data set; an OOD data set is a data set that does not follow the probability distribution of the training data. Indeed, it is critical for the generalizability of a machine learning model to correctly predict unseen data samples that do not necessarily follow the training distribution \cite{caro2023out}. As usual in machine learning, we used the training data to train the model and the test data to evaluate its accuracy (ACC). In addition, the OOD data set was used to evaluate the resulting estimates of the joint probability density of inputs and outputs.

\textit{One-dimensional data set:} The one-dimensional data set consisted of a sample of points distributed from a mixture of two Gaussians for class 0 and a single Gaussian for class 1. The training and test data sets had sizes of 900 and 100, respectively. And both partitions had roughly a the relative frequency of samples of two-thirds for class 0 and one-third for class 1. In addition, the OOD data set corresponded to 500 points sampled from a uniform distribution in the range $[-7, 14]$.

\textit{Two-dimensional data sets:} We evaluated three two-dimensional data sets called \textit{Moons}, \textit{Circles}, and \textit{Spirals}, shown respectively from top to bottom in Fig. \ref{fig: 2-Dimensional Classification}. The \textit{Moons} and the \textit{Circles} data sets had 1800 samples for training and 200 samples for testing, while the \textit{Spirals} data set was divided into 900 training and 100 test data samples; all partitions of the data sets had approximately the same number of samples for each class. Furthermore, for all configurations, the OOD data set corresponded to 400 points uniformly distributed in the two-dimensional space $[0, 1]\times[0, 1]$.

\begin{figure*}
\centering
\includegraphics[scale=0.51]{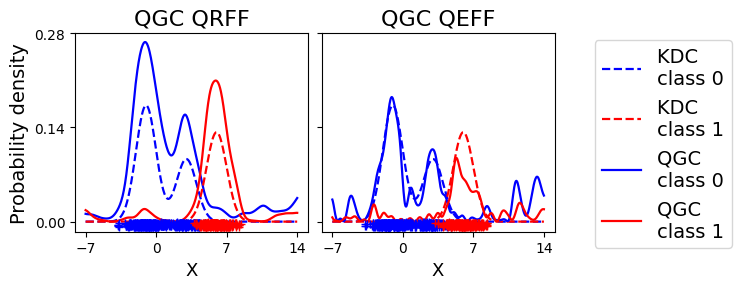}
\caption{Quantum generative classification on the one-dimensional binary data set. The estimated joint density with the QGC is shown with solid lines, while the true joint probability density computed with the classical KDC method is depicted with dashed lines.}{\label{fig: 1-Dimensional Classification}}
\end{figure*}

\subsubsection{Setup}

All one-dimensional and two-dimensional quantum demonstrations had similar setups. For instance, we used eight qubits to build the training and test quantum circuits (see Figs. \ref{fig: QMC training quantum circuit} and \ref{fig: QMC testing quantum circuit}) using a quantum simulator of the \textit{tensor-circuit} library \cite{Zhang2023tensorcircuit}. These circuits were divided into one qubit for the labels, five qubits for the features, corresponding to 32 QEFF components, and two auxiliary qubits. To construct the variational training density matrix, we explored the use of the hardware-efficient ansatz \cite{Kandala2017HEA}; this ansatz had $2\times 8 \times (31+1) = 512$ parameters, corresponding to the application of $31$ HEA quantum layers to the eight qubits. The bandwidth of the Gaussian kernel $h$ of the \textit{1-Dimensional}, \textit{Moons}, \textit{Circles}, and \textit{Spirals} data sets was set to $2^{-3/2}$, $2^{-4}$, $2^{-7/2}$, and $2^{-9/2}$, respectively. In addition, to estimate the true joint probability density of inputs and outputs, we applied to each configuration the classical kernel density classification algorithm (see Sect. \ref{sec: Classification with RKHS}) using the same Gaussian bandwidths of the QGC. As mentioned before we used the test data to determine the accuracy of the models and the OOD data set to evaluate the joint probability density values. For the latter, we compared the results of the QGC algorithm with the results of the KDC method. The metrics used were accuracy, mean absolute error (MAE), and spearman correlation per class (SPC($\cdot$)), where the dot indicates the class. The MAE measures the distance between two probability densities and the SPC($\cdot$) evaluates their relative order by assessing their monotonic relationship.

\begin{table}
  \centering
  \caption{Parameters of the QGC model for both one‑ and two‑dimensional data sets. The parameters of the QGC with the A‑ZZFM apply only to the 2D data. The total number of qubits is \(n_{_\mathcb{T}}\) and \((n_{_\mathcb{A}},n_{_\mathcb{X}},n_{_\mathcb{Y}})\) is the number of ancilla, input, and output qubits. The total circuit depth is given by the number of CNot gates, and Depth (FM, HEA) indicates the depth of the quantum map and of the HEA ansatz, respectively. Also, the number of trainable parameters (Train.\ params.) of the state \(\rho_{_{\mathcb{X}}, _\mathcb{Y}}(\boldsymbol\theta)\) depends on the number of HEA layers \(T\) by \(\lvert\boldsymbol\theta\rvert = 2n_{_\mathcb{T}}(T+1)\).}
  \label{table: params 1D2D classification}
  \begin{tabular}{llll}
    \toprule
    Parameters                       & QGC A‑ZZFM     & QGC QRFF       & QGC QEFF       \\
    \midrule
    \(n_{_\mathcb{T}}\)               & 8              & 8              & 8              \\
    \((n_{_\mathcb{A}},n_{_\mathcb{X}},n_{_\mathcb{Y}})\)
                                     & (2,5,1)        & (2,5,1)        & (2,5,1)        \\
    HEA layers                       & 31             & 31             & 31             \\
    Dim.\ QRFF\&QEFF                 & –              & 32             & 32             \\
    Total depth                      & 257            & 243            & 243            \\
    Depth (FM, HEA)                  & (40,217)       & (26,217)       & (26,217)       \\
    Train.\ params.                  & 512            & 512            & 512            \\
    \bottomrule
  \end{tabular}
\end{table}

\subsubsection{Baseline setup}

In addition to the QGC with QEFF model, we implemented, as baseline, the same QGC strategy but using two other quantum encodings of the state-of-the-art: the QRFF \cite{useche2022quantum} and the ZZ feature map \cite{havlivcek2019supervised}. Given the similarities between QRFF and QEFF embeddings, both models were set with the same number of parameters, as shown in Table \ref{table: params 1D2D classification}; in particular, we also used 5 qubits for QRFF mapping. Regarding the ZZFM benchmarking model, we did not implement it for the one-dimensional case, because in this scenario the model is not comparable to the other two quantum models. In fact, the number of qubits required to construct the ZZFM corresponds to the dimension of the input data (one qubit in this case), see Appendix \ref{app: baselines ZZFM}. Nevertheless, for the two-dimensional data sets, we developed the augmented ZZFM (A-ZZFM) which allows a fair evaluation of this mapping with the QRFF and QEFF. Unlike the standard ZZFM, which uses two qubits to prepare a two-dimensional input feature $( x_1, x_2)$, the A-ZZFM performs the following transformation $(x_1, x_2) \mapsto ( x_1, x_2, {x_1}^{2}, {x_2}^{2}, x_1x_2)$ which increases the input dimension to $5$ and allows the application of the standard ZZFM over the augmented feature space. Besides enriching the feature space, this extension allows to adapt the ZZ feature map to the same $5$ qubits used in the preparation of the quantum Fourier mappings. Additionally, the other set-up parameters of the A-ZZFM and QEFF methods were also similar. For instance, in both models we also used the same number of qubits for the ancilla and for the labels, and the same structure and number of trainable parameters for the HEA ansatz, as shown in Table \ref{table: params 1D2D classification}.

\begin{figure*}
\centering
\includegraphics[scale=0.65]{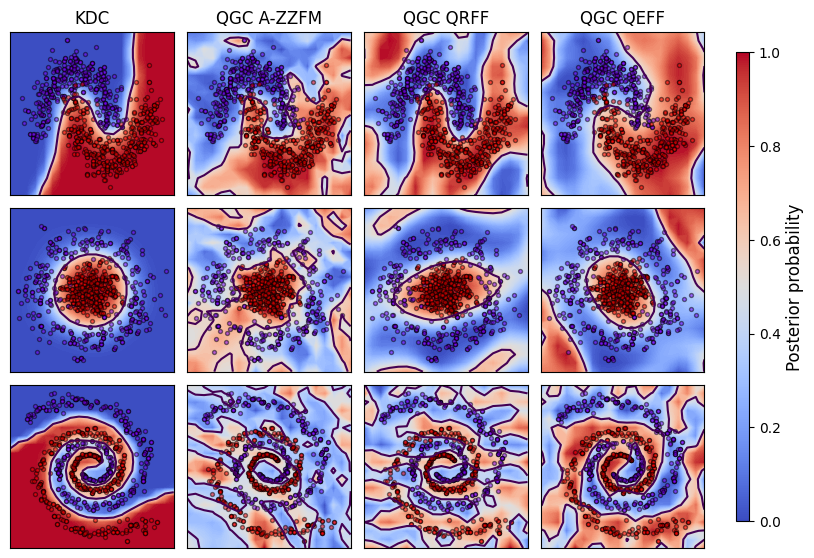}
\caption{Posterior probability $\hat p(\boldsymbol{y^*}|\boldsymbol{x^*})$, derived from the joint $\hat p(\boldsymbol{x^*}, \boldsymbol{y^*})$ of the quantum generative classification algorithm on the two-dimensional binary data sets. From top to bottom, the data sets correspond to \textit{Moons}, \textit{Circles} and \textit{Spirals}. The proposed QGC with the QEFF strategy is shown in the right column, while the QGC with the A-ZZFM \cite{havlivcek2019supervised}, QRFF \cite{useche2022quantum} baseline mappings are shown in the middle figures. The expected distribution calculated with the classical KDC is shown in the left column. For readability, each plot includes only 35\% of the data samples.}
{\label{fig: 2-Dimensional Classification}}
\end{figure*}

\begin{table}
  \centering
  \caption{Binary classification accuracy (ACC) on the test partition of the one‑dimensional data set. The QGC algorithm with both QRFF and QEFF was built using 2 ancilla qubits, 5 input qubits, and 1 output qubit. The best result is shown in bold.}
  \label{tab:one-dim classification}
  \begin{tabular}{lll}
    \toprule
     & QGC QRFF     & QGC QEFF     \\
    \midrule
    ACC & \(\mathbf{0.980}\) & 0.970         \\
    \bottomrule
  \end{tabular}
\end{table}

\begin{table}
  \centering
  \caption{Joint probability density estimation results over the one‑dimensional data set using the QGC method with the QRFF and QEFF. The estimations were evaluated using the one‑dimensional out‑of‑distribution data set. The target distribution was calculated using the classical KDC. The QGC model was built using 2 ancilla qubits, 5 input qubits, and 1 output qubit for the class. The metrics used were the mean average error (MAE) and the Spearman’s correlation per class (SPC(\(\cdot\))). The best results are shown in bold.}
  \label{tab:one-dim density estimation}
  \begin{tabular}{lll}
    \toprule
            & QGC QRFF      & QGC QEFF      \\
    \midrule
    MAE     & 0.027         & \(\mathbf{0.013}\) \\
    SPC(0)  & \(\mathbf{0.731}\) & 0.515         \\
    SPC(1)  & 0.484         & \(\mathbf{0.561}\) \\
    \bottomrule
  \end{tabular}
\end{table}

\subsubsection{Results and discussion}

We present the results of the quantum generative classification method in Figs. \ref{fig: 1-Dimensional Classification} and \ref{fig: 2-Dimensional Classification} and Tables \ref{tab:one-dim classification}, \ref{tab:one-dim density estimation}, \ref{tab: two-dim classification}, and \ref{tab: two-dimensional density estimation}. The results show that the proposed QGC algorithm is capable of estimating the joint probability density of features and labels for the classification of a low dimensional data set.

Starting with the one-dimensional data set, for each quantum Fourier encoding, we plot the class-wise joint probability densities obtained by the QGC method using solid lines and the true joint probability densities estimated by the KDC method with dashed lines. Tables \ref{tab:one-dim classification} and \ref{tab:one-dim density estimation} show that the QGC model achieved good accuracy results and was able to approximate the joint probability density of the characteristics and labels. When comparing the two approaches, both mappings yielded similar performance in terms of the joint probability density estimation, however the QRFF map had a better classification accuracy.

\begin{table}
  \centering
  \caption{Binary classification accuracy (ACC) on the test partition of the two‑dimensional data sets. The demonstrations of QGC strategy for all three quantum maps were built using 2 ancilla qubits, 5 qubits for the input space, and 1 qubit for the classes. The best results are shown in bold.}
  \label{tab: two-dim classification}
  \begin{tabular}{llll}
    \toprule
           & QGC A‑ZZFM            & QGC QRFF                        & QGC QEFF                       \\
    \midrule
    ACC    & \multirow{2}{*}{0.915}        & \multirow{2}{*}{\(\mathbf{0.960}\)} & \multirow{2}{*}{0.955}       \\
    \textit{Moons} &                      &                                 &                                \\
    \addlinespace
    ACC    & \multirow{2}{*}{0.915}        & \multirow{2}{*}{0.905}         & \multirow{2}{*}{\(\mathbf{0.945}\)} \\
    \textit{Circles} &                    &                                 &                                \\
    \addlinespace
    ACC    & \multirow{2}{*}{0.740}        & \multirow{2}{*}{0.800}         & \multirow{2}{*}{\(\mathbf{0.940}\)} \\
    \textit{Spirals} &                    &                                 &                                \\
    \bottomrule
  \end{tabular}
\end{table}

Moreover, Fig. \ref{fig: 2-Dimensional Classification} illustrates the classification boundaries and the conditional probability densities $p(\boldsymbol{y}^*|\boldsymbol x^*)$ obtained from the joint $p(\boldsymbol x^*, \boldsymbol{y}^*)$ for the three two-dimensional data sets using the QGC strategy and the three previously discussed quantum embeddings; this figure also shows the posterior probabilities of the classical KDC algorithm. These figures along with the ACC and SPC($\cdot$) metrics presented in Tables \ref{tab: two-dim classification} and \ref{tab: two-dimensional density estimation} show that the proposed QGC with QEFF model generally outperformed the two other quantum encodings for both the classification and joint probability density estimation tasks. We should highlight that the QGC method with the A-ZZFM achieved better results in terms of MAE, indicating that the probability density functions obtained with the quantum Fourier embeddings were not fully normalized. This normalization could have been improved by increasing the number of Fourier features; nevertheless, we were limited by the computational resources. However, for many machine learning applications \cite{Gallego2023demande, useche2022quantum, Bustos2023addmkde}, it is rather important to obtain an appropriate monotonic relationship between the true and estimated probability density functions, which is captured by the Spearman's correlation.

%\begingroup
%  \scriptsize
%  \setlength{\tabcolsep}{1pt}
%  \renewcommand{\arraystretch}{0.7}
%\begin{table*}
%  \caption{\label{tab: two-dimensional density estimation}%
\begin{table*}
  \centering
  \caption{Results of the joint probability density estimation over the two‑dimensional data sets using the QGC strategy with the A‑ZZFM \cite{havlivcek2019supervised}, QRFF \cite{useche2022quantum}, and QEFF mappings. For each data set, we evaluated the model using a two‑dimensional out‑of‑distribution data set. The target density estimation was calculated using the classical KDC. The QGC quantum demonstrations were built using 2 ancilla qubits, 5 qubits for the input space, and 1 qubit for the classes. We used as metrics the mean average error (MAE) and the Spearman’s correlation per class (SPC(\(\cdot\))). The best results are shown in bold.}
  \label{tab: two-dimensional density estimation}
  \begin{tabular}{@{}*{10}{l}@{}}
        \toprule

        %{2}{*}
        %\multirow{2}{*}{\(\mathbf{0.351}\)}
               & \multicolumn{3}{c}{\textit{Moons}}   & \multicolumn{3}{c}{\textit{Circles}} & \multicolumn{3}{c}{\textit{Spirals}} \\
               \cmidrule(lr){2-4} \cmidrule(lr){5-7}  \cmidrule(lr){8-10}
             %& &               &                   \\
               & MAE &  SPC(0) & SPC(1)
               & MAE & SPC(0) & SPC(1)
               & MAE  & SPC(0) & SPC(1) \\
        %\ns     & \crule{3}                    & \crule{3}                    & \crule{3}                    \\
        \midrule
       QGC     & \multirow{2}{*}{\(\mathbf{0.351}\)} & \multirow{2}{*}{0.608} & \multirow{2}{*}{0.592}

                 & \multirow{2}{*}{\(\mathbf{0.351}\)} & \multirow{2}{*}{0.368} & \multirow{2}{*}{0.238}
                 & \multirow{2}{*}{\(\mathbf{0.371}\)} & \multirow{2}{*}{\(-0.032\)}& \multirow{2}{*}{0.102} \\
        A‑ZZFM &  &  & 
         &  & & 
         & & &  \\
    %\ns     & \crule{3}                    & \crule{3}                    & \crule{3}                    \\
        \midrule
        QGC  &  \multirow{2}{*}{0.631}&\multirow{2}{*}{0.588} & \multirow{2}{*}{0.503} 

                 & \multirow{2}{*}{0.866} & \multirow{2}{*}{0.814} & \multirow{2}{*}{0.416} 
                 & \multirow{2}{*}{3.480} & \multirow{2}{*}{0.409} & \multirow{2}{*}{0.409} \\

      QRFF  &  &  & 
         &  & & 
         & & &  \\
    %\ns     & \crule{3}                    & \crule{3}                    & \crule{3}                    \\
        \midrule
        QGC    & \multirow{2}{*}{1.219} & \multirow{2}{*}{\(\mathbf{0.682}\)} & \multirow{2}{*}{\(\mathbf{0.696}\)} 
                 & \multirow{2}{*}{0.564}  & \multirow{2}{*}{\(\mathbf{0.844}\)}  & \multirow{2}{*}{\(\mathbf{0.568}\)}
                 & \multirow{2}{*}{1.461} & \multirow{2}{*}{\(\mathbf{0.607}\)} & \multirow{2}{*}{\(\mathbf{0.613}\)} \\

        QEFF &  &  & 
         &  & & 
         & & &  \\
        \bottomrule
      \end{tabular}%
\end{table*}
    %}% end \resizebox
  %}% end \makebox
%\end{table*}
%\endgroup

\subsection{4x4 MNIST and 4x4 Fashion-MNIST binary classification}

We also evaluated the proposed quantum generative classification algorithm on the down-sampled 4x4 MNIST and 4x4 Fashion-MNIST (F-MNIST) binary data sets and compared it with two other quantum models from the literature \cite{farhi, dilip2022data}.

\subsubsection{Data sets}

We used the quantum generative classification strategy to perform binary classification on the MNIST and the Fashion-MNIST data sets; both data sets consist of images of 28x28 pixels. The MNIST data set corresponds to a set of handwritten numbers between $0$ and $9$, while Fashion-MNIST data set corresponds to images of ten types of clothing. We performed binary classification by chosing the numbers $3$ and $6$ for MNIST data set and the classes \textit{T-shirt/top} and \textit{Trouser} for Fashion-MNIST data set. Each data set was balanced and corresponded to 12000 images for training and 2000 images for test. To compare the model with other state-of-the-art quantum algorithms \cite{farhi, dilip2022data}, we down-sampled the images from 28x28 to 4x4 pixels using the bilinear interpolation function of the \textit{tensorflow} python library and flattened each resulting image into a vector of $16$ components.

\begin{table}
  \centering
  \caption{Parameters of the QGC and the two baseline models for binary classification of the 4×4 MNIST and 4×4 Fashion‑MNIST data sets. The total number of qubits is denoted by \(n_{_\mathcb{T}}\), and the number of ancilla, input, and output qubits by \((n_{_\mathcb{A}},n_{_\mathcb{X}},n_{_\mathcb{Y}})\). Depth (FM, AS) indicates respectively the depth of the feature map and of the ansatz. CCE and ANL correspond respectively to categorical cross entropy and average negative log‑likelihood. The table also includes the number of ansatz layers, the loss, the number of trainable parameters (Train.\ params.), and the kernel bandwidth of the QGC.}
  \label{tab: params 4x4 MNIST classification}
  \begin{tabular}{llll}
    \toprule
    Parameters                       & Farhi \textit{et al.} \cite{farhi} & Dilip \textit{et al.} \cite{dilip2022data} & QGC QEFF \\
    \midrule
    \(n_{_\mathcb{T}}\)              & 17                                  & 5                                           & 8        \\
    \((n_{_\mathcb{A}},n_{_\mathcb{X}},n_{_\mathcb{Y}})\)
                                     & (0,16,1)                          & (0,5,1)                                   & (1,6,1) \\
    Ansatz layers                    & 3                                   & 2                                           & 6        \\
    Loss                             & Hinge                               & CCE                                         & ANL      \\
    Dim.\ QEFF                       & –                                   & –                                           & 64       \\
    Bandwidth \(h\)                  & –                                   & –                                           & 2.0      \\
    Total depth                      & 96                                  & 76                                          & 99       \\
    Depth (FM, AS)                   & (0,96)                             & (52,24)                                    & (57,42) \\
    Train.\ params.                  & 96                                  & 120                                         & 112      \\
    \bottomrule
  \end{tabular}
\end{table}

\subsubsection{Setup}

We used 8 qubits to build the quantum generative classification algorithm, corresponding to 1 ancilla qubit, 6 qubits for the inputs, and 1 qubit for the labels, as shown in Table \ref{tab: params 4x4 MNIST classification}. The 6-qubits of the QEFF encoding mapped the $16$-dimensional image vectors to a $64$-dimensional feature space, drawing the Fourier weights from a normal distribution $\{\boldsymbol\vartheta_{\alpha_{n_6}, \cdots, \alpha_1}\}_{\alpha_{n_{6}}, \cdots, \alpha_1 \in \{0, 1\}} \sim \mathcal{N}(\mathbf{0}, 4/(2^6-1)\mathbf{I}_{16})$; although we could have used other symmetric sampling distributions (e.g. a uniform distribution). Furthermore, we used $6$ HEA layers to construct the training density matrix, corresponding to $112$ trainable parameters and a total circuit depth of $97$ CNOT gates. Finally, we set the Gaussian kernel bandwidth to $h = 2.0$.

\subsubsection{Baseline setup}

We have implemented two variational quantum algorithms from Farhi et al. \cite{farhi} and Dilip et al. \cite{dilip2022data} to benchmark the proposed quantum generative classification model. See Appendix \ref{app: baselines} for a detailed description of these two quantum models.

Both baseline models follow the usual variational quantum approach, i.e., a quantum feature map of the data and a variational quantum ansatz that is optimized using an appropriate loss function. In the first place, the quantum circuit model of Farhi \textit{et al.} \cite{farhi} uses a total of $17$ qubits, divided into $16$ qubits for the quantum basis encoding and $1$ readout qubit for the two-class discrimination. To construct the variational quantum ansatz, we used $3$ quantum layers, resulting in a quantum circuit with a total depth of $96$ CNOT gates and $96$ trainable parameters. In addition, the algorithm makes use of the hinge loss function for its optimization. Secondly, although the quantum model of Dilip \textit{et al.} \cite{dilip2022data} was originally developed to classify 28x28 pixel images, we adapted the algorithm for the classification of the 4x4 downsampled images. This quantum circuit had a total of $5$ qubits, of which all $5$ qubits were used to encode the $16$-dimensional image vectors using FRQI quantum mapping \cite{dilip2022data} and a single qubit was measured to obtain the predictions. Following the original implementation, we constructed the variational ansatz using 2 quantum layers based on matrix product states (MPS) \cite{huggins2019towards}. The quantum circuit had a total depth of $76$ CNOT gates, $120$ trainable parameters, and was optimized using the categorical cross-entropy (CCE) loss function.

Table \ref{tab: params 4x4 MNIST classification} summarizes the parameters of the baseline methods and the proposed QGC algorithm. For a fair comparison, some hyper-parameters were chosen to ensure that the models had similar circuit depth and number of trainable parameters.

\begin{table}
  \centering
  \caption{Binary classification results of the 4×4 MNIST and 4×4 Fashion‑MNIST data sets. The original images were downsampled from 28×28 pixels to 4×4 pixels for both data sets. We show the accuracy (ACC) on the test partition of the proposed QGC strategy and the two benchmarking quantum algorithms \cite{farhi,dilip2022data}. The best results are shown in bold.}
  \label{tab: results 4x4 MNIST FMNIST classification}
  \begin{tabular}{llll}
    \toprule
           & Farhi \textit{et al.} \cite{farhi} & Dilip \textit{et al.} \cite{dilip2022data} & QGC QEFF \\
    \midrule
    ACC     & \multirow{2}{*}{0.909}              & \multirow{2}{*}{\(\mathbf{0.911}\)}         & \multirow{2}{*}{0.893}       \\
    MNIST   &                                     &                                             &                              \\
    \addlinespace
    ACC     & \multirow{2}{*}{0.686}              & \multirow{2}{*}{\(\mathbf{0.725}\)}         & \multirow{2}{*}{0.686}       \\
    F‑MNIST &                                     &                                             &                              \\
    \bottomrule
  \end{tabular}
\end{table}

\subsubsection{Results and discussion}

In Table \ref{tab: results 4x4 MNIST FMNIST classification}, we present the results of the binary classification of the 4x4 MNIST and 4x4 Fashion-MNIST data sets. These quantum demonstrations illustrate that the proposed quantum generative classification algorithm is competitive against other quantum algorithms of the state-of-the-art \cite{farhi, dilip2022data}. Note that we did not evaluate the joint probability density of the models because the baseline methods were developed for the discriminative task, while the QGC algorithm uses the generative learning approach. Thus, the main advantage of the QGC algorithm is that it is able to achieve both a competitive classification accuracy and an estimate of the joint probability density of inputs and outputs.

\subsection{MNIST and Fashion-MNIST 10-classes classification}

We also tested the deep quantum-classical generative classification algorithm using the proposed QEFF encoding on the ten classes of the MNIST and Fashion-MNIST data sets. The D-QGC method fuses a deep neural network with the QGC strategy for generative learning. To benchmark the proposed model, we implemented the same D-QGC strategy but using the ZZFM \cite{havlivcek2019supervised}. Results show that D-QGC with QEFF model is suitable for high-dimensional data sets, achieving competitive performance in the classification and density estimation tasks.

\subsubsection{Data sets}

In these quantum demonstrations, we use the full MNIST and Fashion-MNIST data sets, which, as mentioned earlier, correspond, respectively, to 28x28 pixel grayscale images of handwritten numbers and clothing types. Both data sets are divided into $10$ classes, with $60000$ images for training and $10000$ images for testing.

To evaluate performance for the joint density estimation task, for each the MNIST and Fashion-MNIST data sets, we built a synthetic data set, which we term latent out-of-distribution (L-OOD) data set. Each corresponded to 10000 instances in $\mathbb{R}^4$ sampled from a uniform distribution in the latent space. For their construction, we recall that the D-QGC algorithm uses a deep neural network to map the raw images ($\mathbb{R}^{28\times 28}$) to a latent space in $\mathbb{R}^D$, $\boldsymbol x_j \mapsto \Phi_{_{\boldsymbol{\mathit{X}}}}^j(\boldsymbol\zeta_{\text{op}})$, in our case the latent space had a dimension of $D = 4$. Henceforth, the L-OOD data set was built by taking samples from the distribution $\bigotimes_{i=0}^{D-1}U[a_i, b_i]$, where $a_i =\min_{0 \le j\le N-1} \Phi_{_{\boldsymbol{\mathit{X}}}}^{i, j} (\boldsymbol\zeta_{\text{op}})$, $b_i =\max_{0 \le j\le N-1} \Phi_{_{\boldsymbol{\mathit{X}}}}^{i, j}(\boldsymbol\zeta_{\text{op}})$, and $\Phi_{_{\boldsymbol{\mathit{X}}}}^{i, j}(\boldsymbol\zeta_{\text{op}})$ is the $i^\text{th}$ component of the latent sample $\Phi_{_{\boldsymbol{\mathit{X}}}}^j(\boldsymbol\zeta_{\text{op}})$. This data set is guaranteed to not follow the distribution of the latent features of the training data set.

\begin{table}
  \centering
  \caption{Parameters of the D‑QGC on the ten‑class MNIST and Fashion‑MNIST data sets. \(n_{_\mathcb{T}}\) denotes the total number of qubits, \((n_{_\mathcb{A}},n_{_\mathcb{X}},n_{_\mathcb{Y}})\) the number of ancilla, input, and output qubits, and Depth (FM, HEA) the depth of the feature map and of the HEA. We also include the size of the latent space, the number of HEA layers, the dimension of the QEFF, the kernel bandwidth, the total circuit depth, and the number of trainable parameters (Train.\ params.) for the classical (CNN) and the quantum neural network (QGC).}
  \label{tab: params 10-classes MNIST classification}
  \begin{tabular}{lll}
    \toprule
    Parameters               & QGC ZZFM & QGC QEFF \\
    \midrule
    \(n_{_\mathcb{T}}\)       & 9        & 9         \\
    \((n_{_\mathcb{A}},n_{_\mathcb{X}},n_{_\mathcb{Y}})\)
                              & (1,4,4) & (1,4,4)  \\
    Latent space             & 4        & 4         \\
    HEA layers               & 56       & 56        \\
    Dim.\ QEFF               & –       & 16        \\
    Bandwidth \(h\)          & –       & 2.0       \\
    Total depth              & 472      & 459       \\
    Depth (FM, HEA)          & (24,448)& (11,448) \\
    CNN Train.\ params.      & 234128   & 234128    \\
    QGC Train.\ params.      & 1026     & 1026      \\
    \bottomrule
  \end{tabular}
\end{table}

\subsubsection{Setup}

As explained in Sect. \ref{sec: D-QGC}, the D-QGC strategy with the QEFF quantum map has two successive training phases: a discriminative and a generative learning phase. For the discriminative step, we built an end-to-end quantum-classical neural network. It consisted of a modified classical LeNet network, see Appendix \ref{app: Model architecture}, followed by the QEFF mapping and the QGC training quantum circuit. This hybrid network transformed the data into the form $\mathbb R^{28\times 28} \rightarrow \mathbb R^{4} \rightarrow  \mathbb C^{16} \rightarrow \mathbb R^{10}$. In fact, the deep classical neural network embedded the $28\times 28$ images in a latent space of dimension $4$. These features were then assigned to a quantum state corresponding to $16$ QEFF, which fed the quantum discriminator (see Sect. 2). This quantum model then optimizes the likelihood of the predicted conditional probabilities $p(\mathbf{y}|\mathbf{x})$ of the ten classes. After obtaining the parameters $\boldsymbol \zeta_{\text{op}}$ and $\boldsymbol \theta_{\text{op}}$ that optimize the hybrid network for the discriminative task, the second training phase consisted of extracting the four-dimensional latent space, and using it to fine-tune the model using the quantum generative strategy, resulting in the fine-tuned parameters $\boldsymbol \theta_{\text{op}}^{\text{ft}}$. We used the test samples to calculate the accuracy of the model and the L-OOD data set to evaluate the performance of the joint probability density estimator. To validate the density estimator, we propose the use of the mean Spearman's correlation given by
\begin{equation}
    MSPC = \frac{1}{L}\sum_{j=0}^{L-1}SPC(j), 
\end{equation}
where $SPC(j)$ is the Spearman's correlation for class $j$ of the L-OOD data set. To evaluate this metric, we estimated the joint probability density of the L-OOD data set using both the D-QGC model and the classical KDC algorithm (see Sect. \ref{sec: Classification with RKHS}); the latter was used as a reference for the true probability density function.

Table \ref{tab: params 10-classes MNIST classification} summarizes the parameters of the hybrid network. In fact, the number of parameters for the classical and quantum neural networks was 234128 and 1026, respectively. The details of the deep neural network used can be found in the Appendix \ref{app: Model architecture}. Regarding the variational quantum algorithm, the circuit had a total of $9$ qubits, including $1$ ancilla qubit, $4$ qubits for the QEFF, and $4 = \lceil\log{10}\rceil$ qubits for the $10$ classes. We also chose 56 HEA layers, resulting in a quantum circuit with a total depth of 459 CNOT gates, of which 448 gates correspond to the HEA ansatz and $11$ gates correspond to the QEFF mapping. In addition, after the hyperparameter optimization, we selected a kernel bandwidth equal to $2.00$ for both the MNIST and F-MNIST data sets. And finally, we chose a kernel bandwidth of $1/16$ of the size of the largest dimension of the latent space to compute the true probability density using the KDC algorithm.

\begin{table}
  \centering
  \caption{Classification results of the ten classes of the MNIST and Fashion‑MNIST data sets. For each data set, we show the accuracy (ACC) of the D‑QGC strategy on the test partition, using as quantum maps the proposed QEFF and the baseline ZZFM \cite{havlivcek2019supervised}. The best results are shown in bold.}
  \label{tab: 10-classes MNIST F-MNIST classification}
  \begin{tabular}{lll}
    \toprule
           & D‑QGC ZZFM                  & D‑QGC QEFF                  \\
    \midrule
    ACC     & \multirow{2}{*}{\(\mathbf{0.986}\)} & \multirow{2}{*}{0.979} \\
    MNIST   &                                &                             \\
    \addlinespace
    ACC     & \multirow{2}{*}{0.891}        & \multirow{2}{*}{\(\mathbf{0.894}\)} \\
    F‑MNIST &                                &                             \\
    \bottomrule
  \end{tabular}
\end{table}

\subsubsection{Baseline setup}

As mentioned above, we also implemented as a baseline the D-QGC strategy using the ZZFM   \cite{havlivcek2019supervised}. This model followed the same training and testing methodology from the proposed QGC with the QEFF, sharing the same classical and quantum neural architectures. In particular, they used the same deep neural network to map the input images to a latent space of size 4, had the same HEA structure, had the same number of trainable parameters, and shared the same number of ancilla, input, and output qubits; see Table \ref{tab: params 10-classes MNIST classification}. Furthemore, we also used $4$ qubits for the ZZFM, since this embedding requires the same number of qubits as the dimension of the input data (in this case, the size of the latent space). The resulting depth the D-QGC circuit was $472$ CNOT gates, of which $24$ CNOT gates corresponded to the ZZ feature mapping. We also choose a kernel bandwidth of $1/16$ of the largest dimension of the latent space to estimate the true probability density with the KDC method.

\begin{table}
  \centering
  \caption{Results of the joint density estimation of the ten‑class MNIST and ten‑class Fashion‑MNIST data sets. For each data set, we show the mean Spearman’s correlation (MSPC) between the D‑QGC model and the classical KDC on the out‑of‑distribution data set, using as quantum encodings the QEFF and the baseline ZZFM \cite{havlivcek2019supervised}. The best results are shown in bold.}
  \label{tab: 10-classes MNIST F-MNIST density estimation}
  \begin{tabular}{lll}
    \toprule
           & D‑QGC ZZFM                  & D‑QGC QEFF                  \\
    \midrule
    MSPC    & \multirow{2}{*}{0.377}      & \multirow{2}{*}{\(\mathbf{0.543}\)} \\
    MNIST   &                             &                              \\
    \addlinespace
    MSPC    & \multirow{2}{*}{0.302}      & \multirow{2}{*}{\(\mathbf{0.657}\)} \\
    F‑MNIST &                             &                              \\
    \bottomrule
  \end{tabular}
\end{table}

\subsubsection{Results and discussion}

Table \ref{tab: 10-classes MNIST F-MNIST classification} shows the classification results of the D-QGC algorithm on the test partition of the MNIST and Fashion-MNIST data sets. The model achieved good classification performance with both ZZFM and QEFF mappings, illustrating that the proposed hybrid network is suitable for multiclass classification of high-dimensional data sets. We also present in Table \ref{tab: 10-classes MNIST F-MNIST density estimation} the performance of the models to learn the joint probability density function of the latent out-of-distribution samples of both the MNIST and F-MNIST data sets. The results demonstrate that the D-QGC model with both quantum mappings is a viable strategy for generative learning. However, comparing the two quantum mappings, the use of QEFF resulted in a higher mean Spearman's correlation, showing that this embedding produces a good approximation of the Gaussian kernel. It should be noticed that we did not evaluate the QRFF mapping as a part of the baseline. Indeed, we found that in a quantum-classical neural network, it is not trivial to optimize the parameters and evaluate its derivatives when using a quantum map based on amplitude encoding. This is in contrast to the QEFF and ZZFM, whose latent features could be directly embedded into the angles of the $R_z$ quantum gates. In summary, the previous results illustrate that compared to the other two quantum maps, the QEFF embedding is a more suitable strategy for variational hybrid generative classification.

\begin{table}
  \centering
  \caption{Accuracy (ACC) of the two‑dimensional binary \textit{Moons} data set using the QGC algorithm with the QEFF mapping. The predictions of the model were performed using the noiseless \textit{AerSimulator} and the noisy \textit{FakePerth} IBM quantum simulators. The demonstrations used 2 ancilla qubits, 4 qubits for the QEFF mapping, and 1 qubit for the classes. The best result is shown in bold.}
  \label{tab:2D FakePerth Aer Classification}
  \begin{tabular}{lll}
    \toprule
     & QGC QEFF (AerSimulator) & QGC QEFF (FakePerth) \\
    \midrule
    ACC & \(\mathbf{0.965}\) & 0.745 \\
    \bottomrule
  \end{tabular}
\end{table}

\begin{table}
  \centering
  \caption{Results of the joint probability density of the binary two‑dimensional \textit{Moons} data set using the QGC strategy and the QEFF feature map. We used the noiseless \textit{AerSimulator} and the noisy \textit{FakePerth} IBM \textit{qiskit} quantum simulators. The models were constructed using 2 ancilla qubits, 5 qubits for the input, and 1 qubit for the output. We used as metrics the mean average error (MAE) and the Spearman’s correlation per class (SPC(\(\cdot\))). The best results are shown in bold.}
  \label{tab: 2-Dimensional Conditional Error FakePerth AerSim}
  \begin{tabular}{lll}
    \toprule
    Metric & QGC QRFF (AerSimulator) & QGC QEFF (FakePerth) \\
    \midrule
    MAE    & 1.576                   & \(\mathbf{1.052}\) \\
    SPC(0) & 0.638                   & \(\mathbf{0.662}\) \\
    SPC(1) & \(\mathbf{0.647}\)      & 0.632 \\
    \bottomrule
  \end{tabular}
\end{table}

\subsection{\label{sec: STD estimation results} Two-dimensional variance estimation and the effects of noise}

As described in Sect. \ref{sec: Conditional pred and error estimation}, the probabilistic characteristics of the quantum generative classification algorithm allows us to estimate the uncertainty of the predictions. Henceforth, in this section we present two quantum demonstrations that illustrate the variance of the predictions of the two-dimensional \textit{Moons} data set, see Sect \ref{sec: Two-dimensional data sets}. Furthermore, we assess the effects of noise of the classification strategy by estimating the posterior probability and its variance on both a noiseless and a noisy quantum simulator of the IBM \textit{qiskit} library.

\subsubsection{Setup}

We evaluated the expectation value of the conditional probability $\mathbb E[\hat p(\boldsymbol y^*|\boldsymbol x^*)]$  of the QGC strategy obtained from the joint $\hat p(\boldsymbol x^*, \boldsymbol y^*)$  and its variance $\text{Var}(\hat p(\boldsymbol y^*|\boldsymbol x^*))$ on the previously described two-dimensional $\textit{Moons}$ data set. For such, we performed two quantum demonstrations using the $\textit{tensor-circuit}$ \cite{Zhang2023tensorcircuit} quantum library for training and the IBM $\textit{qiskit}$ \cite{qiskit2024} quantum package for testing. For the prediction phase, we used the IBM $\textit{AerSimulator}$ and the $\textit{FakePerth}$ quantum simulators; the $\textit{AerSimulator}$ simulates the probabilistic sampling of a noiseless quantum computer, while the  $\textit{FakePerth}$ corresponds to a noisy simulator whose gate decoherence times resemble the real 7-qubits IBM-$\textit{Perth}$ quantum computer. For both $\textit{qiskit}$ simulators, we set a total of  $S=10000$ shots and $R=10$ experiments, see Sect \ref{sec: Conditional pred and error estimation}. In contrast to the demonstrations presented in Sect. \ref{sec: Results two-dimensional data sets}, we used a total of $7$ qubits corresponding to $1$ qubits for the two classes, $4$ qubits for the QEFF mapping, and $2$ ancilla qubits. As previously, we select a kernel bandwidth parameter $h = 2^{-4}$ and used the training data to train the model, the test data to estimate the accuracy, and the out-of-distribution data set to evaluate the resulting joint probability density.

\begin{figure}
\centering
\includegraphics[scale=0.68]{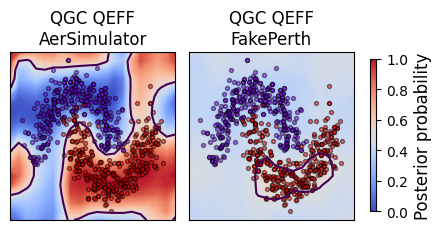}
\caption{Posterior probability $\hat p(\boldsymbol{y^*}|\boldsymbol{x^*})$ obtained from the joint $\hat p(\boldsymbol x^*, \boldsymbol y^*)$ of the quantum generative classification algorithm with the QEFF mapping. The training phase was performed using the \textit{tensor-circuit} library, while the prediction phase was performed using the noiseless \textit{AerSimulator} and the noisy \textit{FakePerth} IBM \textit{qiskit} quantum simulators. Both quantum simulators produced satisfactory results in the classification task.}
{\label{fig: 2-Dimensional FakePerth AerSim Classification}}
\end{figure}

\subsubsection{Results and discussion}

In figures \ref{fig: 2-Dimensional FakePerth AerSim Classification} and \ref{tab: 2-Dimensional Conditional Error FakePerth AerSim}, we plot respectively the conditional probability density and its standard deviation (STD), i.e., the square root of the variance, of the quantum generative model. Tables \ref{tab:2D FakePerth Aer Classification} and \ref{tab: 2-Dimensional Conditional Error FakePerth AerSim} also present respectively the model's accuracy and its joint probability density estimation performance. Firstly, it is worth mentioning that the predictions on the noiseless \textit{AerSimulator} mimic the results of the \textit{tensor-circuit} quantum simulator. This was expected as we carried out a relatively high number of shots. Still, the advantage of using the \textit{qiskit} simulator lies on its stochasticity which allows to estimate the variance of the model, in contrast to the \textit{tensor-circuit} simulator which predicts the exact estimate of the conditional probability density. Regarding the noisy \textit {FakePerth} quantum simulator, it achieved satisfactory results on the density estimation task, but it was not as competitive on the classification task. Nonetheless, these results suggest that the QGC algorithm could be potentially implemented on noisy intermediate-scale quantum computers (NISQ) \cite{Preskill2018quantumcomputingin}. What's more, Fig. \ref{tab: 2-Dimensional Conditional Error FakePerth AerSim} depicts the standard deviation of the posterior probability, illustrating a correlation between the uncertainty of the predictions and the density of training samples; indeed, the figure shows that the QGC model is more certain in regions with a higher density of data. However, figure \ref{tab: 2-Dimensional Conditional Error FakePerth AerSim} shows that the noise from the \textit{FakePerth} simulator was detrimental for the variance calculation. The capability of estimating the uncertainty of the predictions has multiple applications, including the classification of medical images \cite{ZHU2024101506}, indeed, it is especially relevant in a diagnostic procedure to quantify the level of confidence in the predictions \cite{TOLEDOCORTES2022105472}.

\begin{figure}
  \centering
  \includegraphics[scale=0.68]{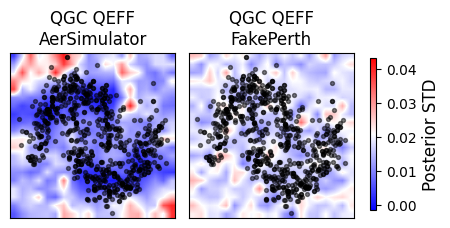}
  \caption{Standard deviation (STD) of the posterior probability density \(\hat p(\boldsymbol{y}^*|\boldsymbol{x}^*)\), derived from the joint $\hat p(\boldsymbol x^*, \boldsymbol y^*)$, of the quantum generative classification algorithm with QEFF encoding. The model was trained using the \textit{tensor-circuit} library, and the prediction was carried out using the noiseless \textit{AerSimulator} and the noisy \textit{FakePerth} IBM \textit{qiskit} quantum simulators. The results show that, especially in the noiseless regime, the model is more confident in regions with a higher density of data.}
  \label{fig:2-DimensionalConditionalErrorFakePerthAerSim}
\end{figure}

\section{\label{sec: Conclusions}Conclusions}

In this paper, we presented a variational quantum algorithm for generative classification based on density matrices and Fourier features. The model uses a variational quantum circuit to prepare a mixed state that summarizes the joint probability distribution of features and labels of a data set, and performs supervised classification of a test sample based on the label with the largest joint probability density. We also introduced the quantum-enhanced Fourier features mapping, which leverages quantum superposition for approximating a Gaussian kernel through an inner product in a low-dimensional feature space. We showed that the model can be viewed as a Gaussian mixture classifier whose Gaussian components are centered on the training samples and a method for constructing a function in a reproducing kernel Hilbert space. We performed various quantum demonstrations on several low- and high-dimensional data sets, illustrating that the QGC with QEFF strategy is competitive against other quantum algorithms of the state-of-the-art \cite{farhi, dilip2022data, havlivcek2019supervised, useche2022quantum}.

In contrast to previous generative models \cite{amin2018quantum, Zoufal2021}, the quantum demonstrations show that the QGC algorithm can be used to prepare in quantum hardware both continuous and discrete joint probability distributions through mixed quantum states. Indeed, we showed that we can apply the variational principle to construct a quantum circuit for generative classification; this approach improves the flexibility of previous nonparametric density estimation methods based on kernel density estimation and quantum state averaging \cite{Useche2021QuantumQudits, vargas2022optimisation, useche2022quantum, Gonzalez2021Classification}. We also demonstrated, through a noiseless and a noisy quantum simulator, that we may use quantum resources to provide probabilistic predictions rather than point-wise estimates. This enabled us to estimate the variance of the classification, which could be applied in areas where estimating confidence is crucial, such as in the classification of medical images \cite{TOLEDOCORTES2022105472}. Furthermore, this optimization-based strategy can be integrated with deep neural networks, allowing its application to high-dimensional data sets. For instance, we applied the hybrid quantum-classical model for the generative classification of the 10 classes MNIST and 10 classes Fashion-MNIST data sets, exhibiting good classification and density estimation performance and illustrating that the QEFF mapping is competitive against other previously proposed quantum encodings \cite{useche2022quantum, havlivcek2019supervised}. 

Considering the multiple applications of generative learning, the proposed quantum generative classification strategy lays the groundwork for approaching other quantum machine learning problems, including multiclass anomaly detection \cite{NEURIPS2022_1d774c11}, statistical inference \cite{held2014applied}, and synthetic data generation \cite{harshvardhan2020comprehensive}. Some open problems of the QGC model to be investigated for future work include its implementation on real quantum computers, the study of the effects of the barren plateau in the variational training, the evaluation of the effects of the exponential concentration \cite{thanasilp2024exponential}, and the error correction \cite{babbush2021errorcorrection}.

\section*{Limitations}\label{sec: limitations}

We have described some of the key characteristics of our quantum generative classification algorithm; nonetheless, some aspects of the model require further review. For instance, it remains to explore the performance of quantum strategy against other classical generative models. In particular, the QGC strategy could potentially be dequantized. Therefore, it would be worthwhile to evaluate the algorithm as a purely classical method. Additionally, it is lacking a thorough review of the model’s scalability in terms of the number of qubits and circuit depth. Indeed, the depth of the QEFF mapping increases exponentially with the number of qubits, and the size of the Hilbert space increases with the number of ancilla qubits. Evaluating this scalability issue is essential for addressing two important barriers of variational quantum algorithms: the barren plateau problem \cite{mcclean2018barren} and the exponential concentration of the kernel \cite{thanasilp2024exponential}.

\section*{Code availability}\label{sec: code section}

The source code for the quantum generative classification algorithm is made available on the following GitHub repository \href{https://github.com/diegour1/QGC}{https://github.com/diegour1/QGC}.

\section*{Acknowledgments}\label{sec: acknowledgments}

D.H.U. acknowledges Google Research for the support through the Google Ph.D. Fellowship. J.E.A.-G. acknowledges funding from the project ``Aprendizaje de máquina para sistemas cuánticos'', HERMES code 57792, UNAL. Additionally, F.A.G. acknowledges the support of ``Centro de excelencia en tecnologías cuánticas y sus aplicaciones a metrología'', Universidad Nacional de Colombia, and ``Ampliación del uso de la mecánica cuántica desde el punto de vista experimental y su relacion con la teoría, generando desarrollos en tecnologías cuánticas útiles para metrología y computación cuántica a nivel nacional'', Ministerio de Ciencia, Tecnología e Innovación de Colombia.

\bibliography{mainref}% Produces the bibliography via BibTeX.

%apsrev4-2.bst 2019-01-14 (MD) hand-edited version of apsrev4-1.bst
%Control: key (0)
%Control: author (8) initials jnrlst
%Control: editor formatted (1) identically to author
%Control: production of article title (0) allowed
%Control: page (0) single
%Control: year (1) truncated
%Control: production of eprint (0) enabled
\providecommand{\noopsort}[1]{}\providecommand{\singleletter}[1]{#1}%
\begin{thebibliography}{78}%
\makeatletter
\providecommand \@ifxundefined [1]{%
 \@ifx{#1\undefined}
}%
\providecommand \@ifnum [1]{%
 \ifnum #1\expandafter \@firstoftwo
 \else \expandafter \@secondoftwo
 \fi
}%
\providecommand \@ifx [1]{%
 \ifx #1\expandafter \@firstoftwo
 \else \expandafter \@secondoftwo
 \fi
}%
\providecommand \natexlab [1]{#1}%
\providecommand \enquote  [1]{``#1''}%
\providecommand \bibnamefont  [1]{#1}%
\providecommand \bibfnamefont [1]{#1}%
\providecommand \citenamefont [1]{#1}%
\providecommand \href@noop [0]{\@secondoftwo}%
\providecommand \href [0]{\begingroup \@sanitize@url \@href}%
\providecommand \@href[1]{\@@startlink{#1}\@@href}%
\providecommand \@@href[1]{\endgroup#1\@@endlink}%
\providecommand \@sanitize@url [0]{\catcode `\\12\catcode `\$12\catcode `\&12\catcode `\#12\catcode `\^12\catcode `\_12\catcode `\%12\relax}%
\providecommand \@@startlink[1]{}%
\providecommand \@@endlink[0]{}%
\providecommand \url  [0]{\begingroup\@sanitize@url \@url }%
\providecommand \@url [1]{\endgroup\@href {#1}{\urlprefix }}%
\providecommand \urlprefix  [0]{URL }%
\providecommand \Eprint [0]{\href }%
\providecommand \doibase [0]{https://doi.org/}%
\providecommand \selectlanguage [0]{\@gobble}%
\providecommand \bibinfo  [0]{\@secondoftwo}%
\providecommand \bibfield  [0]{\@secondoftwo}%
\providecommand \translation [1]{[#1]}%
\providecommand \BibitemOpen [0]{}%
\providecommand \bibitemStop [0]{}%
\providecommand \bibitemNoStop [0]{.\EOS\space}%
\providecommand \EOS [0]{\spacefactor3000\relax}%
\providecommand \BibitemShut  [1]{\csname bibitem#1\endcsname}%
\let\auto@bib@innerbib\@empty
%</preamble>
\bibitem [{\citenamefont {Boser}\ \emph {et~al.}(1992)\citenamefont {Boser}, \citenamefont {Guyon},\ and\ \citenamefont {Vapnik}}]{boser1992training}%
  \BibitemOpen
  \bibfield  {author} {\bibinfo {author} {\bibfnamefont {B.~E.}\ \bibnamefont {Boser}}, \bibinfo {author} {\bibfnamefont {I.~M.}\ \bibnamefont {Guyon}},\ and\ \bibinfo {author} {\bibfnamefont {V.~N.}\ \bibnamefont {Vapnik}},\ }\bibfield  {title} {\bibinfo {title} {A training algorithm for optimal margin classifiers},\ }in\ \href {https://doi.org/10.1145/130385.130401} {\emph {\bibinfo {booktitle} {Proceedings of the Fifth Annual Workshop on Computational Learning Theory}}},\ \bibinfo {series and number} {COLT '92}\ (\bibinfo  {publisher} {Association for Computing Machinery},\ \bibinfo {address} {New York, NY, USA},\ \bibinfo {year} {1992})\ p.\ \bibinfo {pages} {144–152}\BibitemShut {NoStop}%
\bibitem [{\citenamefont {Rumelhart}\ \emph {et~al.}(1986)\citenamefont {Rumelhart}, \citenamefont {Hinton},\ and\ \citenamefont {Williams}}]{rumelhart1986learning}%
  \BibitemOpen
  \bibfield  {author} {\bibinfo {author} {\bibfnamefont {D.~E.}\ \bibnamefont {Rumelhart}}, \bibinfo {author} {\bibfnamefont {G.~E.}\ \bibnamefont {Hinton}},\ and\ \bibinfo {author} {\bibfnamefont {R.~J.}\ \bibnamefont {Williams}},\ }\bibfield  {title} {\bibinfo {title} {Learning representations by back-propagating errors},\ }\href@noop {} {\bibfield  {journal} {\bibinfo  {journal} {nature}\ }\textbf {\bibinfo {volume} {323}},\ \bibinfo {pages} {533} (\bibinfo {year} {1986})}\BibitemShut {NoStop}%
\bibitem [{\citenamefont {LeCun}\ \emph {et~al.}(1998)\citenamefont {LeCun}, \citenamefont {Bottou}, \citenamefont {Bengio},\ and\ \citenamefont {Haffner}}]{lecun1998gradient}%
  \BibitemOpen
  \bibfield  {author} {\bibinfo {author} {\bibfnamefont {Y.}~\bibnamefont {LeCun}}, \bibinfo {author} {\bibfnamefont {L.}~\bibnamefont {Bottou}}, \bibinfo {author} {\bibfnamefont {Y.}~\bibnamefont {Bengio}},\ and\ \bibinfo {author} {\bibfnamefont {P.}~\bibnamefont {Haffner}},\ }\bibfield  {title} {\bibinfo {title} {Gradient-based learning applied to document recognition},\ }\href@noop {} {\bibfield  {journal} {\bibinfo  {journal} {Proceedings of the IEEE}\ }\textbf {\bibinfo {volume} {86}},\ \bibinfo {pages} {2278} (\bibinfo {year} {1998})}\BibitemShut {NoStop}%
\bibitem [{\citenamefont {Berkson}(1944)}]{Berkson1944Logistic}%
  \BibitemOpen
  \bibfield  {author} {\bibinfo {author} {\bibfnamefont {J.}~\bibnamefont {Berkson}},\ }\bibfield  {title} {\bibinfo {title} {Application of the logistic function to bio-assay},\ }\href {https://doi.org/10.1080/01621459.1944.10500699} {\bibfield  {journal} {\bibinfo  {journal} {Journal of the American Statistical Association}\ }\textbf {\bibinfo {volume} {39}},\ \bibinfo {pages} {357} (\bibinfo {year} {1944})},\ \Eprint {https://arxiv.org/abs/https://doi.org/10.1080/01621459.1944.10500699} {https://doi.org/10.1080/01621459.1944.10500699} \BibitemShut {NoStop}%
\bibitem [{\citenamefont {Breiman}(2001)}]{Breiman2001Forest}%
  \BibitemOpen
  \bibfield  {author} {\bibinfo {author} {\bibfnamefont {L.}~\bibnamefont {Breiman}},\ }\bibfield  {title} {\bibinfo {title} {Random forests},\ }\href {https://doi.org/10.1023/A:1010933404324} {\bibfield  {journal} {\bibinfo  {journal} {Machine Learning}\ }\textbf {\bibinfo {volume} {45}},\ \bibinfo {pages} {5} (\bibinfo {year} {2001})}\BibitemShut {NoStop}%
\bibitem [{\citenamefont {Bernardo}\ \emph {et~al.}(2007)\citenamefont {Bernardo}, \citenamefont {Bayarri}, \citenamefont {Berger}, \citenamefont {Dawid}, \citenamefont {Heckerman}, \citenamefont {Smith},\ and\ \citenamefont {West}}]{bernardo2007generative}%
  \BibitemOpen
  \bibfield  {author} {\bibinfo {author} {\bibfnamefont {J.}~\bibnamefont {Bernardo}}, \bibinfo {author} {\bibfnamefont {M.}~\bibnamefont {Bayarri}}, \bibinfo {author} {\bibfnamefont {J.}~\bibnamefont {Berger}}, \bibinfo {author} {\bibfnamefont {A.}~\bibnamefont {Dawid}}, \bibinfo {author} {\bibfnamefont {D.}~\bibnamefont {Heckerman}}, \bibinfo {author} {\bibfnamefont {A.}~\bibnamefont {Smith}},\ and\ \bibinfo {author} {\bibfnamefont {M.}~\bibnamefont {West}},\ }\bibfield  {title} {\bibinfo {title} {Generative or discriminative? getting the best of both worlds},\ }\href@noop {} {\bibfield  {journal} {\bibinfo  {journal} {Bayesian statistics}\ }\textbf {\bibinfo {volume} {8}},\ \bibinfo {pages} {3} (\bibinfo {year} {2007})}\BibitemShut {NoStop}%
\bibitem [{\citenamefont {Harshvardhan}\ \emph {et~al.}(2020)\citenamefont {Harshvardhan}, \citenamefont {Gourisaria}, \citenamefont {Pandey},\ and\ \citenamefont {Rautaray}}]{harshvardhan2020comprehensive}%
  \BibitemOpen
  \bibfield  {author} {\bibinfo {author} {\bibfnamefont {G.}~\bibnamefont {Harshvardhan}}, \bibinfo {author} {\bibfnamefont {M.~K.}\ \bibnamefont {Gourisaria}}, \bibinfo {author} {\bibfnamefont {M.}~\bibnamefont {Pandey}},\ and\ \bibinfo {author} {\bibfnamefont {S.~S.}\ \bibnamefont {Rautaray}},\ }\bibfield  {title} {\bibinfo {title} {A comprehensive survey and analysis of generative models in machine learning},\ }\href@noop {} {\bibfield  {journal} {\bibinfo  {journal} {Computer Science Review}\ }\textbf {\bibinfo {volume} {38}},\ \bibinfo {pages} {100285} (\bibinfo {year} {2020})}\BibitemShut {NoStop}%
\bibitem [{\citenamefont {Kingma}\ and\ \citenamefont {Welling}(2022)}]{kingma2022autoencodingvariationalbayes}%
  \BibitemOpen
  \bibfield  {author} {\bibinfo {author} {\bibfnamefont {D.~P.}\ \bibnamefont {Kingma}}\ and\ \bibinfo {author} {\bibfnamefont {M.}~\bibnamefont {Welling}},\ }\href {https://arxiv.org/abs/1312.6114} {\bibinfo {title} {Auto-encoding variational bayes}} (\bibinfo {year} {2022}),\ \Eprint {https://arxiv.org/abs/1312.6114} {arXiv:1312.6114 [stat.ML]} \BibitemShut {NoStop}%
\bibitem [{\citenamefont {Ackley}\ \emph {et~al.}(1985)\citenamefont {Ackley}, \citenamefont {Hinton},\ and\ \citenamefont {Sejnowski}}]{Ackley1985boltzmann}%
  \BibitemOpen
  \bibfield  {author} {\bibinfo {author} {\bibfnamefont {D.~H.}\ \bibnamefont {Ackley}}, \bibinfo {author} {\bibfnamefont {G.~E.}\ \bibnamefont {Hinton}},\ and\ \bibinfo {author} {\bibfnamefont {T.~J.}\ \bibnamefont {Sejnowski}},\ }\bibfield  {title} {\bibinfo {title} {A learning algorithm for boltzmann machines},\ }\href {https://doi.org/https://doi.org/10.1016/S0364-0213(85)80012-4} {\bibfield  {journal} {\bibinfo  {journal} {Cognitive Science}\ }\textbf {\bibinfo {volume} {9}},\ \bibinfo {pages} {147} (\bibinfo {year} {1985})}\BibitemShut {NoStop}%
\bibitem [{\citenamefont {Papamakarios}\ \emph {et~al.}(2021)\citenamefont {Papamakarios}, \citenamefont {Nalisnick}, \citenamefont {Rezende}, \citenamefont {Mohamed},\ and\ \citenamefont {Lakshminarayanan}}]{Papamakarios2021Normalizing}%
  \BibitemOpen
  \bibfield  {author} {\bibinfo {author} {\bibfnamefont {G.}~\bibnamefont {Papamakarios}}, \bibinfo {author} {\bibfnamefont {E.}~\bibnamefont {Nalisnick}}, \bibinfo {author} {\bibfnamefont {D.~J.}\ \bibnamefont {Rezende}}, \bibinfo {author} {\bibfnamefont {S.}~\bibnamefont {Mohamed}},\ and\ \bibinfo {author} {\bibfnamefont {B.}~\bibnamefont {Lakshminarayanan}},\ }\bibfield  {title} {\bibinfo {title} {Normalizing flows for probabilistic modeling and inference},\ }\href {http://jmlr.org/papers/v22/19-1028.html} {\bibfield  {journal} {\bibinfo  {journal} {Journal of Machine Learning Research}\ }\textbf {\bibinfo {volume} {22}},\ \bibinfo {pages} {1} (\bibinfo {year} {2021})}\BibitemShut {NoStop}%
\bibitem [{\citenamefont {Rasmussen}(2004)}]{Rasmussen2004}%
  \BibitemOpen
  \bibfield  {author} {\bibinfo {author} {\bibfnamefont {C.~E.}\ \bibnamefont {Rasmussen}},\ }\bibinfo {title} {Gaussian processes in machine learning},\ in\ \href {https://doi.org/10.1007/978-3-540-28650-9_4} {\emph {\bibinfo {booktitle} {Advanced Lectures on Machine Learning: ML Summer Schools 2003, Canberra, Australia, February 2 - 14, 2003, T{\"u}bingen, Germany, August 4 - 16, 2003, Revised Lectures}}}\ (\bibinfo  {publisher} {Springer Berlin Heidelberg},\ \bibinfo {address} {Berlin, Heidelberg},\ \bibinfo {year} {2004})\ pp.\ \bibinfo {pages} {63--71}\BibitemShut {NoStop}%
\bibitem [{\citenamefont {Goodfellow}\ \emph {et~al.}(2014)\citenamefont {Goodfellow}, \citenamefont {Pouget-Abadie}, \citenamefont {Mirza}, \citenamefont {Xu}, \citenamefont {Warde-Farley}, \citenamefont {Ozair}, \citenamefont {Courville},\ and\ \citenamefont {Bengio}}]{2014GoodfellowGAN}%
  \BibitemOpen
  \bibfield  {author} {\bibinfo {author} {\bibfnamefont {I.}~\bibnamefont {Goodfellow}}, \bibinfo {author} {\bibfnamefont {J.}~\bibnamefont {Pouget-Abadie}}, \bibinfo {author} {\bibfnamefont {M.}~\bibnamefont {Mirza}}, \bibinfo {author} {\bibfnamefont {B.}~\bibnamefont {Xu}}, \bibinfo {author} {\bibfnamefont {D.}~\bibnamefont {Warde-Farley}}, \bibinfo {author} {\bibfnamefont {S.}~\bibnamefont {Ozair}}, \bibinfo {author} {\bibfnamefont {A.}~\bibnamefont {Courville}},\ and\ \bibinfo {author} {\bibfnamefont {Y.}~\bibnamefont {Bengio}},\ }\bibfield  {title} {\bibinfo {title} {Generative adversarial nets},\ }in\ \href {https://proceedings.neurips.cc/paper_files/paper/2014/file/5ca3e9b122f61f8f06494c97b1afccf3-Paper.pdf} {\emph {\bibinfo {booktitle} {Advances in Neural Information Processing Systems}}},\ Vol.~\bibinfo {volume} {27},\ \bibinfo {editor} {edited by\ \bibinfo {editor} {\bibfnamefont {Z.}~\bibnamefont {Ghahramani}}, \bibinfo {editor} {\bibfnamefont {M.}~\bibnamefont {Welling}}, \bibinfo {editor}
  {\bibfnamefont {C.}~\bibnamefont {Cortes}}, \bibinfo {editor} {\bibfnamefont {N.}~\bibnamefont {Lawrence}},\ and\ \bibinfo {editor} {\bibfnamefont {K.}~\bibnamefont {Weinberger}}}\ (\bibinfo  {publisher} {Curran Associates, Inc.},\ \bibinfo {year} {2014})\BibitemShut {NoStop}%
\bibitem [{\citenamefont {Baum}\ and\ \citenamefont {Petrie}(1966)}]{BaumMarkov1966}%
  \BibitemOpen
  \bibfield  {author} {\bibinfo {author} {\bibfnamefont {L.~E.}\ \bibnamefont {Baum}}\ and\ \bibinfo {author} {\bibfnamefont {T.}~\bibnamefont {Petrie}},\ }\bibfield  {title} {\bibinfo {title} {{Statistical Inference for Probabilistic Functions of Finite State Markov Chains}},\ }\href {https://doi.org/10.1214/aoms/1177699147} {\bibfield  {journal} {\bibinfo  {journal} {The Annals of Mathematical Statistics}\ }\textbf {\bibinfo {volume} {37}},\ \bibinfo {pages} {1554 } (\bibinfo {year} {1966})}\BibitemShut {NoStop}%
\bibitem [{\citenamefont {Jebara}(2004)}]{Jebara2004Generative}%
  \BibitemOpen
  \bibfield  {author} {\bibinfo {author} {\bibfnamefont {T.}~\bibnamefont {Jebara}},\ }\bibinfo {title} {Generative versus discriminative learning},\ in\ \href {https://doi.org/10.1007/978-1-4419-9011-2_2} {\emph {\bibinfo {booktitle} {Machine Learning: Discriminative and Generative}}}\ (\bibinfo  {publisher} {Springer US},\ \bibinfo {address} {Boston, MA},\ \bibinfo {year} {2004})\ pp.\ \bibinfo {pages} {17--60}\BibitemShut {NoStop}%
\bibitem [{\citenamefont {Schuld}\ \emph {et~al.}(2020)\citenamefont {Schuld}, \citenamefont {Bocharov}, \citenamefont {Svore},\ and\ \citenamefont {Wiebe}}]{schuld2020circuit}%
  \BibitemOpen
  \bibfield  {author} {\bibinfo {author} {\bibfnamefont {M.}~\bibnamefont {Schuld}}, \bibinfo {author} {\bibfnamefont {A.}~\bibnamefont {Bocharov}}, \bibinfo {author} {\bibfnamefont {K.~M.}\ \bibnamefont {Svore}},\ and\ \bibinfo {author} {\bibfnamefont {N.}~\bibnamefont {Wiebe}},\ }\bibfield  {title} {\bibinfo {title} {Circuit-centric quantum classifiers},\ }\href@noop {} {\bibfield  {journal} {\bibinfo  {journal} {Physical Review A}\ }\textbf {\bibinfo {volume} {101}},\ \bibinfo {pages} {032308} (\bibinfo {year} {2020})}\BibitemShut {NoStop}%
\bibitem [{\citenamefont {Farhi}\ and\ \citenamefont {Neven}(2018)}]{farhi2018classification}%
  \BibitemOpen
  \bibfield  {author} {\bibinfo {author} {\bibfnamefont {E.}~\bibnamefont {Farhi}}\ and\ \bibinfo {author} {\bibfnamefont {H.}~\bibnamefont {Neven}},\ }\bibfield  {title} {\bibinfo {title} {Classification with quantum neural networks on near term processors},\ }\href@noop {} {\bibfield  {journal} {\bibinfo  {journal} {arXiv preprint arXiv:1802.06002}\ } (\bibinfo {year} {2018})}\BibitemShut {NoStop}%
\bibitem [{\citenamefont {Watkins}\ \emph {et~al.}(2023)\citenamefont {Watkins}, \citenamefont {Chen},\ and\ \citenamefont {Yoo}}]{watkins2023quantum}%
  \BibitemOpen
  \bibfield  {author} {\bibinfo {author} {\bibfnamefont {W.~M.}\ \bibnamefont {Watkins}}, \bibinfo {author} {\bibfnamefont {S.~Y.-C.}\ \bibnamefont {Chen}},\ and\ \bibinfo {author} {\bibfnamefont {S.}~\bibnamefont {Yoo}},\ }\bibfield  {title} {\bibinfo {title} {Quantum machine learning with differential privacy},\ }\href@noop {} {\bibfield  {journal} {\bibinfo  {journal} {Scientific Reports}\ }\textbf {\bibinfo {volume} {13}},\ \bibinfo {pages} {2453} (\bibinfo {year} {2023})}\BibitemShut {NoStop}%
\bibitem [{\citenamefont {Altares-L{\'o}pez}\ \emph {et~al.}(2021)\citenamefont {Altares-L{\'o}pez}, \citenamefont {Ribeiro},\ and\ \citenamefont {Garc{\'\i}a-Ripoll}}]{altares2021automatic}%
  \BibitemOpen
  \bibfield  {author} {\bibinfo {author} {\bibfnamefont {S.}~\bibnamefont {Altares-L{\'o}pez}}, \bibinfo {author} {\bibfnamefont {A.}~\bibnamefont {Ribeiro}},\ and\ \bibinfo {author} {\bibfnamefont {J.~J.}\ \bibnamefont {Garc{\'\i}a-Ripoll}},\ }\bibfield  {title} {\bibinfo {title} {Automatic design of quantum feature maps},\ }\href@noop {} {\bibfield  {journal} {\bibinfo  {journal} {Quantum Science and Technology}\ }\textbf {\bibinfo {volume} {6}},\ \bibinfo {pages} {045015} (\bibinfo {year} {2021})}\BibitemShut {NoStop}%
\bibitem [{\citenamefont {Havl{\'\i}{\v{c}}ek}\ \emph {et~al.}(2019)\citenamefont {Havl{\'\i}{\v{c}}ek}, \citenamefont {C{\'o}rcoles}, \citenamefont {Temme}, \citenamefont {Harrow}, \citenamefont {Kandala}, \citenamefont {Chow},\ and\ \citenamefont {Gambetta}}]{havlivcek2019supervised}%
  \BibitemOpen
  \bibfield  {author} {\bibinfo {author} {\bibfnamefont {V.}~\bibnamefont {Havl{\'\i}{\v{c}}ek}}, \bibinfo {author} {\bibfnamefont {A.~D.}\ \bibnamefont {C{\'o}rcoles}}, \bibinfo {author} {\bibfnamefont {K.}~\bibnamefont {Temme}}, \bibinfo {author} {\bibfnamefont {A.~W.}\ \bibnamefont {Harrow}}, \bibinfo {author} {\bibfnamefont {A.}~\bibnamefont {Kandala}}, \bibinfo {author} {\bibfnamefont {J.~M.}\ \bibnamefont {Chow}},\ and\ \bibinfo {author} {\bibfnamefont {J.~M.}\ \bibnamefont {Gambetta}},\ }\bibfield  {title} {\bibinfo {title} {Supervised learning with quantum-enhanced feature spaces},\ }\href@noop {} {\bibfield  {journal} {\bibinfo  {journal} {Nature}\ }\textbf {\bibinfo {volume} {567}},\ \bibinfo {pages} {209} (\bibinfo {year} {2019})}\BibitemShut {NoStop}%
\bibitem [{\citenamefont {Park}\ \emph {et~al.}(2023)\citenamefont {Park}, \citenamefont {Huh},\ and\ \citenamefont {Park}}]{park2023variational}%
  \BibitemOpen
  \bibfield  {author} {\bibinfo {author} {\bibfnamefont {G.}~\bibnamefont {Park}}, \bibinfo {author} {\bibfnamefont {J.}~\bibnamefont {Huh}},\ and\ \bibinfo {author} {\bibfnamefont {D.~K.}\ \bibnamefont {Park}},\ }\bibfield  {title} {\bibinfo {title} {Variational quantum one-class classifier},\ }\href@noop {} {\bibfield  {journal} {\bibinfo  {journal} {Machine Learning: Science and Technology}\ }\textbf {\bibinfo {volume} {4}},\ \bibinfo {pages} {015006} (\bibinfo {year} {2023})}\BibitemShut {NoStop}%
\bibitem [{\citenamefont {Benedetti}\ \emph {et~al.}(2017)\citenamefont {Benedetti}, \citenamefont {Realpe-G\'omez}, \citenamefont {Biswas},\ and\ \citenamefont {Perdomo-Ortiz}}]{Bendetti2017quantumassisted}%
  \BibitemOpen
  \bibfield  {author} {\bibinfo {author} {\bibfnamefont {M.}~\bibnamefont {Benedetti}}, \bibinfo {author} {\bibfnamefont {J.}~\bibnamefont {Realpe-G\'omez}}, \bibinfo {author} {\bibfnamefont {R.}~\bibnamefont {Biswas}},\ and\ \bibinfo {author} {\bibfnamefont {A.}~\bibnamefont {Perdomo-Ortiz}},\ }\bibfield  {title} {\bibinfo {title} {Quantum-assisted learning of hardware-embedded probabilistic graphical models},\ }\href {https://doi.org/10.1103/PhysRevX.7.041052} {\bibfield  {journal} {\bibinfo  {journal} {Phys. Rev. X}\ }\textbf {\bibinfo {volume} {7}},\ \bibinfo {pages} {041052} (\bibinfo {year} {2017})}\BibitemShut {NoStop}%
\bibitem [{\citenamefont {Zoufal}\ \emph {et~al.}(2021)\citenamefont {Zoufal}, \citenamefont {Lucchi},\ and\ \citenamefont {Woerner}}]{Zoufal2021}%
  \BibitemOpen
  \bibfield  {author} {\bibinfo {author} {\bibfnamefont {C.}~\bibnamefont {Zoufal}}, \bibinfo {author} {\bibfnamefont {A.}~\bibnamefont {Lucchi}},\ and\ \bibinfo {author} {\bibfnamefont {S.}~\bibnamefont {Woerner}},\ }\bibfield  {title} {\bibinfo {title} {Variational quantum boltzmann machines},\ }\href {https://doi.org/10.1007/s42484-020-00033-7} {\bibfield  {journal} {\bibinfo  {journal} {Quantum Machine Intelligence}\ }\textbf {\bibinfo {volume} {3}},\ \bibinfo {pages} {7} (\bibinfo {year} {2021})}\BibitemShut {NoStop}%
\bibitem [{\citenamefont {Chaudhary}\ \emph {et~al.}(2023)\citenamefont {Chaudhary}, \citenamefont {Huembeli}, \citenamefont {MacCormack}, \citenamefont {Patti}, \citenamefont {Kossaifi},\ and\ \citenamefont {Galda}}]{Chaudhary_2023}%
  \BibitemOpen
  \bibfield  {author} {\bibinfo {author} {\bibfnamefont {S.}~\bibnamefont {Chaudhary}}, \bibinfo {author} {\bibfnamefont {P.}~\bibnamefont {Huembeli}}, \bibinfo {author} {\bibfnamefont {I.}~\bibnamefont {MacCormack}}, \bibinfo {author} {\bibfnamefont {T.~L.}\ \bibnamefont {Patti}}, \bibinfo {author} {\bibfnamefont {J.}~\bibnamefont {Kossaifi}},\ and\ \bibinfo {author} {\bibfnamefont {A.}~\bibnamefont {Galda}},\ }\bibfield  {title} {\bibinfo {title} {Towards a scalable discrete quantum generative adversarial neural network},\ }\href {https://doi.org/10.1088/2058-9565/acc4e4} {\bibfield  {journal} {\bibinfo  {journal} {Quantum Science and Technology}\ }\textbf {\bibinfo {volume} {8}},\ \bibinfo {pages} {035002} (\bibinfo {year} {2023})}\BibitemShut {NoStop}%
\bibitem [{\citenamefont {Useche}\ \emph {et~al.}(2021)\citenamefont {Useche}, \citenamefont {Giraldo-Carvajal}, \citenamefont {Zuluaga-Bucheli}, \citenamefont {Jaramillo-Villegas},\ and\ \citenamefont {Gonz{\'{a}}lez}}]{Useche2021QuantumQudits}%
  \BibitemOpen
  \bibfield  {author} {\bibinfo {author} {\bibfnamefont {D.~H.}\ \bibnamefont {Useche}}, \bibinfo {author} {\bibfnamefont {A.}~\bibnamefont {Giraldo-Carvajal}}, \bibinfo {author} {\bibfnamefont {H.~M.}\ \bibnamefont {Zuluaga-Bucheli}}, \bibinfo {author} {\bibfnamefont {J.~A.}\ \bibnamefont {Jaramillo-Villegas}},\ and\ \bibinfo {author} {\bibfnamefont {F.~A.}\ \bibnamefont {Gonz{\'{a}}lez}},\ }\bibfield  {title} {\bibinfo {title} {{Quantum measurement classification with qudits}},\ }\href {https://doi.org/10.1007/S11128-021-03363-Y} {\bibfield  {journal} {\bibinfo  {journal} {Quantum Information Processing 2021 21:1}\ }\textbf {\bibinfo {volume} {21}},\ \bibinfo {pages} {1} (\bibinfo {year} {2021})}\BibitemShut {NoStop}%
\bibitem [{\citenamefont {Vargas-Calder{\'o}n}\ \emph {et~al.}(2022)\citenamefont {Vargas-Calder{\'o}n}, \citenamefont {Gonz{\'a}lez},\ and\ \citenamefont {Vinck-Posada}}]{vargas2022optimisation}%
  \BibitemOpen
  \bibfield  {author} {\bibinfo {author} {\bibfnamefont {V.}~\bibnamefont {Vargas-Calder{\'o}n}}, \bibinfo {author} {\bibfnamefont {F.~A.}\ \bibnamefont {Gonz{\'a}lez}},\ and\ \bibinfo {author} {\bibfnamefont {H.}~\bibnamefont {Vinck-Posada}},\ }\bibfield  {title} {\bibinfo {title} {Optimisation-free density estimation and classification with quantum circuits},\ }\href@noop {} {\bibfield  {journal} {\bibinfo  {journal} {Quantum Machine Intelligence}\ }\textbf {\bibinfo {volume} {4}},\ \bibinfo {pages} {1} (\bibinfo {year} {2022})}\BibitemShut {NoStop}%
\bibitem [{\citenamefont {Harrow}\ \emph {et~al.}(2009)\citenamefont {Harrow}, \citenamefont {Hassidim},\ and\ \citenamefont {Lloyd}}]{harrow2009quantum}%
  \BibitemOpen
  \bibfield  {author} {\bibinfo {author} {\bibfnamefont {A.~W.}\ \bibnamefont {Harrow}}, \bibinfo {author} {\bibfnamefont {A.}~\bibnamefont {Hassidim}},\ and\ \bibinfo {author} {\bibfnamefont {S.}~\bibnamefont {Lloyd}},\ }\bibfield  {title} {\bibinfo {title} {Quantum algorithm for linear systems of equations},\ }\href {https://doi.org/10.1103/PhysRevLett.103.150502} {\bibfield  {journal} {\bibinfo  {journal} {Phys. Rev. Lett.}\ }\textbf {\bibinfo {volume} {103}},\ \bibinfo {pages} {150502} (\bibinfo {year} {2009})}\BibitemShut {NoStop}%
\bibitem [{\citenamefont {Shor}(1994)}]{shor1994algorithms}%
  \BibitemOpen
  \bibfield  {author} {\bibinfo {author} {\bibfnamefont {P.}~\bibnamefont {Shor}},\ }\bibfield  {title} {\bibinfo {title} {Algorithms for quantum computation: discrete logarithms and factoring},\ }in\ \href {https://doi.org/10.1109/SFCS.1994.365700} {\emph {\bibinfo {booktitle} {Proceedings 35th Annual Symposium on Foundations of Computer Science}}}\ (\bibinfo {year} {1994})\ pp.\ \bibinfo {pages} {124--134}\BibitemShut {NoStop}%
\bibitem [{\citenamefont {González}\ \emph {et~al.}(2024)\citenamefont {González}, \citenamefont {Ramos-Pollán},\ and\ \citenamefont {Gallego-Mejia}}]{gonzález2024kdm}%
  \BibitemOpen
  \bibfield  {author} {\bibinfo {author} {\bibfnamefont {F.~A.}\ \bibnamefont {González}}, \bibinfo {author} {\bibfnamefont {R.}~\bibnamefont {Ramos-Pollán}},\ and\ \bibinfo {author} {\bibfnamefont {J.~A.}\ \bibnamefont {Gallego-Mejia}},\ }\href {https://arxiv.org/abs/2305.18204} {\bibinfo {title} {Kernel density matrices for probabilistic deep learning}} (\bibinfo {year} {2024}),\ \Eprint {https://arxiv.org/abs/2305.18204} {arXiv:2305.18204 [cs.LG]} \BibitemShut {NoStop}%
\bibitem [{\citenamefont {Useche}\ \emph {et~al.}(2024)\citenamefont {Useche}, \citenamefont {Bustos-Brinez}, \citenamefont {Gallego-Mejia},\ and\ \citenamefont {Gonz\'alez}}]{useche2022quantum}%
  \BibitemOpen
  \bibfield  {author} {\bibinfo {author} {\bibfnamefont {D.~H.}\ \bibnamefont {Useche}}, \bibinfo {author} {\bibfnamefont {O.~A.}\ \bibnamefont {Bustos-Brinez}}, \bibinfo {author} {\bibfnamefont {J.~A.}\ \bibnamefont {Gallego-Mejia}},\ and\ \bibinfo {author} {\bibfnamefont {F.~A.}\ \bibnamefont {Gonz\'alez}},\ }\bibfield  {title} {\bibinfo {title} {Quantum density estimation with density matrices: Application to quantum anomaly detection},\ }\href {https://doi.org/10.1103/PhysRevA.109.032418} {\bibfield  {journal} {\bibinfo  {journal} {Phys. Rev. A}\ }\textbf {\bibinfo {volume} {109}},\ \bibinfo {pages} {032418} (\bibinfo {year} {2024})}\BibitemShut {NoStop}%
\bibitem [{\citenamefont {Schuld}\ and\ \citenamefont {Petruccione}(2021)}]{Schuld2021quantumkernel}%
  \BibitemOpen
  \bibfield  {author} {\bibinfo {author} {\bibfnamefont {M.}~\bibnamefont {Schuld}}\ and\ \bibinfo {author} {\bibfnamefont {F.}~\bibnamefont {Petruccione}},\ }\bibinfo {title} {Quantum models as kernel methods},\ in\ \href {https://doi.org/10.1007/978-3-030-83098-4_6} {\emph {\bibinfo {booktitle} {Machine Learning with Quantum Computers}}}\ (\bibinfo  {publisher} {Springer International Publishing},\ \bibinfo {address} {Cham},\ \bibinfo {year} {2021})\ pp.\ \bibinfo {pages} {217--245}\BibitemShut {NoStop}%
\bibitem [{\citenamefont {Rahimi}\ and\ \citenamefont {Recht}(2009)}]{Rahimi2009RandomMachines}%
  \BibitemOpen
  \bibfield  {author} {\bibinfo {author} {\bibfnamefont {A.}~\bibnamefont {Rahimi}}\ and\ \bibinfo {author} {\bibfnamefont {B.}~\bibnamefont {Recht}},\ }\bibfield  {title} {\bibinfo {title} {{Random features for large-scale kernel machines}},\ }in\ \href@noop {} {\emph {\bibinfo {booktitle} {Advances in Neural Information Processing Systems 20 - Proceedings of the 2007 Conference}}}\ (\bibinfo {year} {2009})\BibitemShut {NoStop}%
\bibitem [{\citenamefont {Reynolds}\ \emph {et~al.}(2009)\citenamefont {Reynolds} \emph {et~al.}}]{reynolds2009gaussian}%
  \BibitemOpen
  \bibfield  {author} {\bibinfo {author} {\bibfnamefont {D.~A.}\ \bibnamefont {Reynolds}} \emph {et~al.},\ }\bibfield  {title} {\bibinfo {title} {Gaussian mixture models.},\ }\href@noop {} {\bibfield  {journal} {\bibinfo  {journal} {Encyclopedia of biometrics}\ }\textbf {\bibinfo {volume} {741}} (\bibinfo {year} {2009})}\BibitemShut {NoStop}%
\bibitem [{\citenamefont {Toledo-Cortés}\ \emph {et~al.}(2022)\citenamefont {Toledo-Cortés}, \citenamefont {Useche}, \citenamefont {Müller},\ and\ \citenamefont {González}}]{TOLEDOCORTES2022105472}%
  \BibitemOpen
  \bibfield  {author} {\bibinfo {author} {\bibfnamefont {S.}~\bibnamefont {Toledo-Cortés}}, \bibinfo {author} {\bibfnamefont {D.~H.}\ \bibnamefont {Useche}}, \bibinfo {author} {\bibfnamefont {H.}~\bibnamefont {Müller}},\ and\ \bibinfo {author} {\bibfnamefont {F.~A.}\ \bibnamefont {González}},\ }\bibfield  {title} {\bibinfo {title} {Grading diabetic retinopathy and prostate cancer diagnostic images with deep quantum ordinal regression},\ }\href {https://doi.org/https://doi.org/10.1016/j.compbiomed.2022.105472} {\bibfield  {journal} {\bibinfo  {journal} {Computers in Biology and Medicine}\ }\textbf {\bibinfo {volume} {145}},\ \bibinfo {pages} {105472} (\bibinfo {year} {2022})}\BibitemShut {NoStop}%
\bibitem [{\citenamefont {Zhang}\ \emph {et~al.}(2023)\citenamefont {Zhang}, \citenamefont {Allcock}, \citenamefont {Wan}, \citenamefont {Liu}, \citenamefont {Sun}, \citenamefont {Yu}, \citenamefont {Yang}, \citenamefont {Qiu}, \citenamefont {Ye}, \citenamefont {Chen}, \citenamefont {Lee}, \citenamefont {Zheng}, \citenamefont {Jian}, \citenamefont {Yao}, \citenamefont {Hsieh},\ and\ \citenamefont {Zhang}}]{Zhang2023tensorcircuit}%
  \BibitemOpen
  \bibfield  {author} {\bibinfo {author} {\bibfnamefont {S.-X.}\ \bibnamefont {Zhang}}, \bibinfo {author} {\bibfnamefont {J.}~\bibnamefont {Allcock}}, \bibinfo {author} {\bibfnamefont {Z.-Q.}\ \bibnamefont {Wan}}, \bibinfo {author} {\bibfnamefont {S.}~\bibnamefont {Liu}}, \bibinfo {author} {\bibfnamefont {J.}~\bibnamefont {Sun}}, \bibinfo {author} {\bibfnamefont {H.}~\bibnamefont {Yu}}, \bibinfo {author} {\bibfnamefont {X.-H.}\ \bibnamefont {Yang}}, \bibinfo {author} {\bibfnamefont {J.}~\bibnamefont {Qiu}}, \bibinfo {author} {\bibfnamefont {Z.}~\bibnamefont {Ye}}, \bibinfo {author} {\bibfnamefont {Y.-Q.}\ \bibnamefont {Chen}}, \bibinfo {author} {\bibfnamefont {C.-K.}\ \bibnamefont {Lee}}, \bibinfo {author} {\bibfnamefont {Y.-C.}\ \bibnamefont {Zheng}}, \bibinfo {author} {\bibfnamefont {S.-K.}\ \bibnamefont {Jian}}, \bibinfo {author} {\bibfnamefont {H.}~\bibnamefont {Yao}}, \bibinfo {author} {\bibfnamefont {C.-Y.}\ \bibnamefont {Hsieh}},\ and\ \bibinfo {author} {\bibfnamefont {S.}~\bibnamefont {Zhang}},\
  }\bibfield  {title} {\bibinfo {title} {Tensor{C}ircuit: a {Q}uantum {S}oftware {F}ramework for the {NISQ} {E}ra},\ }\href {https://doi.org/10.22331/q-2023-02-02-912} {\bibfield  {journal} {\bibinfo  {journal} {{Quantum}}\ }\textbf {\bibinfo {volume} {7}},\ \bibinfo {pages} {912} (\bibinfo {year} {2023})}\BibitemShut {NoStop}%
\bibitem [{\citenamefont {Zoufal}\ \emph {et~al.}(2019)\citenamefont {Zoufal}, \citenamefont {Lucchi},\ and\ \citenamefont {Woerner}}]{zoufal2019quantum}%
  \BibitemOpen
  \bibfield  {author} {\bibinfo {author} {\bibfnamefont {C.}~\bibnamefont {Zoufal}}, \bibinfo {author} {\bibfnamefont {A.}~\bibnamefont {Lucchi}},\ and\ \bibinfo {author} {\bibfnamefont {S.}~\bibnamefont {Woerner}},\ }\bibfield  {title} {\bibinfo {title} {Quantum generative adversarial networks for learning and loading random distributions},\ }\href@noop {} {\bibfield  {journal} {\bibinfo  {journal} {npj Quantum Information}\ }\textbf {\bibinfo {volume} {5}},\ \bibinfo {pages} {103} (\bibinfo {year} {2019})}\BibitemShut {NoStop}%
\bibitem [{\citenamefont {Romero}\ and\ \citenamefont {Aspuru-Guzik}(2021)}]{romero2021variational}%
  \BibitemOpen
  \bibfield  {author} {\bibinfo {author} {\bibfnamefont {J.}~\bibnamefont {Romero}}\ and\ \bibinfo {author} {\bibfnamefont {A.}~\bibnamefont {Aspuru-Guzik}},\ }\bibfield  {title} {\bibinfo {title} {Variational quantum generators: Generative adversarial quantum machine learning for continuous distributions},\ }\href@noop {} {\bibfield  {journal} {\bibinfo  {journal} {Advanced Quantum Technologies}\ }\textbf {\bibinfo {volume} {4}},\ \bibinfo {pages} {2000003} (\bibinfo {year} {2021})}\BibitemShut {NoStop}%
\bibitem [{\citenamefont {Benedetti}\ \emph {et~al.}(2019)\citenamefont {Benedetti}, \citenamefont {Garcia-Pintos}, \citenamefont {Perdomo}, \citenamefont {Leyton-Ortega}, \citenamefont {Nam},\ and\ \citenamefont {Perdomo-Ortiz}}]{benedetti2019generative}%
  \BibitemOpen
  \bibfield  {author} {\bibinfo {author} {\bibfnamefont {M.}~\bibnamefont {Benedetti}}, \bibinfo {author} {\bibfnamefont {D.}~\bibnamefont {Garcia-Pintos}}, \bibinfo {author} {\bibfnamefont {O.}~\bibnamefont {Perdomo}}, \bibinfo {author} {\bibfnamefont {V.}~\bibnamefont {Leyton-Ortega}}, \bibinfo {author} {\bibfnamefont {Y.}~\bibnamefont {Nam}},\ and\ \bibinfo {author} {\bibfnamefont {A.}~\bibnamefont {Perdomo-Ortiz}},\ }\bibfield  {title} {\bibinfo {title} {A generative modeling approach for benchmarking and training shallow quantum circuits},\ }\href@noop {} {\bibfield  {journal} {\bibinfo  {journal} {npj Quantum Information}\ }\textbf {\bibinfo {volume} {5}},\ \bibinfo {pages} {45} (\bibinfo {year} {2019})}\BibitemShut {NoStop}%
\bibitem [{\citenamefont {Miyahara}\ and\ \citenamefont {Roychowdhury}(2023)}]{miyahara2023quantum}%
  \BibitemOpen
  \bibfield  {author} {\bibinfo {author} {\bibfnamefont {H.}~\bibnamefont {Miyahara}}\ and\ \bibinfo {author} {\bibfnamefont {V.}~\bibnamefont {Roychowdhury}},\ }\bibfield  {title} {\bibinfo {title} {Quantum advantage in variational bayes inference},\ }\href@noop {} {\bibfield  {journal} {\bibinfo  {journal} {Proceedings of the National Academy of Sciences}\ }\textbf {\bibinfo {volume} {120}},\ \bibinfo {pages} {e2212660120} (\bibinfo {year} {2023})}\BibitemShut {NoStop}%
\bibitem [{\citenamefont {Kerenidis}\ \emph {et~al.}(2020)\citenamefont {Kerenidis}, \citenamefont {Luongo},\ and\ \citenamefont {Prakash}}]{kerenidis2020quantum}%
  \BibitemOpen
  \bibfield  {author} {\bibinfo {author} {\bibfnamefont {I.}~\bibnamefont {Kerenidis}}, \bibinfo {author} {\bibfnamefont {A.}~\bibnamefont {Luongo}},\ and\ \bibinfo {author} {\bibfnamefont {A.}~\bibnamefont {Prakash}},\ }\bibfield  {title} {\bibinfo {title} {Quantum expectation-maximization for gaussian mixture models},\ }in\ \href@noop {} {\emph {\bibinfo {booktitle} {International Conference on Machine Learning}}}\ (\bibinfo {organization} {PMLR},\ \bibinfo {year} {2020})\ pp.\ \bibinfo {pages} {5187--5197}\BibitemShut {NoStop}%
\bibitem [{\citenamefont {Amin}\ \emph {et~al.}(2018)\citenamefont {Amin}, \citenamefont {Andriyash}, \citenamefont {Rolfe}, \citenamefont {Kulchytskyy},\ and\ \citenamefont {Melko}}]{amin2018quantum}%
  \BibitemOpen
  \bibfield  {author} {\bibinfo {author} {\bibfnamefont {M.~H.}\ \bibnamefont {Amin}}, \bibinfo {author} {\bibfnamefont {E.}~\bibnamefont {Andriyash}}, \bibinfo {author} {\bibfnamefont {J.}~\bibnamefont {Rolfe}}, \bibinfo {author} {\bibfnamefont {B.}~\bibnamefont {Kulchytskyy}},\ and\ \bibinfo {author} {\bibfnamefont {R.}~\bibnamefont {Melko}},\ }\bibfield  {title} {\bibinfo {title} {Quantum boltzmann machine},\ }\href@noop {} {\bibfield  {journal} {\bibinfo  {journal} {Physical Review X}\ }\textbf {\bibinfo {volume} {8}},\ \bibinfo {pages} {021050} (\bibinfo {year} {2018})}\BibitemShut {NoStop}%
\bibitem [{\citenamefont {Gonz\'{a}lez}\ \emph {et~al.}(2021)\citenamefont {Gonz\'{a}lez}, \citenamefont {Vargas-Calder\'{o}n},\ and\ \citenamefont {Vinck-Posada}}]{Gonzalez2021Classification}%
  \BibitemOpen
  \bibfield  {author} {\bibinfo {author} {\bibfnamefont {F.~A.}\ \bibnamefont {Gonz\'{a}lez}}, \bibinfo {author} {\bibfnamefont {V.}~\bibnamefont {Vargas-Calder\'{o}n}},\ and\ \bibinfo {author} {\bibfnamefont {H.}~\bibnamefont {Vinck-Posada}},\ }\bibfield  {title} {\bibinfo {title} {Classification with quantum measurements},\ }\href {https://doi.org/10.7566/JPSJ.90.044002} {\bibfield  {journal} {\bibinfo  {journal} {Journal of the Physical Society of Japan}\ }\textbf {\bibinfo {volume} {90}},\ \bibinfo {pages} {044002} (\bibinfo {year} {2021})},\ \Eprint {https://arxiv.org/abs/https://doi.org/10.7566/JPSJ.90.044002} {https://doi.org/10.7566/JPSJ.90.044002} \BibitemShut {NoStop}%
\bibitem [{\citenamefont {Gonz{\'a}lez}\ \emph {et~al.}(2022)\citenamefont {Gonz{\'a}lez}, \citenamefont {Gallego}, \citenamefont {Toledo-Cort{\'e}s},\ and\ \citenamefont {Vargas-Calder{\'o}n}}]{gonzalez2022learning}%
  \BibitemOpen
  \bibfield  {author} {\bibinfo {author} {\bibfnamefont {F.~A.}\ \bibnamefont {Gonz{\'a}lez}}, \bibinfo {author} {\bibfnamefont {A.}~\bibnamefont {Gallego}}, \bibinfo {author} {\bibfnamefont {S.}~\bibnamefont {Toledo-Cort{\'e}s}},\ and\ \bibinfo {author} {\bibfnamefont {V.}~\bibnamefont {Vargas-Calder{\'o}n}},\ }\bibfield  {title} {\bibinfo {title} {Learning with density matrices and random features},\ }\href@noop {} {\bibfield  {journal} {\bibinfo  {journal} {Quantum Machine Intelligence}\ }\textbf {\bibinfo {volume} {4}},\ \bibinfo {pages} {23} (\bibinfo {year} {2022})}\BibitemShut {NoStop}%
\bibitem [{\citenamefont {Schölkopf}\ and\ \citenamefont {Smola}(2001)}]{scholkopf2002learning}%
  \BibitemOpen
  \bibfield  {author} {\bibinfo {author} {\bibfnamefont {B.}~\bibnamefont {Schölkopf}}\ and\ \bibinfo {author} {\bibfnamefont {A.~J.}\ \bibnamefont {Smola}},\ }\bibfield  {title} {\bibinfo {title} {Learning with kernels: Support vector machines, regularization, optimization, and beyond},\ }\bibfield  {journal} {\bibinfo  {journal} {Proceedings of 2002 International Conference on Machine Learning and Cybernetics}\ }\textbf {\bibinfo {volume} {1}},\ \href {https://doi.org/10.7551/MITPRESS/4175.001.0001} {10.7551/MITPRESS/4175.001.0001} (\bibinfo {year} {2001})\BibitemShut {NoStop}%
\bibitem [{\citenamefont {Sadowski}(2019)}]{sadowski2019machine}%
  \BibitemOpen
  \bibfield  {author} {\bibinfo {author} {\bibfnamefont {P.}~\bibnamefont {Sadowski}},\ }\bibfield  {title} {\bibinfo {title} {Machine learning kernel method from a quantum generative model},\ }\href@noop {} {\bibfield  {journal} {\bibinfo  {journal} {arXiv preprint arXiv:1907.05103}\ } (\bibinfo {year} {2019})}\BibitemShut {NoStop}%
\bibitem [{\citenamefont {Peters}\ and\ \citenamefont {Schuld}(2023)}]{Peters2023generalization}%
  \BibitemOpen
  \bibfield  {author} {\bibinfo {author} {\bibfnamefont {E.}~\bibnamefont {Peters}}\ and\ \bibinfo {author} {\bibfnamefont {M.}~\bibnamefont {Schuld}},\ }\bibfield  {title} {\bibinfo {title} {Generalization despite overfitting in quantum machine learning models},\ }\href {https://doi.org/10.22331/q-2023-12-20-1210} {\bibfield  {journal} {\bibinfo  {journal} {{Quantum}}\ }\textbf {\bibinfo {volume} {7}},\ \bibinfo {pages} {1210} (\bibinfo {year} {2023})}\BibitemShut {NoStop}%
\bibitem [{\citenamefont {Landman}\ \emph {et~al.}(2022)\citenamefont {Landman}, \citenamefont {Thabet}, \citenamefont {Dalyac}, \citenamefont {Mhiri},\ and\ \citenamefont {Kashefi}}]{landman2022classically}%
  \BibitemOpen
  \bibfield  {author} {\bibinfo {author} {\bibfnamefont {J.}~\bibnamefont {Landman}}, \bibinfo {author} {\bibfnamefont {S.}~\bibnamefont {Thabet}}, \bibinfo {author} {\bibfnamefont {C.}~\bibnamefont {Dalyac}}, \bibinfo {author} {\bibfnamefont {H.}~\bibnamefont {Mhiri}},\ and\ \bibinfo {author} {\bibfnamefont {E.}~\bibnamefont {Kashefi}},\ }\bibfield  {title} {\bibinfo {title} {Classically approximating variational quantum machine learning with random fourier features},\ }\href@noop {} {\bibfield  {journal} {\bibinfo  {journal} {arXiv preprint arXiv:2210.13200}\ } (\bibinfo {year} {2022})}\BibitemShut {NoStop}%
\bibitem [{\citenamefont {Sweke}\ \emph {et~al.}(2023)\citenamefont {Sweke}, \citenamefont {Recio}, \citenamefont {Jerbi}, \citenamefont {Gil-Fuster}, \citenamefont {Fuller}, \citenamefont {Eisert},\ and\ \citenamefont {Meyer}}]{sweke2023potential}%
  \BibitemOpen
  \bibfield  {author} {\bibinfo {author} {\bibfnamefont {R.}~\bibnamefont {Sweke}}, \bibinfo {author} {\bibfnamefont {E.}~\bibnamefont {Recio}}, \bibinfo {author} {\bibfnamefont {S.}~\bibnamefont {Jerbi}}, \bibinfo {author} {\bibfnamefont {E.}~\bibnamefont {Gil-Fuster}}, \bibinfo {author} {\bibfnamefont {B.}~\bibnamefont {Fuller}}, \bibinfo {author} {\bibfnamefont {J.}~\bibnamefont {Eisert}},\ and\ \bibinfo {author} {\bibfnamefont {J.~J.}\ \bibnamefont {Meyer}},\ }\bibfield  {title} {\bibinfo {title} {Potential and limitations of random fourier features for dequantizing quantum machine learning},\ }\href@noop {} {\bibfield  {journal} {\bibinfo  {journal} {arXiv preprint arXiv:2309.11647}\ } (\bibinfo {year} {2023})}\BibitemShut {NoStop}%
\bibitem [{\citenamefont {Ardila-Garc{\'\i}a}\ \emph {et~al.}(2025)\citenamefont {Ardila-Garc{\'\i}a}, \citenamefont {Vargas-Calder{\'o}n}, \citenamefont {Gonz{\'a}lez}, \citenamefont {Useche},\ and\ \citenamefont {Vinck-Posada}}]{ardila2025memo}%
  \BibitemOpen
  \bibfield  {author} {\bibinfo {author} {\bibfnamefont {J.~E.}\ \bibnamefont {Ardila-Garc{\'\i}a}}, \bibinfo {author} {\bibfnamefont {V.}~\bibnamefont {Vargas-Calder{\'o}n}}, \bibinfo {author} {\bibfnamefont {F.~A.}\ \bibnamefont {Gonz{\'a}lez}}, \bibinfo {author} {\bibfnamefont {D.~H.}\ \bibnamefont {Useche}},\ and\ \bibinfo {author} {\bibfnamefont {H.}~\bibnamefont {Vinck-Posada}},\ }\bibfield  {title} {\bibinfo {title} {Memo-qcd: quantum density estimation through memetic optimisation for quantum circuit design},\ }\href@noop {} {\bibfield  {journal} {\bibinfo  {journal} {Quantum Machine Intelligence}\ }\textbf {\bibinfo {volume} {7}},\ \bibinfo {pages} {3} (\bibinfo {year} {2025})}\BibitemShut {NoStop}%
\bibitem [{\citenamefont {Yamasaki}\ \emph {et~al.}(2020)\citenamefont {Yamasaki}, \citenamefont {Subramanian}, \citenamefont {Sonoda},\ and\ \citenamefont {Koashi}}]{Yamasaki2020}%
  \BibitemOpen
  \bibfield  {author} {\bibinfo {author} {\bibfnamefont {H.}~\bibnamefont {Yamasaki}}, \bibinfo {author} {\bibfnamefont {S.}~\bibnamefont {Subramanian}}, \bibinfo {author} {\bibfnamefont {S.}~\bibnamefont {Sonoda}},\ and\ \bibinfo {author} {\bibfnamefont {M.}~\bibnamefont {Koashi}},\ }\bibfield  {title} {\bibinfo {title} {Learning with optimized random features: Exponential speedup by quantum machine learning without sparsity and low-rank assumptions},\ }in\ \href {https://proceedings.neurips.cc/paper_files/paper/2020/file/9ddb9dd5d8aee9a76bf217a2a3c54833-Paper.pdf} {\emph {\bibinfo {booktitle} {Advances in Neural Information Processing Systems}}},\ Vol.~\bibinfo {volume} {33},\ \bibinfo {editor} {edited by\ \bibinfo {editor} {\bibfnamefont {H.}~\bibnamefont {Larochelle}}, \bibinfo {editor} {\bibfnamefont {M.}~\bibnamefont {Ranzato}}, \bibinfo {editor} {\bibfnamefont {R.}~\bibnamefont {Hadsell}}, \bibinfo {editor} {\bibfnamefont {M.}~\bibnamefont {Balcan}},\ and\ \bibinfo {editor} {\bibfnamefont
  {H.}~\bibnamefont {Lin}}}\ (\bibinfo  {publisher} {Curran Associates, Inc.},\ \bibinfo {year} {2020})\ pp.\ \bibinfo {pages} {13674--13687}\BibitemShut {NoStop}%
\bibitem [{\citenamefont {Gil-Fuster}\ \emph {et~al.}(2024)\citenamefont {Gil-Fuster}, \citenamefont {Eisert},\ and\ \citenamefont {Dunjko}}]{Gil-Fuster2024}%
  \BibitemOpen
  \bibfield  {author} {\bibinfo {author} {\bibfnamefont {E.}~\bibnamefont {Gil-Fuster}}, \bibinfo {author} {\bibfnamefont {J.}~\bibnamefont {Eisert}},\ and\ \bibinfo {author} {\bibfnamefont {V.}~\bibnamefont {Dunjko}},\ }\bibfield  {title} {\bibinfo {title} {On the expressivity of embedding quantum kernels},\ }\href {https://doi.org/10.1088/2632-2153/ad2f51} {\bibfield  {journal} {\bibinfo  {journal} {Machine Learning: Science and Technology}\ }\textbf {\bibinfo {volume} {5}},\ \bibinfo {pages} {025003} (\bibinfo {year} {2024})}\BibitemShut {NoStop}%
\bibitem [{\citenamefont {M{\"o}tt{\"o}nen}\ \emph {et~al.}(2005)\citenamefont {M{\"o}tt{\"o}nen}, \citenamefont {Vartiainen}, \citenamefont {Bergholm},\ and\ \citenamefont {Salomaa}}]{mottonen2005transformation}%
  \BibitemOpen
  \bibfield  {author} {\bibinfo {author} {\bibfnamefont {M.}~\bibnamefont {M{\"o}tt{\"o}nen}}, \bibinfo {author} {\bibfnamefont {J.~J.}\ \bibnamefont {Vartiainen}}, \bibinfo {author} {\bibfnamefont {V.}~\bibnamefont {Bergholm}},\ and\ \bibinfo {author} {\bibfnamefont {M.~M.}\ \bibnamefont {Salomaa}},\ }\bibfield  {title} {\bibinfo {title} {Transformation of quantum states using uniformly controlled rotations},\ }\href@noop {} {\bibfield  {journal} {\bibinfo  {journal} {Quantum Information \& Computation}\ }\textbf {\bibinfo {volume} {5}},\ \bibinfo {pages} {467} (\bibinfo {year} {2005})}\BibitemShut {NoStop}%
\bibitem [{\citenamefont {Shende}\ \emph {et~al.}(2006)\citenamefont {Shende}, \citenamefont {Bullock},\ and\ \citenamefont {Markov}}]{shende2006synthesis}%
  \BibitemOpen
  \bibfield  {author} {\bibinfo {author} {\bibfnamefont {V.~V.}\ \bibnamefont {Shende}}, \bibinfo {author} {\bibfnamefont {S.~S.}\ \bibnamefont {Bullock}},\ and\ \bibinfo {author} {\bibfnamefont {I.~L.}\ \bibnamefont {Markov}},\ }\bibfield  {title} {\bibinfo {title} {Synthesis of quantum-logic circuits},\ }\href@noop {} {\bibfield  {journal} {\bibinfo  {journal} {IEEE Transactions on Computer-Aided Design of Integrated Circuits and Systems}\ }\textbf {\bibinfo {volume} {25}},\ \bibinfo {pages} {1000} (\bibinfo {year} {2006})}\BibitemShut {NoStop}%
\bibitem [{\citenamefont {Mitarai}\ \emph {et~al.}(2018)\citenamefont {Mitarai}, \citenamefont {Negoro}, \citenamefont {Kitagawa},\ and\ \citenamefont {Fujii}}]{mitarai2018quantum}%
  \BibitemOpen
  \bibfield  {author} {\bibinfo {author} {\bibfnamefont {K.}~\bibnamefont {Mitarai}}, \bibinfo {author} {\bibfnamefont {M.}~\bibnamefont {Negoro}}, \bibinfo {author} {\bibfnamefont {M.}~\bibnamefont {Kitagawa}},\ and\ \bibinfo {author} {\bibfnamefont {K.}~\bibnamefont {Fujii}},\ }\bibfield  {title} {\bibinfo {title} {Quantum circuit learning},\ }\href@noop {} {\bibfield  {journal} {\bibinfo  {journal} {Physical Review A}\ }\textbf {\bibinfo {volume} {98}},\ \bibinfo {pages} {032309} (\bibinfo {year} {2018})}\BibitemShut {NoStop}%
\bibitem [{\citenamefont {Schuld}\ \emph {et~al.}(2019)\citenamefont {Schuld}, \citenamefont {Bergholm}, \citenamefont {Gogolin}, \citenamefont {Izaac},\ and\ \citenamefont {Killoran}}]{schuld2019evaluating}%
  \BibitemOpen
  \bibfield  {author} {\bibinfo {author} {\bibfnamefont {M.}~\bibnamefont {Schuld}}, \bibinfo {author} {\bibfnamefont {V.}~\bibnamefont {Bergholm}}, \bibinfo {author} {\bibfnamefont {C.}~\bibnamefont {Gogolin}}, \bibinfo {author} {\bibfnamefont {J.}~\bibnamefont {Izaac}},\ and\ \bibinfo {author} {\bibfnamefont {N.}~\bibnamefont {Killoran}},\ }\bibfield  {title} {\bibinfo {title} {Evaluating analytic gradients on quantum hardware},\ }\href@noop {} {\bibfield  {journal} {\bibinfo  {journal} {Physical Review A}\ }\textbf {\bibinfo {volume} {99}},\ \bibinfo {pages} {032331} (\bibinfo {year} {2019})}\BibitemShut {NoStop}%
\bibitem [{\citenamefont {Vapnik}\ and\ \citenamefont {Mukherjee}(1999)}]{vapnik1999support}%
  \BibitemOpen
  \bibfield  {author} {\bibinfo {author} {\bibfnamefont {V.}~\bibnamefont {Vapnik}}\ and\ \bibinfo {author} {\bibfnamefont {S.}~\bibnamefont {Mukherjee}},\ }\bibfield  {title} {\bibinfo {title} {Support vector method for multivariate density estimation},\ }in\ \href {https://proceedings.neurips.cc/paper_files/paper/1999/file/207f88018f72237565570f8a9e5ca240-Paper.pdf} {\emph {\bibinfo {booktitle} {Advances in Neural Information Processing Systems}}},\ Vol.~\bibinfo {volume} {12},\ \bibinfo {editor} {edited by\ \bibinfo {editor} {\bibfnamefont {S.}~\bibnamefont {Solla}}, \bibinfo {editor} {\bibfnamefont {T.}~\bibnamefont {Leen}},\ and\ \bibinfo {editor} {\bibfnamefont {K.}~\bibnamefont {M\"{u}ller}}}\ (\bibinfo  {publisher} {MIT Press},\ \bibinfo {year} {1999})\BibitemShut {NoStop}%
\bibitem [{\citenamefont {Rosenblatt}(1956)}]{rosenblatt1956remarks}%
  \BibitemOpen
  \bibfield  {author} {\bibinfo {author} {\bibfnamefont {M.}~\bibnamefont {Rosenblatt}},\ }\bibfield  {title} {\bibinfo {title} {Remarks on some nonparametric estimates of a density function},\ }\href@noop {} {\bibfield  {journal} {\bibinfo  {journal} {The annals of mathematical statistics}\ ,\ \bibinfo {pages} {832}} (\bibinfo {year} {1956})}\BibitemShut {NoStop}%
\bibitem [{\citenamefont {Parzen}(1962)}]{parzen1962estimation}%
  \BibitemOpen
  \bibfield  {author} {\bibinfo {author} {\bibfnamefont {E.}~\bibnamefont {Parzen}},\ }\bibfield  {title} {\bibinfo {title} {On estimation of a probability density function and mode},\ }\href@noop {} {\bibfield  {journal} {\bibinfo  {journal} {The annals of mathematical statistics}\ }\textbf {\bibinfo {volume} {33}},\ \bibinfo {pages} {1065} (\bibinfo {year} {1962})}\BibitemShut {NoStop}%
\bibitem [{\citenamefont {Hastie}\ \emph {et~al.}(2009)\citenamefont {Hastie}, \citenamefont {Tibshirani}, \citenamefont {Friedman},\ and\ \citenamefont {Friedman}}]{hastie2009elements}%
  \BibitemOpen
  \bibfield  {author} {\bibinfo {author} {\bibfnamefont {T.}~\bibnamefont {Hastie}}, \bibinfo {author} {\bibfnamefont {R.}~\bibnamefont {Tibshirani}}, \bibinfo {author} {\bibfnamefont {J.~H.}\ \bibnamefont {Friedman}},\ and\ \bibinfo {author} {\bibfnamefont {J.~H.}\ \bibnamefont {Friedman}},\ }\href@noop {} {\emph {\bibinfo {title} {The elements of statistical learning: data mining, inference, and prediction}}},\ Vol.~\bibinfo {volume} {2}\ (\bibinfo  {publisher} {Springer},\ \bibinfo {year} {2009})\BibitemShut {NoStop}%
\bibitem [{\citenamefont {Ping}\ \emph {et~al.}(2013)\citenamefont {Ping}, \citenamefont {Li}, \citenamefont {Pan}, \citenamefont {Luo},\ and\ \citenamefont {Zhang}}]{Ping2013OptimalPO}%
  \BibitemOpen
  \bibfield  {author} {\bibinfo {author} {\bibfnamefont {Y.}~\bibnamefont {Ping}}, \bibinfo {author} {\bibfnamefont {H.}~\bibnamefont {Li}}, \bibinfo {author} {\bibfnamefont {X.}~\bibnamefont {Pan}}, \bibinfo {author} {\bibfnamefont {M.}~\bibnamefont {Luo}},\ and\ \bibinfo {author} {\bibfnamefont {Z.}~\bibnamefont {Zhang}},\ }\bibfield  {title} {\bibinfo {title} {Optimal purification of arbitrary quantum mixed states},\ }\href {https://api.semanticscholar.org/CorpusID:122400833} {\bibfield  {journal} {\bibinfo  {journal} {International Journal of Theoretical Physics}\ }\textbf {\bibinfo {volume} {52}},\ \bibinfo {pages} {4367} (\bibinfo {year} {2013})}\BibitemShut {NoStop}%
\bibitem [{\citenamefont {Holmes}\ \emph {et~al.}(2022)\citenamefont {Holmes}, \citenamefont {Sharma}, \citenamefont {Cerezo},\ and\ \citenamefont {Coles}}]{HolmesExpressibility2022}%
  \BibitemOpen
  \bibfield  {author} {\bibinfo {author} {\bibfnamefont {Z.}~\bibnamefont {Holmes}}, \bibinfo {author} {\bibfnamefont {K.}~\bibnamefont {Sharma}}, \bibinfo {author} {\bibfnamefont {M.}~\bibnamefont {Cerezo}},\ and\ \bibinfo {author} {\bibfnamefont {P.~J.}\ \bibnamefont {Coles}},\ }\bibfield  {title} {\bibinfo {title} {Connecting ansatz expressibility to gradient magnitudes and barren plateaus},\ }\href {https://doi.org/10.1103/PRXQuantum.3.010313} {\bibfield  {journal} {\bibinfo  {journal} {PRX Quantum}\ }\textbf {\bibinfo {volume} {3}},\ \bibinfo {pages} {010313} (\bibinfo {year} {2022})}\BibitemShut {NoStop}%
\bibitem [{\citenamefont {McClean}\ \emph {et~al.}(2018)\citenamefont {McClean}, \citenamefont {Boixo}, \citenamefont {Smelyanskiy}, \citenamefont {Babbush},\ and\ \citenamefont {Neven}}]{mcclean2018barren}%
  \BibitemOpen
  \bibfield  {author} {\bibinfo {author} {\bibfnamefont {J.~R.}\ \bibnamefont {McClean}}, \bibinfo {author} {\bibfnamefont {S.}~\bibnamefont {Boixo}}, \bibinfo {author} {\bibfnamefont {V.~N.}\ \bibnamefont {Smelyanskiy}}, \bibinfo {author} {\bibfnamefont {R.}~\bibnamefont {Babbush}},\ and\ \bibinfo {author} {\bibfnamefont {H.}~\bibnamefont {Neven}},\ }\bibfield  {title} {\bibinfo {title} {Barren plateaus in quantum neural network training landscapes},\ }\href@noop {} {\bibfield  {journal} {\bibinfo  {journal} {Nature communications}\ }\textbf {\bibinfo {volume} {9}},\ \bibinfo {pages} {4812} (\bibinfo {year} {2018})}\BibitemShut {NoStop}%
\bibitem [{\citenamefont {Kandala}\ \emph {et~al.}(2017)\citenamefont {Kandala}, \citenamefont {Mezzacapo}, \citenamefont {Temme}, \citenamefont {Takita}, \citenamefont {Brink}, \citenamefont {Chow},\ and\ \citenamefont {Gambetta}}]{Kandala2017HEA}%
  \BibitemOpen
  \bibfield  {author} {\bibinfo {author} {\bibfnamefont {A.}~\bibnamefont {Kandala}}, \bibinfo {author} {\bibfnamefont {A.}~\bibnamefont {Mezzacapo}}, \bibinfo {author} {\bibfnamefont {K.}~\bibnamefont {Temme}}, \bibinfo {author} {\bibfnamefont {M.}~\bibnamefont {Takita}}, \bibinfo {author} {\bibfnamefont {M.}~\bibnamefont {Brink}}, \bibinfo {author} {\bibfnamefont {J.~M.}\ \bibnamefont {Chow}},\ and\ \bibinfo {author} {\bibfnamefont {J.~M.}\ \bibnamefont {Gambetta}},\ }\bibfield  {title} {\bibinfo {title} {Hardware-efficient variational quantum eigensolver for small molecules and quantum magnets},\ }\href {https://doi.org/10.1038/nature23879} {\bibfield  {journal} {\bibinfo  {journal} {Nature}\ }\textbf {\bibinfo {volume} {549}},\ \bibinfo {pages} {242} (\bibinfo {year} {2017})}\BibitemShut {NoStop}%
\bibitem [{\citenamefont {Leone}\ \emph {et~al.}(2024)\citenamefont {Leone}, \citenamefont {Oliviero}, \citenamefont {Cincio},\ and\ \citenamefont {Cerezo}}]{Leone2024practicalusefulness}%
  \BibitemOpen
  \bibfield  {author} {\bibinfo {author} {\bibfnamefont {L.}~\bibnamefont {Leone}}, \bibinfo {author} {\bibfnamefont {S.~F.}\ \bibnamefont {Oliviero}}, \bibinfo {author} {\bibfnamefont {L.}~\bibnamefont {Cincio}},\ and\ \bibinfo {author} {\bibfnamefont {M.}~\bibnamefont {Cerezo}},\ }\bibfield  {title} {\bibinfo {title} {On the practical usefulness of the {H}ardware {E}fficient {A}nsatz},\ }\href {https://doi.org/10.22331/q-2024-07-03-1395} {\bibfield  {journal} {\bibinfo  {journal} {{Quantum}}\ }\textbf {\bibinfo {volume} {8}},\ \bibinfo {pages} {1395} (\bibinfo {year} {2024})}\BibitemShut {NoStop}%
\bibitem [{\citenamefont {Livni}\ \emph {et~al.}(2014)\citenamefont {Livni}, \citenamefont {Shalev-Shwartz},\ and\ \citenamefont {Shamir}}]{livni2014computational}%
  \BibitemOpen
  \bibfield  {author} {\bibinfo {author} {\bibfnamefont {R.}~\bibnamefont {Livni}}, \bibinfo {author} {\bibfnamefont {S.}~\bibnamefont {Shalev-Shwartz}},\ and\ \bibinfo {author} {\bibfnamefont {O.}~\bibnamefont {Shamir}},\ }\bibfield  {title} {\bibinfo {title} {On the computational efficiency of training neural networks},\ }\href@noop {} {\bibfield  {journal} {\bibinfo  {journal} {Advances in neural information processing systems}\ }\textbf {\bibinfo {volume} {27}} (\bibinfo {year} {2014})}\BibitemShut {NoStop}%
\bibitem [{\citenamefont {Babbush}\ \emph {et~al.}(2021)\citenamefont {Babbush}, \citenamefont {McClean}, \citenamefont {Newman}, \citenamefont {Gidney}, \citenamefont {Boixo},\ and\ \citenamefont {Neven}}]{babbush2021errorcorrection}%
  \BibitemOpen
  \bibfield  {author} {\bibinfo {author} {\bibfnamefont {R.}~\bibnamefont {Babbush}}, \bibinfo {author} {\bibfnamefont {J.~R.}\ \bibnamefont {McClean}}, \bibinfo {author} {\bibfnamefont {M.}~\bibnamefont {Newman}}, \bibinfo {author} {\bibfnamefont {C.}~\bibnamefont {Gidney}}, \bibinfo {author} {\bibfnamefont {S.}~\bibnamefont {Boixo}},\ and\ \bibinfo {author} {\bibfnamefont {H.}~\bibnamefont {Neven}},\ }\bibfield  {title} {\bibinfo {title} {Focus beyond quadratic speedups for error-corrected quantum advantage},\ }\href {https://doi.org/10.1103/PRXQuantum.2.010103} {\bibfield  {journal} {\bibinfo  {journal} {PRX Quantum}\ }\textbf {\bibinfo {volume} {2}},\ \bibinfo {pages} {010103} (\bibinfo {year} {2021})}\BibitemShut {NoStop}%
\bibitem [{\citenamefont {Gallego-Mejia}\ \emph {et~al.}(2024)\citenamefont {Gallego-Mejia}, \citenamefont {Bustos-Brinez},\ and\ \citenamefont {González}}]{gallegomejia2024latentanomalydetectiondensity}%
  \BibitemOpen
  \bibfield  {author} {\bibinfo {author} {\bibfnamefont {J.}~\bibnamefont {Gallego-Mejia}}, \bibinfo {author} {\bibfnamefont {O.}~\bibnamefont {Bustos-Brinez}},\ and\ \bibinfo {author} {\bibfnamefont {F.~A.}\ \bibnamefont {González}},\ }\href {https://arxiv.org/abs/2408.07623} {\bibinfo {title} {Latent anomaly detection through density matrices}} (\bibinfo {year} {2024}),\ \Eprint {https://arxiv.org/abs/2408.07623} {arXiv:2408.07623 [cs.LG]} \BibitemShut {NoStop}%
\bibitem [{\citenamefont {Caro}\ \emph {et~al.}(2023)\citenamefont {Caro}, \citenamefont {Huang}, \citenamefont {Ezzell}, \citenamefont {Gibbs}, \citenamefont {Sornborger}, \citenamefont {Cincio}, \citenamefont {Coles},\ and\ \citenamefont {Holmes}}]{caro2023out}%
  \BibitemOpen
  \bibfield  {author} {\bibinfo {author} {\bibfnamefont {M.~C.}\ \bibnamefont {Caro}}, \bibinfo {author} {\bibfnamefont {H.-Y.}\ \bibnamefont {Huang}}, \bibinfo {author} {\bibfnamefont {N.}~\bibnamefont {Ezzell}}, \bibinfo {author} {\bibfnamefont {J.}~\bibnamefont {Gibbs}}, \bibinfo {author} {\bibfnamefont {A.~T.}\ \bibnamefont {Sornborger}}, \bibinfo {author} {\bibfnamefont {L.}~\bibnamefont {Cincio}}, \bibinfo {author} {\bibfnamefont {P.~J.}\ \bibnamefont {Coles}},\ and\ \bibinfo {author} {\bibfnamefont {Z.}~\bibnamefont {Holmes}},\ }\bibfield  {title} {\bibinfo {title} {Out-of-distribution generalization for learning quantum dynamics},\ }\href@noop {} {\bibfield  {journal} {\bibinfo  {journal} {Nature Communications}\ }\textbf {\bibinfo {volume} {14}},\ \bibinfo {pages} {3751} (\bibinfo {year} {2023})}\BibitemShut {NoStop}%
\bibitem [{\citenamefont {Gallego-Mejia}\ and\ \citenamefont {González}(2023)}]{Gallego2023demande}%
  \BibitemOpen
  \bibfield  {author} {\bibinfo {author} {\bibfnamefont {J.~A.}\ \bibnamefont {Gallego-Mejia}}\ and\ \bibinfo {author} {\bibfnamefont {F.~A.}\ \bibnamefont {González}},\ }\bibfield  {title} {\bibinfo {title} {Demande: Density matrix neural density estimation},\ }\href {https://doi.org/10.1109/ACCESS.2023.3279123} {\bibfield  {journal} {\bibinfo  {journal} {IEEE Access}\ }\textbf {\bibinfo {volume} {11}},\ \bibinfo {pages} {53062} (\bibinfo {year} {2023})}\BibitemShut {NoStop}%
\bibitem [{\citenamefont {Bustos-Brinez}\ \emph {et~al.}(2023)\citenamefont {Bustos-Brinez}, \citenamefont {Gallego-Mejia},\ and\ \citenamefont {Gonz{\'a}lez}}]{Bustos2023addmkde}%
  \BibitemOpen
  \bibfield  {author} {\bibinfo {author} {\bibfnamefont {O.~A.}\ \bibnamefont {Bustos-Brinez}}, \bibinfo {author} {\bibfnamefont {J.~A.}\ \bibnamefont {Gallego-Mejia}},\ and\ \bibinfo {author} {\bibfnamefont {F.~A.}\ \bibnamefont {Gonz{\'a}lez}},\ }\bibfield  {title} {\bibinfo {title} {Ad-dmkde: Anomaly detection through density matrices and fourier features},\ }in\ \href@noop {} {\emph {\bibinfo {booktitle} {Information Technology and Systems}}},\ \bibinfo {editor} {edited by\ \bibinfo {editor} {\bibfnamefont {{\'A}.}~\bibnamefont {Rocha}}, \bibinfo {editor} {\bibfnamefont {C.}~\bibnamefont {Ferr{\'a}s}},\ and\ \bibinfo {editor} {\bibfnamefont {W.}~\bibnamefont {Ibarra}}}\ (\bibinfo  {publisher} {Springer International Publishing},\ \bibinfo {address} {Cham},\ \bibinfo {year} {2023})\ pp.\ \bibinfo {pages} {327--338}\BibitemShut {NoStop}%
\bibitem [{\citenamefont {{Farhi}}\ and\ \citenamefont {{Neven}}(2018)}]{farhi}%
  \BibitemOpen
  \bibfield  {author} {\bibinfo {author} {\bibfnamefont {E.}~\bibnamefont {{Farhi}}}\ and\ \bibinfo {author} {\bibfnamefont {H.}~\bibnamefont {{Neven}}},\ }\bibfield  {title} {\bibinfo {title} {{Classification with Quantum Neural Networks on Near Term Processors}},\ }\href {https://doi.org/10.48550/arXiv.1802.06002} {\bibfield  {journal} {\bibinfo  {journal} {arXiv e-prints}\ ,\ \bibinfo {eid} {arXiv:1802.06002}} (\bibinfo {year} {2018})},\ \Eprint {https://arxiv.org/abs/1802.06002} {arXiv:1802.06002 [quant-ph]} \BibitemShut {NoStop}%
\bibitem [{\citenamefont {Dilip}\ \emph {et~al.}(2022)\citenamefont {Dilip}, \citenamefont {Liu}, \citenamefont {Smith},\ and\ \citenamefont {Pollmann}}]{dilip2022data}%
  \BibitemOpen
  \bibfield  {author} {\bibinfo {author} {\bibfnamefont {R.}~\bibnamefont {Dilip}}, \bibinfo {author} {\bibfnamefont {Y.-J.}\ \bibnamefont {Liu}}, \bibinfo {author} {\bibfnamefont {A.}~\bibnamefont {Smith}},\ and\ \bibinfo {author} {\bibfnamefont {F.}~\bibnamefont {Pollmann}},\ }\bibfield  {title} {\bibinfo {title} {Data compression for quantum machine learning},\ }\href@noop {} {\bibfield  {journal} {\bibinfo  {journal} {Physical Review Research}\ }\textbf {\bibinfo {volume} {4}},\ \bibinfo {pages} {043007} (\bibinfo {year} {2022})}\BibitemShut {NoStop}%
\bibitem [{\citenamefont {Huggins}\ \emph {et~al.}(2019)\citenamefont {Huggins}, \citenamefont {Patil}, \citenamefont {Mitchell}, \citenamefont {Whaley},\ and\ \citenamefont {Stoudenmire}}]{huggins2019towards}%
  \BibitemOpen
  \bibfield  {author} {\bibinfo {author} {\bibfnamefont {W.}~\bibnamefont {Huggins}}, \bibinfo {author} {\bibfnamefont {P.}~\bibnamefont {Patil}}, \bibinfo {author} {\bibfnamefont {B.}~\bibnamefont {Mitchell}}, \bibinfo {author} {\bibfnamefont {K.~B.}\ \bibnamefont {Whaley}},\ and\ \bibinfo {author} {\bibfnamefont {E.~M.}\ \bibnamefont {Stoudenmire}},\ }\bibfield  {title} {\bibinfo {title} {Towards quantum machine learning with tensor networks},\ }\href@noop {} {\bibfield  {journal} {\bibinfo  {journal} {Quantum Science and technology}\ }\textbf {\bibinfo {volume} {4}},\ \bibinfo {pages} {024001} (\bibinfo {year} {2019})}\BibitemShut {NoStop}%
\bibitem [{\citenamefont {Javadi-Abhari}\ \emph {et~al.}(2024)\citenamefont {Javadi-Abhari}, \citenamefont {Treinish}, \citenamefont {Krsulich}, \citenamefont {Wood}, \citenamefont {Lishman}, \citenamefont {Gacon}, \citenamefont {Martiel}, \citenamefont {Nation}, \citenamefont {Bishop}, \citenamefont {Cross}, \citenamefont {Johnson},\ and\ \citenamefont {Gambetta}}]{qiskit2024}%
  \BibitemOpen
  \bibfield  {author} {\bibinfo {author} {\bibfnamefont {A.}~\bibnamefont {Javadi-Abhari}}, \bibinfo {author} {\bibfnamefont {M.}~\bibnamefont {Treinish}}, \bibinfo {author} {\bibfnamefont {K.}~\bibnamefont {Krsulich}}, \bibinfo {author} {\bibfnamefont {C.~J.}\ \bibnamefont {Wood}}, \bibinfo {author} {\bibfnamefont {J.}~\bibnamefont {Lishman}}, \bibinfo {author} {\bibfnamefont {J.}~\bibnamefont {Gacon}}, \bibinfo {author} {\bibfnamefont {S.}~\bibnamefont {Martiel}}, \bibinfo {author} {\bibfnamefont {P.~D.}\ \bibnamefont {Nation}}, \bibinfo {author} {\bibfnamefont {L.~S.}\ \bibnamefont {Bishop}}, \bibinfo {author} {\bibfnamefont {A.~W.}\ \bibnamefont {Cross}}, \bibinfo {author} {\bibfnamefont {B.~R.}\ \bibnamefont {Johnson}},\ and\ \bibinfo {author} {\bibfnamefont {J.~M.}\ \bibnamefont {Gambetta}},\ }\href {https://doi.org/10.48550/arXiv.2405.08810} {\bibinfo {title} {Quantum computing with {Q}iskit}} (\bibinfo {year} {2024}),\ \Eprint {https://arxiv.org/abs/2405.08810} {arXiv:2405.08810 [quant-ph]}
  \BibitemShut {NoStop}%
\bibitem [{\citenamefont {Preskill}(2018)}]{Preskill2018quantumcomputingin}%
  \BibitemOpen
  \bibfield  {author} {\bibinfo {author} {\bibfnamefont {J.}~\bibnamefont {Preskill}},\ }\bibfield  {title} {\bibinfo {title} {Quantum {C}omputing in the {NISQ} era and beyond},\ }\href {https://doi.org/10.22331/q-2018-08-06-79} {\bibfield  {journal} {\bibinfo  {journal} {{Quantum}}\ }\textbf {\bibinfo {volume} {2}},\ \bibinfo {pages} {79} (\bibinfo {year} {2018})}\BibitemShut {NoStop}%
\bibitem [{\citenamefont {Zhu}\ \emph {et~al.}(2024)\citenamefont {Zhu}, \citenamefont {Pan}, \citenamefont {Mou}, \citenamefont {Deng}, \citenamefont {Zhu}, \citenamefont {Wang}, \citenamefont {Pareek}, \citenamefont {Hyams}, \citenamefont {Carneiro}, \citenamefont {Hadfield}, \citenamefont {El-Deiry}, \citenamefont {Yang}, \citenamefont {Tan}, \citenamefont {Tong}, \citenamefont {Ta}, \citenamefont {Zhu}, \citenamefont {Gao}, \citenamefont {Lai}, \citenamefont {Cheng}, \citenamefont {Chen},\ and\ \citenamefont {Xue}}]{ZHU2024101506}%
  \BibitemOpen
  \bibfield  {author} {\bibinfo {author} {\bibfnamefont {L.}~\bibnamefont {Zhu}}, \bibinfo {author} {\bibfnamefont {J.}~\bibnamefont {Pan}}, \bibinfo {author} {\bibfnamefont {W.}~\bibnamefont {Mou}}, \bibinfo {author} {\bibfnamefont {L.}~\bibnamefont {Deng}}, \bibinfo {author} {\bibfnamefont {Y.}~\bibnamefont {Zhu}}, \bibinfo {author} {\bibfnamefont {Y.}~\bibnamefont {Wang}}, \bibinfo {author} {\bibfnamefont {G.}~\bibnamefont {Pareek}}, \bibinfo {author} {\bibfnamefont {E.}~\bibnamefont {Hyams}}, \bibinfo {author} {\bibfnamefont {B.~A.}\ \bibnamefont {Carneiro}}, \bibinfo {author} {\bibfnamefont {M.~J.}\ \bibnamefont {Hadfield}}, \bibinfo {author} {\bibfnamefont {W.~S.}\ \bibnamefont {El-Deiry}}, \bibinfo {author} {\bibfnamefont {T.}~\bibnamefont {Yang}}, \bibinfo {author} {\bibfnamefont {T.}~\bibnamefont {Tan}}, \bibinfo {author} {\bibfnamefont {T.}~\bibnamefont {Tong}}, \bibinfo {author} {\bibfnamefont {N.}~\bibnamefont {Ta}}, \bibinfo {author} {\bibfnamefont {Y.}~\bibnamefont {Zhu}}, \bibinfo {author}
  {\bibfnamefont {Y.}~\bibnamefont {Gao}}, \bibinfo {author} {\bibfnamefont {Y.}~\bibnamefont {Lai}}, \bibinfo {author} {\bibfnamefont {L.}~\bibnamefont {Cheng}}, \bibinfo {author} {\bibfnamefont {R.}~\bibnamefont {Chen}},\ and\ \bibinfo {author} {\bibfnamefont {W.}~\bibnamefont {Xue}},\ }\bibfield  {title} {\bibinfo {title} {Harnessing artificial intelligence for prostate cancer management},\ }\href {https://doi.org/https://doi.org/10.1016/j.xcrm.2024.101506} {\bibfield  {journal} {\bibinfo  {journal} {Cell Reports Medicine}\ }\textbf {\bibinfo {volume} {5}},\ \bibinfo {pages} {101506} (\bibinfo {year} {2024})}\BibitemShut {NoStop}%
\bibitem [{\citenamefont {You}\ \emph {et~al.}(2022)\citenamefont {You}, \citenamefont {Cui}, \citenamefont {Shen}, \citenamefont {Yang}, \citenamefont {Lu}, \citenamefont {Zheng},\ and\ \citenamefont {Le}}]{NEURIPS2022_1d774c11}%
  \BibitemOpen
  \bibfield  {author} {\bibinfo {author} {\bibfnamefont {Z.}~\bibnamefont {You}}, \bibinfo {author} {\bibfnamefont {L.}~\bibnamefont {Cui}}, \bibinfo {author} {\bibfnamefont {Y.}~\bibnamefont {Shen}}, \bibinfo {author} {\bibfnamefont {K.}~\bibnamefont {Yang}}, \bibinfo {author} {\bibfnamefont {X.}~\bibnamefont {Lu}}, \bibinfo {author} {\bibfnamefont {Y.}~\bibnamefont {Zheng}},\ and\ \bibinfo {author} {\bibfnamefont {X.}~\bibnamefont {Le}},\ }\bibfield  {title} {\bibinfo {title} {A unified model for multi-class anomaly detection},\ }in\ \href {https://proceedings.neurips.cc/paper_files/paper/2022/file/1d774c112926348c3e25ea47d87c835b-Paper-Conference.pdf} {\emph {\bibinfo {booktitle} {Advances in Neural Information Processing Systems}}},\ Vol.~\bibinfo {volume} {35},\ \bibinfo {editor} {edited by\ \bibinfo {editor} {\bibfnamefont {S.}~\bibnamefont {Koyejo}}, \bibinfo {editor} {\bibfnamefont {S.}~\bibnamefont {Mohamed}}, \bibinfo {editor} {\bibfnamefont {A.}~\bibnamefont {Agarwal}}, \bibinfo {editor}
  {\bibfnamefont {D.}~\bibnamefont {Belgrave}}, \bibinfo {editor} {\bibfnamefont {K.}~\bibnamefont {Cho}},\ and\ \bibinfo {editor} {\bibfnamefont {A.}~\bibnamefont {Oh}}}\ (\bibinfo  {publisher} {Curran Associates, Inc.},\ \bibinfo {year} {2022})\ pp.\ \bibinfo {pages} {4571--4584}\BibitemShut {NoStop}%
\bibitem [{\citenamefont {Held}\ and\ \citenamefont {Saban{\'e}s~Bov{\'e}}(2014)}]{held2014applied}%
  \BibitemOpen
  \bibfield  {author} {\bibinfo {author} {\bibfnamefont {L.}~\bibnamefont {Held}}\ and\ \bibinfo {author} {\bibfnamefont {D.}~\bibnamefont {Saban{\'e}s~Bov{\'e}}},\ }\bibfield  {title} {\bibinfo {title} {Applied statistical inference},\ }\href@noop {} {\bibfield  {journal} {\bibinfo  {journal} {Springer, Berlin Heidelberg, doi}\ }\textbf {\bibinfo {volume} {10}},\ \bibinfo {pages} {16} (\bibinfo {year} {2014})}\BibitemShut {NoStop}%
\bibitem [{\citenamefont {Thanasilp}\ \emph {et~al.}(2024)\citenamefont {Thanasilp}, \citenamefont {Wang}, \citenamefont {Cerezo},\ and\ \citenamefont {Holmes}}]{thanasilp2024exponential}%
  \BibitemOpen
  \bibfield  {author} {\bibinfo {author} {\bibfnamefont {S.}~\bibnamefont {Thanasilp}}, \bibinfo {author} {\bibfnamefont {S.}~\bibnamefont {Wang}}, \bibinfo {author} {\bibfnamefont {M.}~\bibnamefont {Cerezo}},\ and\ \bibinfo {author} {\bibfnamefont {Z.}~\bibnamefont {Holmes}},\ }\bibfield  {title} {\bibinfo {title} {Exponential concentration in quantum kernel methods},\ }\href@noop {} {\bibfield  {journal} {\bibinfo  {journal} {Nature communications}\ }\textbf {\bibinfo {volume} {15}},\ \bibinfo {pages} {5200} (\bibinfo {year} {2024})}\BibitemShut {NoStop}%
\end{thebibliography}%

\onecolumngrid
\appendix

\section{\label{sec: representer theorem for ms}Representer theorem for density matrices}

We present a proof of Eq. \ref{eq: variational RKHS}, which corresponds to a special form of the representer theorem \cite{scholkopf2002learning, Schuld2021quantumkernel} in terms of density matrices when the risk to minimize is the average negative log-likelihood of the training data.

\begin{prop}
Consider a data set of $N$ training features and labels $\{(\boldsymbol{x}_j, \boldsymbol{y}_j)\}_{\{0\cdots N-1\}}$ ,a test sample $(\boldsymbol{x}^*, \boldsymbol{y}^*)$, and two quantum feature maps, one for the inputs $\bar \psi_{_\mathcb{X}}:\mathbb{X}\rightarrow\mathcal{X}$, $\boldsymbol{x}' \mapsto \kettwo{\bar\psi_{_\mathcb{X}}'}$ and one for the outputs $\bar\psi_{_\mathcb{Y}}:\mathbb{Y}\rightarrow\mathcal{Y}$, $\boldsymbol{y}' \mapsto \kettwo{\bar\psi_{_\mathcb{Y}}'}$, whose inner products in the Hilbert spaces correspond to the exact shift-invariant kernels in the data space, i.e., for any $\boldsymbol{x}', \boldsymbol{x}''\in \mathbb X$ and any $\boldsymbol{y}', \boldsymbol{y}''\in \mathbb Y$, $\abstwo{\ketbratwo{\bar\psi_{_{\mathcb{X}}}'}{\bar\psi_{_{\mathcb{X}}}''}}^2 = \kappa_{_{\mathbb X}}(\boldsymbol{x}', \boldsymbol{x}'')$, $\abstwo{\ketbratwo{\bar\psi_{_\mathcb{Y}}'}{\bar\psi_{_\mathcb{Y}}''}}^2 = \kappa_{_{\mathbb Y}}(\boldsymbol{y}', \boldsymbol{y}')$. Also, let $\bar\rho_{{_{\mathcb{X}}}, {_\mathcb{Y}}}(\boldsymbol \theta) \in \mathcal{X}\otimes\mathcal{Y}$ be a variational density matrix with parameters $\boldsymbol \theta$, and $\mathcal{M}_{_{\mathbb{X}}} = [\int_{\mathbb{X}}\kappa_{\mathbb{X}}(\boldsymbol x', \boldsymbol{x}^*)\,d \boldsymbol x^*]^{-1}$, $ \mathcal{M}_{_{\mathbb{Y}}} =[\int_{\mathbb{Y}}\kappa_{\mathbb{Y}}(\boldsymbol y', \boldsymbol{y}^*)\,d \boldsymbol y^*]^{-1}$ the normalization constants of the kernels. Then,  the minimization $\boldsymbol \theta_{\text{op}} = \argmin_{\boldsymbol{\theta}}\mathcal{L}(\boldsymbol{\theta})$ of the loss $\mathcal{L}(\boldsymbol{\theta})=-(1/N)\sum_{j=0}^{N-1}\log{f(\boldsymbol x_j, \boldsymbol y_j| \boldsymbol \theta)}$, with $f(\boldsymbol x', \boldsymbol y'| \boldsymbol \theta)= \mathcal{M}_{_{\mathbb{X}}}\mathcal{M}_{_{\mathbb{Y}}}\bratwo{\bar\psi_{_\mathcb{X},_\mathcb{Y}}'}\bar\rho_{{_{\mathcb{X}}}, {_\mathcb{Y}}}(\boldsymbol\theta) \kettwo{\bar\psi_{_\mathcb{X}, _\mathcb{Y}}'}$ and $\kettwo{\bar\psi_{_{\mathcb{X}, \mathcb{Y}}}'} = \kettwo{\bar\psi_{_\mathcb{X}}'}\otimes\kettwo{\bar\psi_{_\mathcb{Y}}'}$ corresponds a function $f(\boldsymbol x^*, \boldsymbol y^*| \boldsymbol \theta_{\text{op}})$ in a RKHS of the form
\begin{equation}
f(\boldsymbol x^*, \boldsymbol y^*| \boldsymbol \theta_{\text{op}}) = \mathcal{M}_{_{\mathbb{X}}}\mathcal{M}_{_{\mathbb{Y}}}\sum_{j=0}^{N-1}q_j(\boldsymbol{\theta}_{\text{op}})\kappa_{_{\mathbb{X}}}(\boldsymbol{x}_j, \boldsymbol{x}^*)\kappa_{_{\mathbb{Y}}}(\boldsymbol{y}_j, \boldsymbol{y}^*),
\end{equation}
with $q_j(\boldsymbol{\theta}_{\text{op}})\in \mathbb R$ and $q_j(\boldsymbol{\theta}_{\text{op}})\ge 0$ for all $j$, and $\sum_jq_j(\boldsymbol{\theta}_{\text{op}})=1$. 
\label{prop: proposition 1}
\end{prop}

\textit{Proof:} Consider the quantum features of the training data set $\{\kettwo{\bar\psi_{_{\mathcb{X}}, _{\mathcb Y}}^j}\}_{0\cdots N-1}$ and a variational density matrix $\bar\rho_{{_{\mathcb{X}}}, {_\mathcb{Y}}}(\boldsymbol \theta)$ that can be written as the convex combination of two density matrices $\bar\rho_{{_{\mathcb{X}}}, {_\mathcb{Y}}}(\boldsymbol \theta) = (1-r(\boldsymbol{\theta}))\bar\rho_{{_{\mathcb{X}}}, {_\mathcb{Y}}}^\perp(\boldsymbol \theta) + r(\boldsymbol{\theta})\bar\rho_{{_{\mathcb{X}}}, {_\mathcb{Y}}}^\parallel(\boldsymbol \theta)$ where $0 \leq r(\boldsymbol{\theta})\leq 1$, and $\bar\rho_{{_{\mathcb{X}}}, {_\mathcb{Y}}}^\perp(\boldsymbol \theta)$, $\bar\rho_{{_{\mathcb{X}}}, {_\mathcb{Y}}}^\parallel(\boldsymbol \theta)$ are mixed states that are, respectively, orthogonal and non-orthogonal to the training quantum features, i.e., $\text{Tr}\big[\bar\rho_{{_{\mathcb{X}}}, {_\mathcb{Y}}}^\perp(\boldsymbol \theta)\kettwo{\bar\psi_{_{\mathcb{X}}, _{\mathcb Y}}^k}\bratwo{\bar\psi_{_{\mathcb{X}}, _{\mathcb Y}}^k}\big] = 0$ for all $k$ and $\text{Tr}\big[\bar\rho_{{_{\mathcb{X}}}, {_\mathcb{Y}}}^\parallel(\boldsymbol \theta)\kettwo{\bar\psi_{_{\mathcb{X}}, _{\mathcb Y}}^k}\bratwo{\bar\psi_{_{\mathcb{X}}, _{\mathcb Y}}^k}\big] \neq 0$ for some $k$. Furthermore, since $\bar\rho_{{_{\mathcb{X}}}, {_\mathcb{Y}}}^\parallel(\boldsymbol \theta)$ is non-orthogonal to the training features, we can write it as a convex combination  of them, 
\begin{equation}
\bar\rho_{{_{\mathcb{X}}}, {_\mathcb{Y}}}^\parallel(\boldsymbol \theta) = \sum_{j=0}^{N-1} q_j(\boldsymbol \theta)\kettwo{\bar\psi_{_{\mathcb{X}}, _{\mathcb Y}}^j}\bratwo{\bar\psi_{_{\mathcb{X}}, _{\mathcb Y}}^j},  	
\end{equation}
with $q_j(\boldsymbol{\theta})\in \mathbb R$, $q_j(\boldsymbol{\theta}_{\text{op}})\ge 0$ for all $j$, and $\sum_jq_j(\boldsymbol{\theta})=1$. Evaluating the function $f$ over some generic training feature $\kettwo{\bar\psi_{_{\mathcb{X}}, _{\mathcb Y}}^k}$ leads to
\begin{align}
f(\boldsymbol x_k, \boldsymbol y_k| \boldsymbol \theta) &=   \mathcal{M}_{_{\mathbb{X}}}\mathcal{M}_{_{\mathbb{Y}}}\bratwo{\bar\psi_{_{\mathcb{X}}, _{\mathcb{Y}}}^k}\bar\rho_{{_{\mathcb{X}}}, {_\mathcb{Y}}}(\boldsymbol{\theta})\kettwo{\bar\psi_{_{\mathcb{X}}, _\mathcb{Y}}^k}  \notag \\ 
&= \mathcal{M}_{_{\mathbb{X}}}\mathcal{M}_{_{\mathbb{Y}}}\bratwo{\bar\psi_{_{\mathcb{X}}, _{\mathcb{Y}}}^k}(1-r(\boldsymbol{\theta}))\bar\rho_{{_{\mathcb{X}}}, {_\mathcb{Y}}}^\perp(\boldsymbol \theta) + r(\boldsymbol{\theta})\bar\rho_{{_{\mathcb{X}}}, {_\mathcb{Y}}}^\parallel(\boldsymbol \theta)\kettwo{\bar\psi_{_{\mathcb{X}}, _\mathcb{Y}}^k}  \notag \\ 
&= r(\boldsymbol{\theta})\mathcal{M}_{_{\mathbb{X}}}\mathcal{M}_{_{\mathbb{Y}}}\sum_{j=0}^{N-1}q_j(\boldsymbol{\theta})\abstwo{\ketbratwo{\bar\psi_{_{\mathcb{X}}, _{\mathcb{Y}}}^j}{\bar\psi_{_{\mathcb{X}}, _{\mathcb{Y}}}^k}}^2 \notag \\
&= r(\boldsymbol{\theta})\mathcal{M}_{_{\mathbb{X}}}\mathcal{M}_{_{\mathbb{Y}}}\sum_{j=0}^{N-1}q_j(\boldsymbol{\theta})\abstwo{\ketbratwo{\bar\psi_{_{\mathcb{X}}}^j}{\bar\psi_{_{\mathcb{X}}}^k}}^2 \abstwo{\ketbratwo{\bar\psi_{_{\mathcb{Y}}}^j}{\bar\psi_{_{\mathcb{Y}}}^k}}^2.
\end{align}
In addition, the optimal value $\boldsymbol{\theta}_{\text{op}}=\argmin_{\boldsymbol{\theta}}\mathcal{L}(\boldsymbol \theta)$ of the minimization of the loss
\begin{align}
   \mathcal L(\boldsymbol{\theta}) &= -\frac{1}{N}\sum_{k=0}^{N-1}\log{\big(\mathcal{M}_{_{\mathbb{X}}}\mathcal{M}_{_{\mathbb{Y}}}\bratwo{\bar\psi_{_{\mathcb{X}}, _{\mathcb {Y}}}^k}\bar\rho_{_{\mathcb{X}}, {_{\mathcb{Y}}}}
(\boldsymbol\theta) \kettwo{\bar\psi_{_{\mathcb{X}}, {_{\mathcb{Y}}}}^k}\big)}\notag \\
    &= -\frac{1}{N}\sum_{k=0}^{N-1}\log{\bigg\{r(\boldsymbol{\theta}) \mathcal{M}_{_{\mathbb{X}}}\mathcal{M}_{_{\mathbb{Y}}}\sum_{j=0}^{N-1}q_j(\boldsymbol{\theta})\abstwo{\ketbratwo{\bar\psi_{_{\mathcb{X}}}^j}{\bar\psi_{_{\mathcb{X}}}^k}}^2 \abstwo{\ketbratwo{\bar\psi_{_{\mathcb{Y}}}^j}{\bar\psi_{_{\mathcb{Y}}}^k}}^2\bigg\}} \notag \\
    &= -\frac{1}{N}\sum_{k=0}^{N-1}\log{\bigg\{\mathcal{M}_{_{\mathbb{X}}}\mathcal{M}_{_{\mathbb{Y}}}\sum_{j=0}^{N-1}q_j(\boldsymbol{\theta})\abstwo{\ketbratwo{\bar\psi_{_{\mathcb{X}}}^j}{\bar\psi_{_{\mathcb{X}}}^k}}^2 \abstwo{\ketbratwo{\bar\psi_{_{\mathcb{Y}}}^j}{\bar\psi_{_{\mathcb{Y}}}^k}}^2\bigg\}} - \log{r(\boldsymbol{\theta})}
\end{align}
guarantees that $r(\boldsymbol{\theta}_{\text{op}}) = 1$, because $- \log{r(\boldsymbol{\theta})}\geq 0$ and $- \log{r(\boldsymbol{\theta}_{\text{op}})} = 0$. Therefore, we would have that evaluating the function $f$ over some test sample  $(\boldsymbol{x}^* , \boldsymbol{y}^* )\mapsto \kettwo{\bar\psi_{_{\mathcb X}}^*}\otimes\kettwo{\bar\psi_{_{\mathcb Y}}^*}=\kettwo{\bar\psi_{_{\mathcb X, \mathcb Y}}^*}$ at the optimal values $\boldsymbol{\theta}_{\text{op}}$ results into
\begin{equation}
f(\boldsymbol x^*, \boldsymbol y^*| \boldsymbol \theta_{\text{op}}) = \mathcal{M}_{_{\mathbb{X}}}\mathcal{M}_{_{\mathbb{Y}}}\sum_{j=0}^{N-1}q_j(\boldsymbol{\theta}_{\text{op}})\abstwo{\ketbratwo{\bar\psi_{_{\mathcb{X}}}^j}{\bar\psi_{_{\mathcb{X}}}^*}}^2 \abstwo{\ketbratwo{\bar\psi_{_{\mathcb{Y}}}^j}{\bar\psi_{_{\mathcb{Y}}}^*}}^2.
\end{equation}
Finally, since the quantum feature maps $ \kettwo{\bar\psi_{_\mathcb{X}}'} \in \mathcal X$ and $\kettwo{\bar\psi_{_\mathcb{Y}}'}\in \mathcal Y$  explicitly induce the shift-invariant kernels $\kappa_{_{\mathbb{X}}}$, $\kappa_{_{\mathbb{Y}}}$ by an inner product in their respective Hilbert spaces, we would have that
\begin{align}
f(\boldsymbol x^*, \boldsymbol y^*| \boldsymbol \theta_{\text{op}}) = \mathcal{M}_{_{\mathbb{X}}}\mathcal{M}_{_{\mathbb{Y}}}\sum_{j=0}^{N-1}q_j(\boldsymbol{\theta}_{\text{op}})\kappa_{_{\mathbb{Y}}}(\boldsymbol{y}_j, \boldsymbol{y}^*)\kappa_{_{\mathbb{X}}}(\boldsymbol{x}_j, \boldsymbol{x}^*),
\end{align} 
where each $q_j(\boldsymbol{\theta}_{\text{op}})\in \mathbb R$ and $q_j(\boldsymbol{\theta}_{\text{op}})\ge 0$, and $\sum_jq_j(\boldsymbol{\theta}_{\text{op}})=1$.

\section{\label{sec: QRFF-QEFF}Interplay between quantum random and quantum enhanced Fourier features}

\begin{prop}

The quantum enhanced Fourier feature encoding   $\psi_{_\mathcb{X}}:\mathbb{R}^D\rightarrow\mathbb{C}^{2^{n_{_{\mathcb{X}}}}}$,  $\boldsymbol{x}'\mapsto \kettwo{\psi_{_{\mathcb{X}}}'}$ given by,
\begin{align}
&	\kettwo{\psi_{_{\mathcb{X}}}'} = \exp \bigg\{-\frac{i}{2}\sum_{\alpha_1, \alpha_2, \cdots, \alpha_{n_{_{\mathcb{X}}}} \in \{0, 1\}}\bigg(\sqrt{\frac{1}{2h^2}}\boldsymbol{\vartheta}_{\alpha_{n_{_{\mathcb{X}}}}, \cdots, \alpha_2, \alpha_1}\cdot\boldsymbol{x}'\bigg) \notag \\
 & \hspace{0.5in}\times \bigg(\sigma^{\alpha_{n_{_{\mathcb{X}}}}} \otimes \cdots \otimes \sigma^{\alpha_2}\otimes \sigma^{\alpha_1}\bigg)\bigg\}H^{\otimes n_{_{\mathcb{X}}}}\kettwo{0_{_{\mathcb{X}}}},
    \label{eq: qeff mapping appendix}
\end{align}
where $\sigma^{0} = \mathbf{I}_2$, $\sigma^{1} = \sigma^{z}$, and the weights $\{\boldsymbol\vartheta_{\alpha_{n_{_{\mathcb{X}}}}, \cdots, \alpha_2, \alpha_1}\} \in \mathbb{R}^D$, except for the all zeros index $\boldsymbol\vartheta_{0, 0, \cdots, 0} = \boldsymbol 0$, are sampled from an arbitrary symmetric probability distribution $\{\boldsymbol\vartheta_{\alpha_{n_{_{\mathcb{X}}}}, \cdots, \alpha_2, \alpha_1}\}\sim\phi(\mathbf{0}, \frac{4}{(2^{n_{_{\mathcb{X}}}})-1}\mathbf{I}_D)$ with mean zero $\boldsymbol{\mu}=\mathbf{0}\in\mathbb R^D$ and covariance matrix $\Sigma = \frac{4}{(2^{n_{_{\mathcb{X}}}})-1}\mathbf{I}_D$ builds the quantum random Fourier feature mapping \cite{useche2022quantum} 
\begin{equation}
    \kettwo{\psi_{_{\mathcb{X}}}'} = \sqrt{\frac{1}{2^{n_{_{\mathcb{X}}}}}}\sum_{k=0}^{2^{n_{_{\mathcb{X}}}} - 1}e^{i\sqrt{\frac{1}{2h^2}}\boldsymbol{w}_k \cdot\boldsymbol{x}'}\kettwo{k_{_{\mathcb{X}}}},\label{eq: QRFF map appendix}
\end{equation}
where $\{\kettwo{k_{_{\mathcb{X}}}}\}_{0 \cdots 2^{n_{_{\mathcb{X}}}} - 1}$ is the canonical basis, and the weights $\{\boldsymbol{w}_k\}_{0 \cdots 2^{n_{_{\mathcb{X}}}} - 1}\in \mathbb R^D$ are normally distributed $\boldsymbol{w}_k \sim \mathcal{N}(\mathbf{0}, \mathbf I_D)$.
\label{prop: proposition 2}
\end{prop}

\textit{Proof:} The initial $n_{_\mathcb{X}}$ Hadamard gates applied to each qubit of the all-zeros state build the following state 
\begin{equation}
\bigotimes_{l=1}^{n_{_{\mathcb{X}}}}\sqrt{\frac{1}{2}}\bigg(\kettwo{0}+\kettwo{1}\bigg) = H^{\otimes n_{_{\mathcb{X}}}}\kettwo{0_{_{\mathcb{X}}}}
\end{equation}
also, applying the complex unitary matrix $U_{_\mathcb X}(\boldsymbol{x}'|{\boldsymbol \Theta})$ of the QEFF embedding, see Eq. \ref{eq: unitary QEFF}, we obtain
\begin{align}
&\sqrt{\frac{1}{2^{n_{_{\mathcb{X}}}}}}\exp \bigg\{-\frac{i}{2}\sum_{\alpha_1, \alpha_2, \cdots, \alpha_{n_{_{\mathcb{X}}}} \in \{0, 1\}}\bigg(\sqrt{\frac{1}{2h^2}}\boldsymbol{\vartheta}_{\alpha_{n_{_{\mathcb{X}}}}, \cdots, \alpha_2, \alpha_1}\cdot\boldsymbol{x}'\bigg) \notag \\ & \hspace{0.5in}\times \bigg(\sigma^{\alpha_{n_{_{\mathcb{X}}}}} \otimes \cdots \otimes \sigma^{\alpha_2}\otimes \sigma^{\alpha_1}\bigg)\bigg\}\bigotimes_{l=1}^{n_{_{\mathcb{X}}}}\bigg(\kettwo{0}+\kettwo{1}\bigg),
\end{align}
considering that the complex exponential is a diagonal matrix, and that $\bigotimes_{l=1}^{n_{_{\mathcb{X}}}}\big(\kettwo{0}+\kettwo{1}\big)$ is the vector of ones $\boldsymbol{1}\in\mathbb{R}^{2^{n_{_{\mathcb{X}}}}}$, the previous matrix-vector multiplication extracts the diagonal terms of the exponential matrix, and therefore, we can write the expression as
\begin{align}
    &\sqrt{\frac{1}{2^{n_{_{\mathcb{X}}}}}}\exp \bigg\{-\frac{i}{2}\bigg[\sum_{\alpha_1, \alpha_2, \cdots, \alpha_{n_{_{\mathcb{X}}}} \in \{0, 1\}}\bigg(\sqrt{\frac{1}{2h^2}}\boldsymbol{\vartheta}_{\alpha_{n_{_{\mathcb{X}}}}, \cdots, \alpha_2, \alpha_1}\cdot\boldsymbol{x}'\bigg) \notag \\ &\hspace{0.5in}\times\bigg(\sigma^{\alpha_{n_{_{\mathcb{X}}}}} \otimes \cdots \otimes \sigma^{\alpha_2}\otimes \sigma^{\alpha_1}\bigg)\bigg]\bigotimes_{l=1}^{n_{_{\mathcb{X}}}}\bigg(\kettwo{0}+\kettwo{1}\bigg)\bigg\} = \notag \\
    &\sqrt{\frac{1}{2^{n_{_{\mathcb{X}}}}}}\exp \bigg\{-\frac{i}{2}\sum_{\alpha_1, \alpha_2, \cdots, \alpha_{n_{_{\mathcb{X}}}} \in \{0, 1\}}\bigg[\bigg(\sqrt{\frac{1}{2h^2}}\boldsymbol{\vartheta}_{\alpha_{n_{_{\mathcb{X}}}}, \cdots, \alpha_2, \alpha_1}\cdot\boldsymbol{x}'\bigg) \notag \\ &\hspace{0.5in}\times\bigg(\sigma^{\alpha_{n_{_{\mathcb{X}}}}} \otimes \cdots \otimes \sigma^{\alpha_2}\otimes \sigma^{\alpha_1}\bigg)\bigotimes_{l=1}^{n_{_{\mathcb{X}}}}\bigg(\kettwo{0}+\kettwo{1}\bigg)\bigg]\bigg\}=\notag \\
    &\sqrt{\frac{1}{2^{n_{_{\mathcb{X}}}}}}\exp \bigg\{-\frac{i}{2}\sum_{\alpha_1, \alpha_2, \cdots, \alpha_{n_{_{\mathcb{X}}}} \in \{0, 1\}}\bigg[\bigg(\sqrt{\frac{1}{2h^2}}\boldsymbol{\vartheta}_{\alpha_{n_{_{\mathcb{X}}}}, \cdots, \alpha_2, \alpha_1}\cdot\boldsymbol{x}'\bigg) \bigotimes_{l=n_{_{\mathcb{X}}}} ^{1}\bigg(\sigma^{\alpha_l}\big(\kettwo{0}+\kettwo{1}\big)\bigg)\bigg]\bigg\}.
\end{align}
When $\alpha_l=0$,. we would have that $\sigma^{\alpha_l} = \mathbf{I}_2$ and the qubit-wise operation over the superposition would lead to $\big(\kettwo{0}+\kettwo{1}\big)$. In the contrary, when $\alpha_l=1$, we would obtain $\sigma^{\alpha_l} = \sigma^{z}$ and there would be a sign shift in the second term of the superposition $\big(\kettwo{0}-\kettwo{1}\big)$. Hence the previous expression would result into
\begin{equation}
    \sqrt{\frac{1}{2^{n_{_{\mathcb{X}}}}}}\exp \bigg\{-\frac{i}{2}\sum_{\alpha_1, \alpha_2, \cdots, \alpha_{n_{_{\mathcb{X}}}} \in \{0, 1\}}\bigg[\bigg(\sqrt{\frac{1}{2h^2}}\boldsymbol{\vartheta}_{\alpha_{n_{_{\mathcb{X}}}}, \cdots, \alpha_2, \alpha_1}\cdot\boldsymbol{x}'\bigg) \bigotimes_{l=n_{_{\mathcb{X}}}} ^{1}\bigg(\kettwo{0}+(-1)^{\displaystyle\delta_{\alpha_l,1}}\kettwo{1}\bigg)\bigg]\bigg\},
\end{equation}
where the $(-1)^{\displaystyle\delta_{\alpha_l,1}}$ is used to indicate the sign shift. Furthermore, we can expand the tensor product as follows
\begin{align}
    &\sqrt{\frac{1}{2^{n_{_{\mathcb{X}}}}}}\exp \bigg\{-\frac{i}{2}\sum_{\alpha_1, \alpha_2, \cdots, \alpha_{n_{_{\mathcb{X}}}} \in \{0, 1\}}\bigg[\bigg(\sqrt{\frac{1}{2h^2}}\boldsymbol{\vartheta}_{\alpha_{n_{_{\mathcb{X}}}}, \cdots, \alpha_2, \alpha_1}\cdot\boldsymbol{x}'\bigg) \bigotimes_{l=n_{_{\mathcb{X}}}}^{1}\bigg(\sum_{b_l={0}}^1(-1)^{{\displaystyle\delta_{\alpha_l,1}}{\displaystyle\delta_{b_l,1}}}\kettwo{b_l}\bigg)\bigg]\bigg\} =\notag \\
    & \sqrt{\frac{1}{2^{n_{_{\mathcb{X}}}}}}\exp \bigg\{-\frac{i}{2}\sum_{\alpha_1, \alpha_2, \cdots, \alpha_{n_{_{\mathcb{X}}}} \in \{0, 1\}}\bigg[\bigg(\sqrt{\frac{1}{2h^2}}\boldsymbol{\vartheta}_{\alpha_{n_{_{\mathcb{X}}}}, \cdots, \alpha_2, \alpha_1}\cdot\boldsymbol{x}'\bigg) \notag \\
    &\hspace{0.5in}\times \bigg(\sum_{b_1, b_2, \cdots, b_{n_{_{\mathcb{X}}}} \in \{0, 1\}} (-1)^{\displaystyle{\delta_{\alpha_{n_{_{\mathcb{X}}}}  ,1}}\displaystyle{\delta_{b_{n_{_{\mathcb{X}}}},1}}}\cdots(-1)^{\displaystyle{\delta_{\alpha_2,1}}\displaystyle{\delta_{b_2,1}}}(-1)^{\displaystyle\delta_{\alpha_1,1}\displaystyle{\delta_{b_{1},1}}}\kettwo{b_{n_{_{\mathcb{X}}}}\cdots b_2b_1}\bigg)\bigg]\bigg\} = \notag \\
    &\sqrt{\frac{1}{2^{n_{_{\mathcb{X}}}}}}\exp \bigg\{-\frac{i}{2}\sum_{b_1, b_2, \cdots, b_{n_{_{\mathcb{X}}}} \in \{0, 1\}}\notag \\
    &\hspace{0.5in}\times \bigg[ \sum_{\alpha_1, \alpha_2, \cdots, \alpha_{n_{_{\mathcb{X}}}} \in \{0, 1\}}\bigg(\bigg(\sqrt{\frac{1}{2h^2}}\boldsymbol{\vartheta}_{\alpha_{n_{_{\mathcb{X}}}}, \cdots, \alpha_2, \alpha_1}\cdot\boldsymbol{x}'\bigg)\bigg(\prod_{l=1}^{n_{_{\mathcb{X}}}}(-1)^{\displaystyle\delta_{\alpha_l,1}\displaystyle{\delta_{b_{l},1}}}\bigg)\bigg)\kettwo{b_{n_{_{\mathcb{X}}}}\cdots b_2b_1}\bigg]\bigg\} = \notag \\
    & \sqrt{\frac{1}{2^{n_{_{\mathcb{X}}}}}}\exp \bigg\{i\sum_{b_1, b_2, \cdots, b_{n_{_{\mathcb{X}}}} \in \{0, 1\}}\notag \\
    &\hspace{0.5in}\times \bigg[ \sqrt{\frac{1}{2h^2}}\bigg(\sum_{\alpha_1, \alpha_2, \cdots, \alpha_{n_{_{\mathcb{X}}}} \in \{0, 1\}}\bigg(\frac{1}{2}(-1)\prod_{l=1}^{n_{_{\mathcb{X}}}}(-1)^{\displaystyle\delta_{\alpha_l,1}\displaystyle{\delta_{b_{l},1}}}\boldsymbol{\vartheta}_{\alpha_{n_{_{\mathcb{X}}}}, \cdots, \alpha_2, \alpha_1}\bigg)\cdot\boldsymbol{x}'\bigg)\kettwo{b_{n_{_{\mathcb{X}}}}\cdots b_2b_1}\bigg]\bigg\}. \label{eq: qeff b7}
\end{align}
Note that for each $2^{n_{_{\mathcb{X}}}}$ combination of $\kettwo{b_{n_{_{\mathcb{X}}}}\cdots b_2b_1}$ the expression  
\begin{equation}
    \sum_{\alpha_1, \alpha_2, \cdots, \alpha_{n_{_{\mathcb{X}}}} \in \{0, 1\}}\bigg(\frac{1}{2}(-1)\prod_{l=1}^{n_{_{\mathcb{X}}}}(-1)^{\displaystyle\delta_{\alpha_l,1}\displaystyle{\delta_{b_{l},1}}}\boldsymbol{\vartheta}_{\alpha_{n_{_{\mathcb{X}}}}, \cdots, \alpha_2, \alpha_1}\bigg), \label{eq: summattion QEFF}
\end{equation}
corresponds to a summation of $2^{n_{_{\mathcb{X}}}}-1$ terms because we may ignore the weight vector accounting for the global phase $\boldsymbol\vartheta_{0, 0, \cdots, 0} = \boldsymbol 0$. Also, considering that the distribution $\Phi$ of the weights $  \{\boldsymbol\vartheta_{\alpha_{n_{_{\mathcb{X}}}}, \cdots, \alpha_2, \alpha_1}\}$ is symmetric, the weights of the summation also follow this symmetric arbitrary distribution $  \{\prod(-1)^{\displaystyle\delta_{\alpha_l,1}\displaystyle{\delta_{b_{l},1}}}\boldsymbol\vartheta_{\alpha_{n_{_{\mathcb{X}}}}, \cdots, \alpha_2, \alpha_1}\}\sim\Phi(\mathbf{0}, \frac{4}{(2^{n_{_{\mathcb{X}}}})-1}\mathbf{I}_D)$, with mean $\mathbf{0}$ and covariance matrix $\frac{4}{(2^{n_{_{\mathcb{X}}}})-1}\mathbf{I}_D$. Furthermore, the multivariate central limit theorem guarantees that the sum $S_m = \sum_{i=1}^m -\frac{1}{2}X_i$ of the i.i.d random variables $\{X_1, X_2, \cdots, X_m\}$ obtained from the same arbitrary probability distribution $\Phi(\mathbf{0}, \tau\mathbf{I}_D)$ would be normally distributed from $\mathcal{N}(\mathbf{0}, \frac{m}{4}\tau\mathbf{I}_D)$. Henceforth, if we set the random variables $\{X_1, X_2, \cdots, X_m\}$ to be the $2^{n_{_{\mathcb{X}}}}-1$ weights $\{\prod(-1)^{\displaystyle\delta_{\alpha_l,1}\displaystyle{\delta_{b_{l},1}}}\boldsymbol{\vartheta}_{\alpha_{n_{_{\mathcb{X}}}}, \cdots, \alpha_2, \alpha_1}\}$ we would have that
\begin{equation}
    \boldsymbol{w}_{ b_{n_{_{\mathcb{X}}}}, \cdots , b_2, b_1} = \sum_{\alpha_1, \alpha_2, \cdots, \alpha_{n_{_{\mathcb{X}}}} \in \{0, 1\}}\bigg(\frac{1}{2}(-1)\prod_{l=1}^{n_{_{\mathcb{X}}}}(-1)^{\displaystyle\delta_{\alpha_l,1}\displaystyle{\delta_{b_{l},1}}}\boldsymbol{\vartheta}_{\alpha_{n_{_{\mathcb{X}}}}, \cdots, \alpha_2, \alpha_1}\bigg),
    \label{eq: Pauli-to-canonical QEFF weights}
\end{equation}
would be normally distributed from $\mathcal{N}(\mathbf{0}, \mathbf I_D)$. Hence, we may rewrite Eq. \ref{eq: qeff b7} as, 
\begin{align}
    &\sqrt{\frac{1}{2^{n_{_{\mathcb{X}}}}}}\exp \bigg\{i\sum_{b_1, b_2, \cdots, b_{n_{_{\mathcb{X}}}} \in \{0, 1\}} \bigg[ \sqrt{\frac{1}{2h^2}}\bigg(\boldsymbol{w}_{ b_{n_{_{\mathcb{X}}}}, \cdots , b_2, b_1}\cdot\boldsymbol{x}'\bigg)\kettwo{b_{n_{_{\mathcb{X}}}}\cdots b_2b_1}\bigg]\bigg\} = \notag \\
    &\sqrt{\frac{1}{2^{n_{_{\mathcb{X}}}}}}\sum_{b_1, b_2, \cdots, b_{n_{_{\mathcb{X}}}} \in \{0, 1\}}\exp \bigg\{i\sqrt{\frac{1}{2h^2}}\bigg(\boldsymbol{w}_{ b_{n_{_{\mathcb{X}}}}, \cdots , b_2, b_1}\cdot\boldsymbol{x}'\bigg)\bigg\}\kettwo{b_{n_{_{\mathcb{X}}}}\cdots b_2b_1}=\notag \\
    &\sqrt{\frac{1}{2^{n_{_{\mathcb{X}}}}}}\sum_{k=0}^{2^{n_{_{\mathcb{X}}}} - 1}e^{i\sqrt{\frac{1}{2h^2}}\boldsymbol{w}_k \cdot\boldsymbol{x}'}\kettwo{k_{_{\mathcb{X}}}}.
\end{align}
In the last step, we rewrote the quantum state in terms of the canonical basis in the decimal representation $\{\kettwo{k_{_{\mathcb{X}}}}\}$, where $k = \sum_{l=1}^{n_{_{\mathcb{X}}}}2^{l-1}b_l$, leading to the QRFF mapping which satisfies that $\{\boldsymbol{w}_k\}_{0 \cdots 2^{n_{_{\mathcb{X}}}} - 1}\in \mathbb R^D$ and $\boldsymbol{w}_k \sim \mathcal{N}(\mathbf{0}, \mathbf I_D)$.

\section{\label{app: baselines}Baseline for quantum classification}

Farhi and Neven \cite{farhi} introduced a variational quantum algorithm tailored for binary classification on near-term quantum devices. The authors used 17 qubits for the binary classification of the downsampled 4x4 MNIST handwritten numerical images. This model utilizes parameterized unitary transformations applied to input quantum states, followed by a measurement of a single qubit to infer the predicted label. Their method focused on the classification of classical n-bit data strings mapped to the computational basis, which for the 4x4 pixel images $n = 16$. The algorithm performs the optimization of the parameters via stochastic gradient descent and use the Hinge loss function.

%Schuld et al. Circuit-Centric Quantum Classifiers \cite{schuld2020circuit} propose a circuit-centric quantum classifier (CCQC) that emphasizes low-depth quantum circuits, ensuring that the number of parameters grows only poly-logarithmically with the input size making it suitable for current quantum hardware. Their approach uses amplitude encoding to represent input data as quantum states, which are manipulated through parameterized single and two-qubit gates to generate predictions.
%The model targets supervised classification and uses a hybrid quantum-classical optimization scheme, estimating gradients by slightly altering the circuit structure and minimizing a Hinge type loss function. Experiments are performed with 8 qubits on data sets like MNIST binary classification, Cancer, Sonar, Wine, and Simeion.
%In our baseline implementation, we used 10 qubits to encode (())

Dilip \textit{et al.} \cite{dilip2022data} explored data compression for quantum machine learning by leveraging matrix-product states (MPS) to encode classical data into efficient quantum representations. The method focused on the classification of the 10-classes Fashion-MNIST data set, demonstrating that a MPS can reduce the quantum circuit depth and qubit requirements while maintaining the classification performance. The quantum demonstrations used the FRQI (flexible representation of quantum images) quantum feature map and a Hardware-efficient strategy based on MPS with 11 qubits for the variational ansatz. For training, they minimized the categorical cross entropy loss function using a L2 regularization. The authors claimed that the MPS-based classifier offers a competitive accuracy compared other quantum learning methods based on tensor networks and that it can be implemented on quantum computers with limited resources. Although the algorithm was designed for the classification of the 28x28 pixel images, we adjusted the model for the down-sampled 4x4 pixel images. In our implementation, we used 5 qubits to encode the corresponding $16$-dimensional flattened image vectors using the FRQI quantum map and measured only one qubit for performing the binary classification. As in the original proposal, we used the categorical cross entropy loss.

\section{\label{app: baselines ZZFM}ZZ feature map}

\begin{figure}[H]
\centering
\includegraphics[scale=0.61]{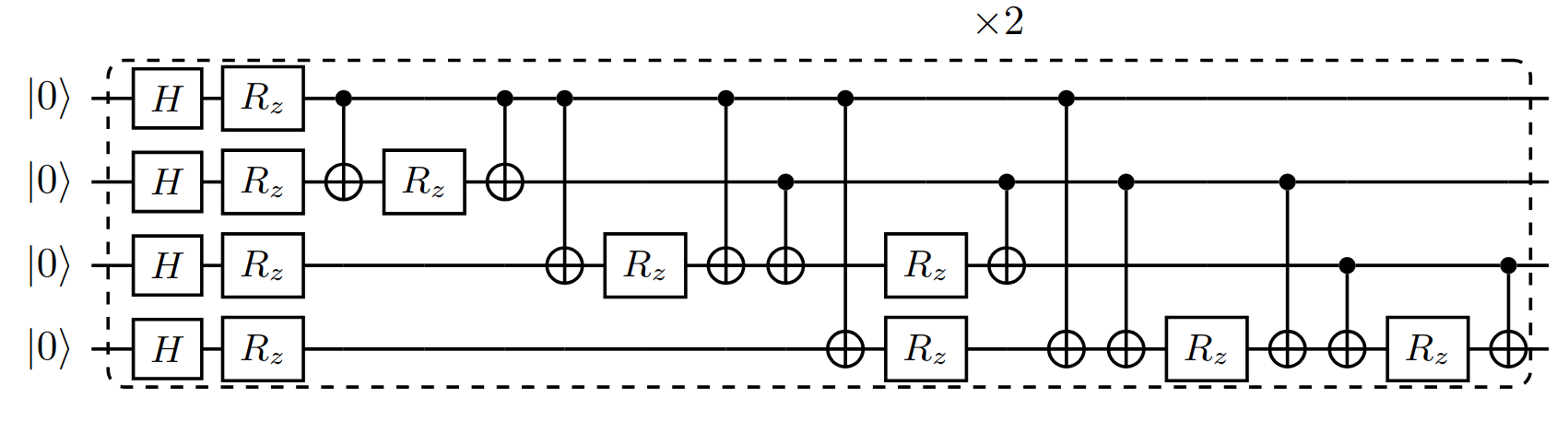}
\caption{Circuit diagram for the $4$-qubit ZZ feature map. The angles of the $R_z$ single gates are not shown for simplicity.}\label{fig: ZZ FeatureMap Quantum circuit}
\end{figure}

Havlicek \textit{et al.} \cite{havlivcek2019supervised} introduced the ZZ feature map, which encodes a data sample $\boldsymbol x' \in [0, 2\pi)^{n_{_\mathcb X}}$ to a $n_{_\mathcb X}$-qubit quantum state $\kettwo{\varphi_{_{\mathcb{X}}}'}\in \mathbb{C}^{2^{n_{_\mathcb X}}}$ corresponding to a second-order evolution along the $z$-axis. The map is given by
\begin{equation}
\kettwo{\varphi_{_{\mathcb{X}}}'} = U_{_\mathcb X}^{\text{Z}}(\boldsymbol{x}')H^{\otimes n_{_{\mathcb{X}}}}U_{_\mathcb X}^{\text{Z}}(\boldsymbol{x}')H^{\otimes n_{_{\mathcb{X}}}}\kettwo{0_{_{\mathcb{X}}}},
\end{equation}
with 
\begin{equation}
U_{_\mathcb X}^{\text{Z}}(\boldsymbol{x}') = \exp\left(i\bigg\{\sum_{i=1}^{n_{_{\mathcb{X}}}} x_i'Z_i + \sum_{j=i+1}^{n_{_{\mathcb{X}}}} \sum_{i=1}^{n_{_{\mathcb{X}}}} (\pi-x_i')(\pi-x_j') Z_i Z_j \bigg\}\right),
\end{equation}
where $Z_i = \big[\bigotimes_{k=i+1}^{n_{_{\mathcb{X}}}}\mathbf{I}_2\big]\otimes\sigma^{z}\otimes\big[\bigotimes_{k=1}^{i-1}\mathbf{I}_2\big]$ is the Pauli-z matrix applied to qubit $i$ and $x_i'$ is the $i^\text{th}$ component of $\boldsymbol x'$. In Fig. \ref{fig: ZZ FeatureMap Quantum circuit}, we illustrate an example of the circuit used to prepare a data sample $\boldsymbol x' \in [0, 2\pi)^4$ using $4$ qubits; for simplicity, the figure does not include the angles $\{x_i'\}, \ \{(\pi -x_i' )( \pi -x_j')\}$ of the $R_z$ rotations.

The repeated application of the circuit $U_{_\mathcb X}^{\text{Z}}(\boldsymbol{x}')H^{\otimes n_{_{\mathcb{X}}}}$ is justified in \cite{havlivcek2019supervised} where they conjectured that the evaluation of inner products generated from circuits with two basis changes and diagonal gates up to an additive error is computationally hard in a classical computer.

\section{\label{app: Model architecture} Neural architecture of the deep quantum-classical generative classification model}    
\label{subsec:architecture}  
The hybrid network architecture, see Table \ref{tab: QEFF architecture}, of the quantum demonstrations with the D-QGC algorithm using the QEFF quantum map had the following sequential components:  
\subsubsection*{Inputs}  
\begin{itemize}  
    \item \textbf{Input layer}: Input layer with the dimension of the input images.
\end{itemize}

\subsubsection*{Encoder}  
\begin{itemize}  
    \item \textbf{Convolutional layer }: A 2D convolutional layer with 20 filters, kernel size 5x5, stride (1,1), and `same' padding.  
    \item \textbf{Average pooling }: An average pooling with a \(2 \times 2\) kernel and stride of 2.  
    \item \textbf{Convolutional layer }: A 2D convolutional layer with 50 filters, 5x5 kernel, stride (1,1), and `same' padding.  
    \item \textbf{Average pooling}: Second average pooling layer with identical \(2 \times 2\) kernel and stride 2.  
    \item \textbf{Flatten}: A flattening of the previous 3D feature map into a 1D vector.
\end{itemize}  

\subsubsection*{Dense feature transformation}  
\begin{itemize}  
    \item \textbf{Dense layer}: Fully connected layer with 84 neurons and ReLU activation for non-linear feature transformation.  
    \item \textbf{Dense layer}: Intermediate dense layer to compress the features to 30 dimensions with a ReLU activation.  
    \item \textbf{Dense layer}: Dense layer to compress the features to a latent space of dimension 4. This last classical layer has no activation.
\end{itemize}  

\subsubsection*{Quantum layer}  
\begin{itemize}  
    \item \textbf{Quantum-enhanced Fourier features layer}:
    The QEFF encoding layer maps the 4-dimensional latent space to a 16-dimensional complex vector using 4 qubits.
    \item \textbf{QGC quantum layer}: This quantum layer integrates the QEFF mapping and the hardware efficient ansatz \cite{Kandala2017HEA} using a total of $n_{_{\mathcb{T}}} = 9$ qubits (\(n_{_{\mathcb{A}}} = 1\), \(n_{_{\mathcb{X}}} = 4\),  \(n_{_{\mathcb{Y}}} = 4\)) and \(1,\!022\) trainable variational parameters. It maps the 16-dimensional QEFF vector into a $10$-dimensional vector with $2^{n_{_{\mathcb{Y}}} - 1} < 10 \le 2^{n_{_{\mathcb{Y}}}}$ representing the joint probability density of the data sample over all the possible classes.  
\end{itemize}
\begin{table}[h!]
  \centering
  \caption{Sequential neural architecture of the D-QGC model using the QEFF encoding. The deep neural network maps the 28$\times$28 pixel images to a latent space of dimension 4, which then feeds the QGC algorithm. The final quantum layer predicts a 10-dimensional vector corresponding to the joint probabilities of a given data sample over all the possible labels. We illustrate the output shape of each classical and quantum layer, the number of trainable and non-trainable parameters (Num. params.), and the activation functions. In addition, ‘no activation’ is used to emphasize that there was no activation in the last classical layer and on the quantum layers, and ‘None’ in the output shape denotes the batch dimension.}
  \label{tab: QEFF architecture}
    \begin{tabular}{llll}
      \toprule
      Layer & Output shape & Num.\ params. & Activation \\
      \midrule
      Input layer            & (None, 28, 28)       & 0              & -- \\
      Convolutional          & (None, 28, 28, 20)   & 520            & ReLU \\
      Average pooling        & (None, 14, 14, 20)   & 0              & -- \\
      Convolutional          & (None, 14, 14, 50)   & 25,050         & ReLU \\
      Average pooling        & (None, 7, 7, 50)     & 0              & -- \\
      Flatten                & (None, 2450)         & 0              & -- \\
      Dense layer            & (None, 84)           & 205,884        & ReLU \\
      Dense layer            & (None, 30)           & 2,550          & ReLU \\
      Dense layer            & (None, 4)            & 124            & No activation \\
      QEFF quantum layer     & (None, 16)           & 60 (non-trainable) & No activation \\
      QGC quantum layer      & (None, 10)           & 1,026          & No activation \\
      \midrule
      Total params.          & --                   & 235,214        & -- \\
      Trainable params.      & --                   & 235,154        & -- \\
      Non-trainable params.  & --                   & 60             & -- \\
      \bottomrule
    \end{tabular}%
\end{table}

As a baseline, we also implemented the same previous neural architecture of the D-QGC algorithm but used the ZZFM quantum mapping instead of the QEFF, see Table \ref{tab: ZZFM architecture}. The ZZFM layer also maps the 4-dimensional latent space to a 16-dimensional complex vector using 4 qubits, which then feeds the QGC quantum circuit. 

\begin{table}
  \centering
  \caption{Sequential architecture of the baseline D-QGC model which uses the ZZFM. The deep neural network maps the 28$\times$28 pixel images to a latent space of dimension 4, and this feature vector is then mapped to the QGC quantum layer. This final layer predicts a 10-dimensional vector corresponding to the joint probabilities of a given data sample over all the possible classes. We illustrate the output shape, the number of parameters (Num. params.), and the activation functions of each classical and quantum layer. Here, we use ‘no activation’ to indicate that there was no activation in the last three layers and ‘None’ in the output shape to show the additional dimension corresponding to the batch.}
  \label{tab: ZZFM architecture}
    \begin{tabular}{llll}
      \toprule
      Layer & Output shape & Num.\ params. & Activation \\
      \midrule
      Input layer            & (None, 28, 28)     & 0              & -- \\
      Convolutional          & (None, 28, 28, 20) & 520            & ReLU \\
      Average pooling        & (None, 14, 14, 20) & 0              & -- \\
      Convolutional          & (None, 14, 14, 50) & 25,050         & ReLU \\
      Average pooling        & (None, 7, 7, 50)   & 0              & -- \\
      Flatten                & (None, 2450)       & 0              & -- \\
      Dense layer            & (None, 84)         & 205,884        & ReLU \\
      Dense layer            & (None, 30)         & 2,550          & ReLU \\
      Dense layer            & (None, 4)          & 124            & No activation \\
      ZZFM quantum layer     & (None, 16)         & 0              & No activation \\
      QGC quantum layer      & (None, 10)         & 1,026          & No activation \\
      \midrule
      Total params.          & --                 & 235,154        & -- \\
      Trainable params.      & --                 & 235,154        & -- \\
      Non‑trainable params.  & --                 & 0              & -- \\
      \bottomrule
    \end{tabular}%
\end{table}

\end{document}